\documentclass[twocolumn]{aastex701}

\usepackage{amsmath}
\graphicspath{{./}{./figures/}}

\newcommand{\nustar}{\textit{NuSTAR}}
\newcommand{\xmm}{\textit{XMM-Newton}}
\newcommand{\chandra}{\textit{Chandra}}
\newcommand{\ixpe}{\textit{IXPE}}
\newcommand{\xrism}{\textit{XRISM}}
\newcommand{\rxte}{\textit{RXTE}}

\newcommand{\swift}{\textit{Swift}}
\newcommand{\suzaku}{\textit{Suzaku}}
\newcommand{\msun}{M_{\odot}}
\newcommand{\risco}{R_{\rm ISCO}}
\newcommand{\rin}{R_{\rm in}}
\newcommand{\afe}{A_{\rm Fe}}

\newcommand{\fcov}{f_{\rm cov}}
\newcommand{\kte}{kT_{e}}
\newcommand{\ktin}{kT_{\rm in}}
\newcommand{\logxi}{\log\xi}
\newcommand{\logn}{\log n_e}

\shorttitle{NuSTAR Spectroscopy of Cygnus X-1}
\shortauthors{Duraphe et al.}

\begin{document}

\title{A NuSTAR Reflection-Spectroscopy Survey of Cygnus X-1}

\author{Kshitij Duraphe}
\affiliation{Boston University, 725 Commonwealth Avenue, Boston, MA 02215, USA}
\email[show]{kshitijd@bu.edu}

\author[orcid=0009-0002-6037-4613]{Kartik Mandar}
\affiliation{Raman Research Institute, Bangalore}
\affiliation{Indian Institute of Science Education and Research Bhopal}
\email{kartik4321mandar@gmail.com}

\author[orcid=0000-0002-0705-6619]{Gopal Bhatta}
\affiliation{Janusz Gil Institute of Astronomy, University of Zielona G\'ora,\\
ul. Szafrana 2, 65-516 Zielona G\'ora, Poland}
\email[show]{g.bhatta@ia.uz.zgora.pl}

\author[orcid=0009-0009-1218-7451]{Chooda Khanal}
\affiliation{Miami Dade College, Miami, FL 33176, USA}
\affiliation{Florida International University, Miami, FL 33199, USA}
\email[show]{ckhanal@fiu.edu}

\begin{abstract}
Relativistic-reflection measurements of Cygnus X-1 disagree on the extent of disk truncation and commonly infer supersolar iron abundances. We analyze a selected sample of 26 archival \nustar{} observations obtained between 2012 and 2024 using two configurations from one reflection-model family, with posterior modes sampled by preconditioned sequential Monte Carlo. In the baseline recovered modes, disk-surface ionization increases with photon index (Pearson $r=+0.59$), with state medians rising from $\logxi\approx3.3$ in the hard state to $\approx3.9$ in the soft state. The corresponding free-emissivity fits give median inner radii of 6.6, 4.4, and $4.0,\risco$ in the hard, intermediate, and soft states. Fixing $q=3$ moves three of eight soft-state observations to the ISCO and one to $2.4,\risco$; four fixed-$q$ fits have lower $\chi^2$ than the sampled free-$q$ solutions, showing that those runs missed higher-likelihood regions. The inferred radii are therefore model- and mode-dependent, and the spectra neither require nor exclude an ISCO disk or $\rin\gtrsim20,\risco$. Baseline fitted abundances span $\afe=1.6$--8.6, with a median of 4.9. Two observations separated by 7.2 hr yield $\afe=1.9\pm0.2$ and $4.5\pm1.0$, indicating that fitted abundance is not a direct composition measurement. Fixing $\afe=1.6$ drives some densities toward the grid boundary and worsens the fits relative to $\afe=4.5$. The wind parameters remain sensitive to the continuum, abundance, and orbital-phase sampling.
\end{abstract}


\section{Introduction} \label{sec:intro}

The optical identification of the bright Cygnus X-ray source with the O9.7\,Iab supergiant HDE\,226868 provided the first strong dynamical evidence for a stellar-mass black hole \citep{Bolton1972, WebsterMurdin1972}. Cygnus\,X-1 is a persistent galactic high-mass X-ray binary (HMXB) and has been monitored extensively for five decades. Current measurements give a black hole mass of $M_{\rm BH} = 21.2 \pm 2.2\,\msun$, a distance of $d = 2.22 \pm 0.18$\,kpc, and an orbital inclination of $27\fdg5 \pm 0\fdg8$ \citep{Orosz2011, MillerJones2021}. Unlike the transient systems that constitute most known black-hole X-ray binaries, Cygnus\,X-1 persistently accretes from the focused wind of its supergiant companion. Repeated observations can therefore sample the accretion flow in different spectral states.

The broadband X-ray spectrum of Cyg\,X-1 has two canonical states and a range of intermediate spectra \citep{RemillardMcClintock2006, Done2007}. In the low/hard state, thermal Comptonization of soft seed photons by a hot ($\kte \sim 50$--200\,keV), optically thin electron population produces a power-law continuum with photon index $\Gamma \sim 1.4$--1.8 and a high-energy rollover \citep{SunyaevTitarchuk1980, Zdziarski1996, PoutanenSvensson1996, Zdziarski2020}. In the high/soft state, quasi-thermal emission from a geometrically thin, optically thick accretion disk dominates below $\sim 5$\,keV \citep{ShakuraSunyaev1973}, and a steeper power-law tail ($\Gamma \gtrsim 2.4$) accounts for the residual hard emission. The source spends most of its time in the hard state but makes transitions to the soft state over weeks to months \citep{Grinberg2013}. Whether the cool disk is truncated in the hard state or extends to the innermost stable circular orbit (ISCO) remains uncertain \citep{Esin1997, Done2007}.

All-sky monitors have provided daily coverage of Cyg\,X-1 for three decades. These include the All-Sky Monitor on \rxte{} \citep{Levine1996}, the Monitor of All-sky X-ray Image \citep{Matsuoka2009}, and the Burst Alert Telescope on \swift{} \citep{Krimm2013}. \citet{Grinberg2013} used these light curves to define operational hard, intermediate, and soft states, and \citet{Deka2021} measured the long-term flux distributions from RXTE-ASM and MAXI data. High-resolution \chandra/HETG spectra have characterized the focused stellar wind and its orbital dependence \citep{Hanke2009, Miskovicova2016, Hirsch2019}. \xmm{}/EPIC observations have measured the orbital and clump-driven variability of the wind absorber \citep{Grinberg2015, Lai2022, Lai2024}, and \xrism/Resolve has resolved its highly ionized iron absorption structure \citep{Yamada2025}.

The relativistic X-ray reflection spectrum produced when hard coronal photons illuminate the disk surface encodes the structure of the inner accretion flow. \citet{Fabian1989}, working specifically on Cyg\,X-1, first identified inner-disk fluorescent Fe\,K$\alpha$ emission as a probe of strong gravity; the modern formalism couples ionized atomic-physics reflection grids \citep{RossFabian2005, Garcia2014} with a relativistic transfer function that imprints gravitational redshift, Doppler boosting, and light bending onto the reflected spectrum \citep{Dauser2014, Dauser2016}. The signatures most accessible to a non-imaging hard-X-ray telescope are the broad, skewed Fe\,K$\alpha$ line at 6.4--6.97\,keV and the Compton backscattering hump peaking near 20--30\,keV; jointly fitting these features with the underlying Comptonized continuum constrains the inner-disk radius, the surface ionization $\logxi$, the iron abundance $\afe$, the disk inclination, and (subject to assumptions about the coronal emissivity profile) the dimensionless spin $a_*$ of the black hole \citep{Garcia2014, Dauser2014, Dauser2016}. Reflection models must also account for Comptonization of the reflected emission by the corona, an effect routinely included in recent versions of the \texttt{relxill} family \citep{Steiner2017}.

The \textit{Nuclear Spectroscopic Telescope Array} \citep[\nustar;][]{Harrison2013} provides focused spectroscopy over 3--79\,keV with sufficient effective area to separate the continuum and reflection components in bright black-hole binaries. Studies of Cyg\,X-1 have combined \nustar{} with \suzaku{} to measure soft-state reflection \citep{Tomsick2014}, constrained the inner disk in the hard state \citep{Parker2015}, reported a disk at the ISCO across several soft-state epochs \citep{Walton2016}, and reported hard-state disk truncation \citep{Basak2017}. A joint \nustar{} and \ixpe{} analysis has also tested gravitational light bending in the soft state with spectroscopy and polarimetry \citep{Steiner2024}. The inferred inner radii range from the ISCO to tens of gravitational radii and depend strongly on the adopted model.

Three unresolved issues motivate this work. First, the truncated-disk model of \citet{Esin1997} predicts that $\rin$ increases as the spectrum hardens and the inner disk is replaced by an advection-dominated flow. Reflection analyses by \citet{Tomsick2014}, \citet{Parker2015}, and \citet{Walton2016} instead place the disk near the ISCO in both the hard and soft states. \citet{Basak2017} report hard-state disk truncation under their adopted continuum and reflection model. \citet{Zdziarski2024} show that soft-state disk and spin inferences depend strongly on the adopted continuum and disk-atmosphere treatment, while \citet{Nosirov2025} quantify related systematic effects.

Second, reflection studies of Cyg\,X-1 infer supersolar iron abundances, typically $A_{\rm Fe}\sim4$--10 \citep{Walton2016,Parker2015,Tomsick2018}, well above the companion-star abundance. This discrepancy may result from the assumed disk density, conventionally fixed at $\logn=15$. Fits with higher densities generally yield lower iron abundances \citep{GarciaDauser2018,Jiang2019,Liu2023,Ding2024,Tomsick2018}, but this has not been tested across the full range of Cyg\,X-1 spectral states.

Third, different coronal geometries can produce similar time-averaged emissivity profiles, so reflection spectra do not distinguish uniquely among compact lamp-post, extended, and truncated-disk coronae \citep{PoutanenSvensson1996,Steiner2017}. \ixpe{} polarimetry favors a corona extended in the disk plane and measures changes through state transitions \citep{Krawczynski2022,Jana2024,Steiner2024}. Reverberation analyses constrain the coronal size and disk--corona separation from X-ray timing measurements \citep{Mastroserio2020,ONeill2025}.

The assumed black hole spin also affects the ISCO location and relativistic line profile. Continuum-fitting studies of the soft state favor $a_*>0.95$ \citep{Gou2011,Zhao2021}, whereas reflection analyses yield a wider range that depends on the coronal geometry and disk density \citep{Draghis2023,Zdziarski2024}. Recent high-resolution XRISM spectroscopy of Cyg\,X-1 reinforces this point: \citet{2026ApJ..1005L..78Z} show that the spin inferred from the Fe\,K band shifts substantially with the adopted continuum and reflection prescription, and \citet{2026NewAR.10201746Z} review the resulting tension between spins measured in X-ray binaries and those inferred from gravitational-wave sources. We fix the spin and report inner-disk radii in units of $\risco$; we do not attempt to measure the spin.

The focused, clumpy wind of HDE\,226868 produces variable ionized absorption along the line of sight \citep{Hanke2009,Miskovicova2016,Hirsch2019,Grinberg2015,Lai2022,Lai2024,Yamada2025}. This absorption must be included when fitting the reflection spectrum. Because \nustar{} cannot resolve individual wind lines, we use an ionized partial-covering absorber \citep{Reeves2008} with parameters $(N_{\rm H,wind},\logxi_{\rm wind},f_{\rm cov,wind})$.

Previous reflection analyses have considered individual observations and used different Comptonization prescriptions, reflection grids, and absorber models. These differences complicate comparisons among spectral states. Here we apply one model family to a selected sample of 26 archival \nustar{} observations of Cyg\,X-1 satisfying the quality criteria in Section~\ref{sec:obs}. We test whether the inner disk recedes as the spectrum hardens, whether high-density reflection models reduce the iron-abundance discrepancy, and whether the fitted wind properties vary with spectral state.

\section{The \nustar{} bandpass} \label{sec:nustar}

\nustar{} has two co-aligned focal plane modules, FPMA and FPMB, whose relative calibration is accurate to a few percent \citep{Madsen2015}. Its 3--79\,keV bandpass covers the soft-disk shoulder, the Fe\,K complex, the 20--30\,keV Compton hump, and part of the continuum rollover. These features constrain the reflection fraction, inner-disk radius, inclination, and coronal electron temperature. Cygnus\,X-1 remains below the bright-source limit in the standard observing mode, and its data can be processed without a separate pile-up correction.

The 3\,keV lower bound lies above the peak of the cool disk in Cygnus\,X-1. The fitted disk temperature $\ktin$ and normalization therefore depend on continuum extrapolation, and the soft-state fits are constrained mainly by the reflection spectrum rather than by the disk peak. In several observations, the uncertainty on $\kte$ extends toward the 300\,keV upper bound because the rollover lies above the observed band. A tear in the FPMA thermal blanket in 2017 introduced a calibration excess at 3--5\,keV \citep{Madsen2020}. We retain the full band and model the residual smooth calibration difference with an energy-dependent detector correction (Section~\ref{sec:obs}). The standard pipeline applies the dead-time correction required at the observed count rates \citep{Bachetti2015}. Finally, \nustar{} does not resolve the wind absorption lines measured with \chandra/HETG \citep{Hanke2009} and \xrism/Resolve \citep{Yamada2025}, so we model the wind as a smooth ionized partial-covering absorber.

\section{Observations} \label{sec:obs}

\subsection{Dataset}

We selected public \nustar{} observations of Cygnus\,X-1 obtained from the launch of the mission through 2024 July. We required an archive issue flag of 0, usable spectra from both focal-plane modules, and a net FPMA exposure greater than 2\,ks. The archive search returned 27 observations with issue flag 0; we excluded ObsID 80902318003, a 0.212\,ks exposure, leaving the 26 observations listed in Table~\ref{tab:obslog}.

\begin{deluxetable}{lccc}
\tablecaption{Observation log for the 26 \nustar{} observations analyzed in
this work, the same dataset as \citet{Duraphe2026}. Observations are grouped
by the baseline photon-index state assignment. Net exposure is the deadtime-corrected
FPMA livetime of the fitted spectrum; FPMB agrees to within a few percent.
The state is assigned from the photon index ($\Gamma < 1.7$ hard,
$1.7 \le \Gamma < 2.1$ intermediate, $\Gamma \ge 2.1$ soft). The configuration
column gives the spectral model adopted for each observation:
M1 $=$ \texttt{thcomp}$\otimes$\texttt{diskbb} $+$ \texttt{relxillCp}
(Comptonization-dominated) and M2 $=$ \texttt{diskbb} $+$ \texttt{relxillCp}
(disk-dominated). Both include Galactic and wind absorption and the
detector-calibration term described in Section~\ref{sec:obs}. \label{tab:obslog}}
\tablewidth{0pt}
\tablehead{
\colhead{ObsID} & \colhead{UT Start} & \colhead{Exp.} & \colhead{Config.} \\
 & (date) & (ks) &
}
\startdata
\cutinhead{Hard state ($\Gamma < 1.7$)}
30002150002 & 2016-05-27 & 50.1 & M1 \\
30002150004 & 2016-05-29 & 50.9 & M1 \\
30002150008 & 2016-06-02 & 28.7 & M1 \\
91002320004 & 2024-07-12 & 13.7 & M1 \\
90101020002 & 2016-02-11 & 13.5 & M1 \\
30202032002 & 2016-07-18 & 13.3 & M1 \\
30901039002 & 2024-04-08 & 16.9 & M1 \\
30702017006 & 2022-05-20 & 12.4 & M1 \\
90802013002 & 2022-06-20 & 13.2 & M1 \\
90802013004 & 2022-06-21 & 13.8 & M1 \\
30001011007 & 2014-05-20 & 34.4 & M1 \\
80502335002 & 2019-08-06 & 13.4 & M1 \\
\cutinhead{Intermediate state ($1.7 \le \Gamma < 2.1$)}
30302019004 & 2018-02-08 & 12.7 & M1 \\
30001011005 & 2014-04-29 & 13.5 & M2 \\
30101022002 & 2015-05-27 & 19.9 & M1 \\
30302019006 & 2018-03-26 & 11.0 & M2 \\
30001011011 & 2015-01-19 & 16.8 & M2 \\
80902318002 & 2023-05-24 & 13.5 & M1 \\
\cutinhead{Soft state ($\Gamma \ge 2.1$)}
80902318004 & 2023-06-14 & 9.3 & M2 \\
30302019010 & 2018-05-27 & 8.2 & M1 \\
80502335006 & 2019-11-13 & 11.9 & M1 \\
30302019002 & 2017-11-04 & 9.4 & M2 \\
30302019012 & 2018-08-11 & 12.1 & M1 \\
30001011002 & 2012-10-31 & 10.4 & M2 \\
10014001001 & 2012-11-01 & 4.2 & M2 \\
30001011009 & 2014-10-04 & 20.4 & M2 \\
\enddata
\end{deluxetable}

We reprocess each observation through the \nustar{} Data Analysis Software (\textsc{nustardas}) inside \textsc{heasoft} v6.35.1 \citep{HEASoft2014} with the current CALDB. We run \texttt{nupipeline} with the relaxed event status expression \texttt{STATUS==b0000xxx00xxxx000}, which prevents the standard noise filter from vetoing source counts at the high focal-plane count rates of Cygnus\,X-1 \citep{Bachetti2015}. We extract source spectra from a 120\arcsec{} circular region centered on the source and background spectra from a source-free region on the same detector, generate the redistribution matrix and ancillary response files with \texttt{nuproducts}, and group the source spectra with \texttt{grppha} to a minimum of 20 counts per bin. We add a 1\% systematic uncertainty uniformly at the \textsc{xspec} \citep{Arnaud1996} stage, consistent with the \nustar{} calibration accuracy for bright sources \citep{Madsen2015, Basak2017}. This calibration floor contributes to the parameter uncertainties and reported reduced $\chi^2$ values, but treats the systematic error as independent between channels. It does not capture correlated effective-area errors; a reduced $\chi^2$ near unity and the quoted statistical intervals should therefore not be read as tests of the complete physical model.

We fit both modules over 3--79\,keV in every observation. For the 14 observations obtained after the 2017 FPMA multi-layer-insulation tear, the CALDB effective-area correction reduces but does not remove a smooth low-energy difference between the modules \citep{Madsen2020}. We model this residual with the multiplicative form $\mathcal{C}_j(E)=A_jE^{-\Delta\Gamma_j}$, implemented with \texttt{plabs} for detector $j$. For FPMA, $A_{\rm A}$ is fixed to unity; for FPMB, $A_{\rm B}$ is free over 0.9--1.1. The saved \textsc{xspec} definitions assign nominal soft limits of $-0.05$ to $+0.05$ to the slope parameters and hard limits of $-0.10$ to $+0.10$. The posterior prior follows the hard limits, $\Delta\Gamma_j\in[-0.10,+0.10]$. We also use this energy-dependent correction for the pre-tear ObsID 30001011011, with $\Delta\Gamma_{\rm A}=0$. For the other 11 observations, $\mathcal{C}_{\rm A}=1$ and $\mathcal{C}_{\rm B}$ is an energy-independent normalization free over 0.9--1.1. All astrophysical parameters are linked between FPMA and FPMB. A common-sign change in the FPMA and FPMB slopes can mimic part of the astrophysical photon index, so the results involving $\Gamma$ assume this calibration prescription.

\subsection{Comparison with timing properties} \label{sec:hardness_comparison}

\begin{figure*}[t]
\centering
\includegraphics[width=\textwidth]{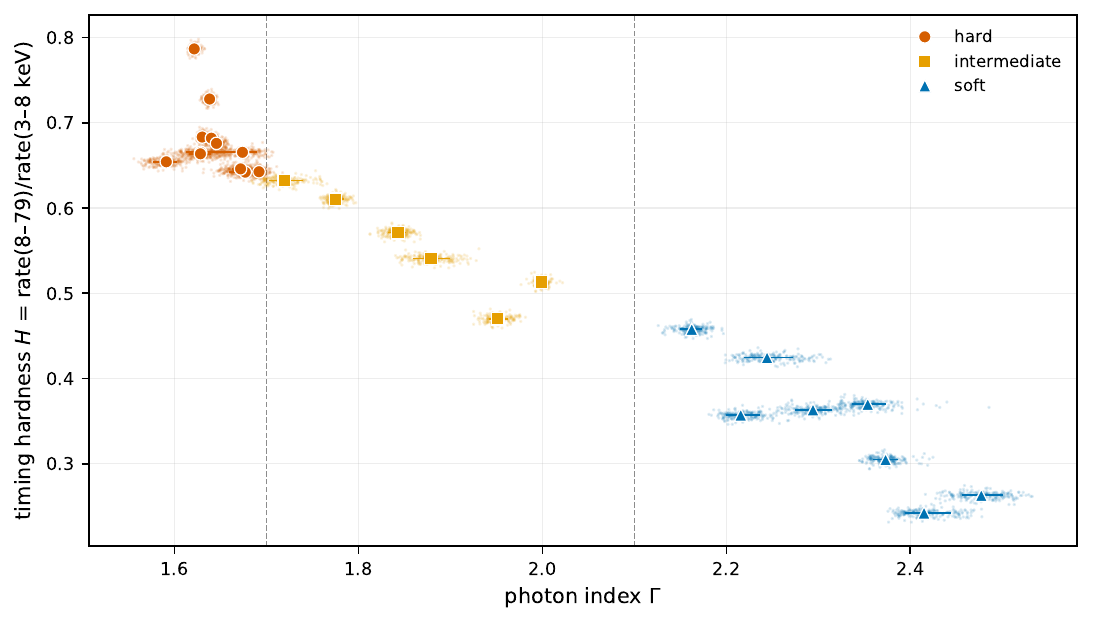}
\caption{Photon index $\Gamma$ against the timing hardness ratio, $H = \mathrm{rate}(8$--$79\,\mathrm{keV}) / \mathrm{rate}(3$--$8\,\mathrm{keV})$, for the 26 \nustar{} observations. Markers show the posterior median of $\Gamma$ and the ratio of the mean 8--79 and 3--8\,keV count rates from the 100\,s light curves. Error bars give the 16th--84th percentile interval for $\Gamma$ and the propagated standard error for $H$ (usually smaller than the marker). For each observation, the light points show 200 posterior samples of $\Gamma$, with a small vertical jitter. Colors indicate the state assigned to each sample. Dashed lines mark the boundaries at $\Gamma=1.7$ and 2.1.}
\label{fig:state_consistency}
\end{figure*}

\citet{Duraphe2026} analyze the timing properties of the same 26 \nustar{} observations and report per-observation hardness ratios $H = \mathrm{rate}(8$--$79\,\mathrm{keV}) / \mathrm{rate}(3$--$8\,\mathrm{keV})$ from 100\,s-binned light curves. Figure~\ref{fig:state_consistency} compares these ratios with the photon indices from our reflection fits. The quantities are strongly anticorrelated, although the ordering is not strictly monotonic for every observation: hard-state observations have $H \approx 0.68$--$0.86$, intermediate observations have $H \approx 0.27$--$0.68$, and soft observations have $H \approx 0.20$--$0.31$. For the ratio of the mean hard- and soft-band rates, the association with the baseline $\Gamma$ values is approximately Pearson $r=-0.95$ and Spearman $\rho_s=-0.92$. We use $\Gamma = 1.7$ and 2.1 as the state boundaries in the analysis below.

\begin{figure*}[t]
\centering
\includegraphics[width=\textwidth]{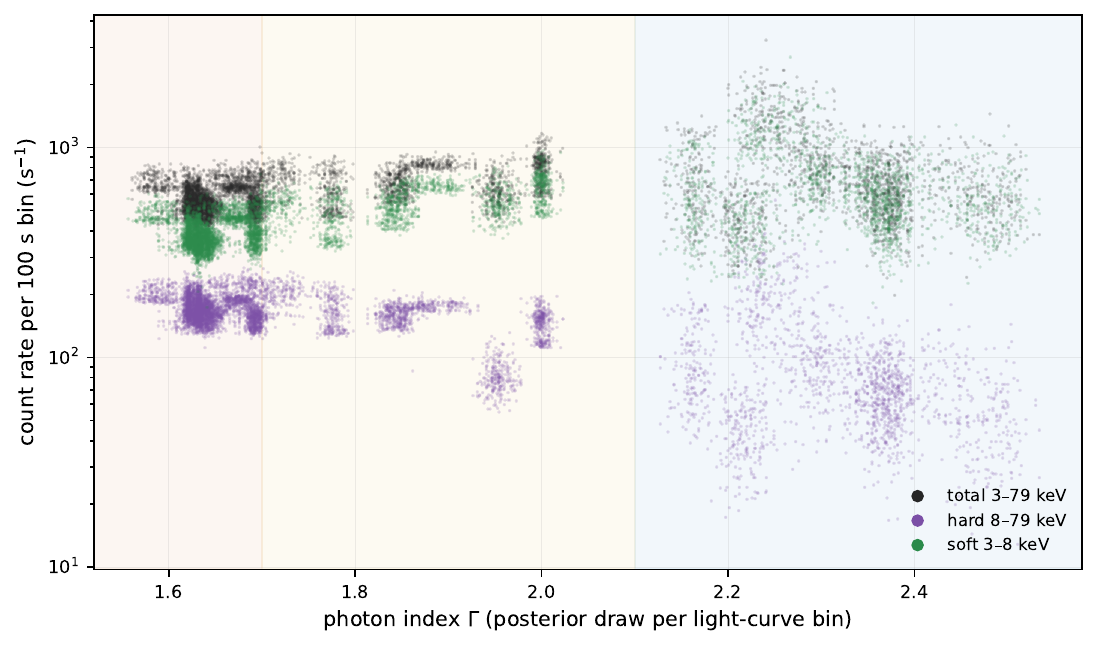}
\caption{\nustar{} count rate per 100\,s light-curve bin against photon index $\Gamma$ for the 26 observations in three bands: total, 3--79\,keV (black); hard, 8--79\,keV (purple); and soft, 3--8\,keV (green). Each rate bin is assigned a $\Gamma$ value drawn from the posterior for its observation. The horizontal width of each distribution represents uncertainty in $\Gamma$, and the vertical width represents variability within the observation. Shading marks the $\Gamma$-based state intervals. From the hard to the soft state, the hard-band rate decreases and the soft-band rate increases, while the total rate changes little. The light curves are from the timing analysis.}
\label{fig:rate_vs_gamma}
\end{figure*}

Figure~\ref{fig:rate_vs_gamma} compares $\Gamma$ with the count rates in the total (3--79\,keV), hard (8--79\,keV), and soft (3--8\,keV) bands. Each light-curve bin is paired with a $\Gamma$ value drawn from the posterior for that observation, giving about 8000 points per band. From the hard to the soft state, the median hard-band rate decreases by a factor of $\sim 2$ and the soft-band rate increases by $\sim 70\%$; the faintest soft-state bins are about ten times fainter in the hard band than the brightest hard-state bins. These changes are consistent with spectral pivoting near 10\,keV \citep{Zdziarski2002, Bock2011}. The total 3--79\,keV rate changes little because the soft band supplies about 85\% of the counts and the changes in the two bands partly cancel.

\begin{figure}[t]
\centering
\includegraphics[width=\columnwidth]{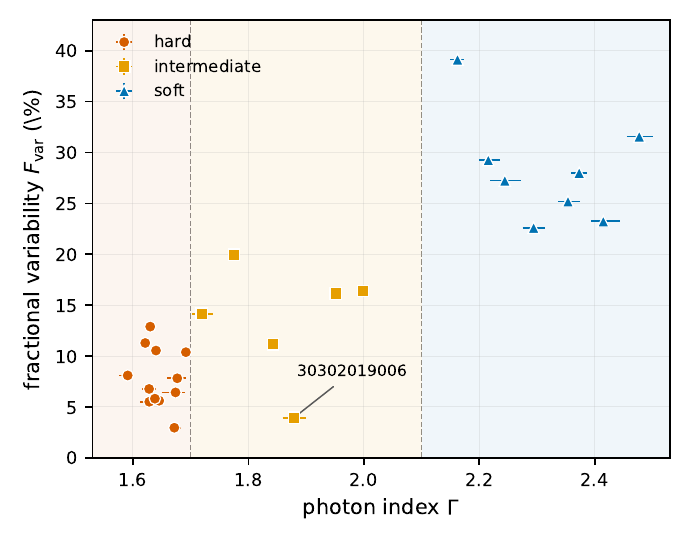}
\caption{Broadband (3--79\,keV) fractional rms against photon index $\Gamma$ for the 26 observations, colored by spectral state. We calculate the excess-variance $F_{\rm var}$ \citep{Vaughan2003} from the 100\,s-binned, barycenter-corrected combined FPMA+FPMB light curves. Horizontal error bars give the 16th--84th percentile interval for $\Gamma$; vertical error bars give the formal measurement-noise uncertainty on $F_{\rm var}$ and are generally smaller than the markers. The soft-state observations have the largest amplitudes. ObsID 30302019006 is labeled because it has the lowest intermediate-state amplitude, $F_{\rm var}=3.94\pm0.08\%$, rather than an upper limit.}
\label{fig:fvar_gamma}
\end{figure}

Figure~\ref{fig:fvar_gamma} compares the broadband (3--79\,keV) fractional rms with the photon index. We calculate $F_{\rm var}$ from the 100\,s-binned, barycenter-corrected combined FPMA+FPMB light curves with the excess-variance estimator of \citet{Vaughan2003}. The quoted errors propagate the measurement noise in that estimator. They do not include residual uncertainty in the count-rate-dependent dead-time correction \citep{Bachetti2015}, so we use these measurements only for a qualitative comparison among states.

The hard-state observations have $F_{\rm var}=3.0$--12.9\%, the intermediate observations have 3.9--19.9\%, and the soft-state observations have 22.6--39.1\%. ObsID 30302019006 has $F_{\rm var}=3.94\pm0.08\%$, the lowest value in the intermediate state, but its excess variance is positive. We do not classify it as a variability non-detection on the basis of these light curves.

\section{Spectral model} \label{sec:model}

We use one \textsc{xspec} model family with two continuum configurations, chosen for each observation from preliminary Levenberg--Marquardt fits. We did not apply a common evidence, information-criterion, posterior-predictive, or cross-validation threshold to this choice. M1 and M2 represent the direct Comptonized continuum and reflection differently, and their use is correlated with spectral state. State comparisons thus also depend on the adopted configuration.

The Comptonization-dominated configuration (hereafter M1) has the form
$\mathcal{C}(E)\times$\allowbreak\,\texttt{TBabs}\,$\times$\allowbreak\,\texttt{zxipcf}\,$\times$\allowbreak\,\texttt{(thcomp}\,$\otimes$\allowbreak\,\texttt{diskbb + relxillCp + zgauss)}.
The photon index and electron temperature of \texttt{thcomp} are linked to the corresponding illuminating-continuum parameters of \texttt{relxillCp}. The \texttt{zgauss} component represents a narrow Fe\,K$\alpha$ emission line where required (Section~\ref{sec:model_components}) and is included in 15 of the 18 M1 fits. M1 is used for all 12 hard-state observations and for 6 of the 14 intermediate- and soft-state observations in which the preliminary LM fit indicates that an explicit Comptonized component is still required, giving a total of 18 observations.

The disk-dominated configuration (hereafter M2) omits the explicit Comptonization component and has the form
$\mathcal{C}(E)\times$\,\texttt{TBabs}$\times$\,\texttt{zxipcf}$\times$\,\texttt{(diskbb + relxillCp)}.
This configuration is used for the remaining eight observations: three intermediate-state and five soft-state observations. In M2, \texttt{relxillCp} supplies both the Comptonized continuum and reflection, in addition to the separate \texttt{diskbb} component. The adopted configuration for each observation is listed in Table~\ref{tab:obslog}. The spectral fits, rather than a threshold in $\Gamma$, determine the configuration. Two soft-state observations (ObsIDs 30302019010 and 80502335006) use M1 but have small \texttt{thcomp} covering fractions ($\sim0.04$--$0.09$); either configuration gives a similar fit in these cases.

\subsection{Model components} \label{sec:model_components}

$\mathcal{C}(E)$ denotes the detector-dependent calibration term described in Section~\ref{sec:obs}; it is either an energy-independent constant or the \texttt{plabs} correction. \texttt{TBabs} \citep{Wilms2000} models the Galactic absorption column along the line of sight. We fix its column density at $N_{\rm H}=0.6\times10^{22}$\,cm$^{-2}$, use the abundance table of \citet{Wilms2000}, and adopt the photoionization cross sections of \citet{Verner1996}. \texttt{zxipcf} \citep{Reeves2008} describes the focused stellar wind as a partially covering ionized absorber with three free parameters: the wind column density $N_{\rm H,wind}$, the ionization parameter $\logxi_{\rm wind}$, and the covering fraction $\fcov$. The absorber redshift is fixed at zero. This three-parameter absorber represents the net attenuation and ionization state but not the line structure resolved with \chandra/HETG (Section~\ref{sec:intro}).

The Comptonized continuum is modeled with \texttt{thcomp}\,$\otimes$\,\texttt{diskbb}. The \texttt{diskbb} model \citep{ShakuraSunyaev1973} describes the multicolor disk emission through the inner-disk temperature $\ktin$ and the disk normalization, while the convolution model \texttt{thcomp} \citep{Zdziarski2020} thermally Comptonizes the seed photons using a corona characterized by the electron temperature $\kte$, asymptotic photon index $\Gamma$, and a covering-fraction parameter that specifies the fraction of disk photons intercepted by the corona. We use \texttt{thcomp} rather than the additive \texttt{nthComp} prescription \citep{Zdziarski1996, PoutanenSvensson1996} because the convolution treatment provides a self-consistent connection between the corona and its seed-photon source and avoids double-counting the disk emission.

The relativistic reflection component is modeled with \texttt{relxillCp} \citep{Garcia2014, Dauser2014, Dauser2016}, with the operating mode depending on the adopted configuration. In M1 the direct continuum is supplied by \texttt{thcomp}, and \texttt{relxillCp} is therefore used in reflection-only mode (\texttt{refl\_frac} $<0$), contributing only the reflected spectrum and avoiding duplication of the coronal continuum \citep{Steiner2017, Dauser2016}. In M2 no explicit \texttt{thcomp} component is present, so \texttt{relxillCp} is used in its standard self-consistent mode (\texttt{refl\_frac} $>0$), in which the model provides both the Comptonized continuum and the associated reflection. In this configuration, \texttt{refl\_frac} is a free parameter, with fitted values ranging from approximately $0.4$ to $5$. In both configurations we use the extended high-density reflection grid of \citet{Ding2024}, which extends the disk-density range from the standard \texttt{relxill} limit of $\logn=15$ to $\logn=20$. This extension allows the strength of the Fe\,K emission to vary with disk density rather than being accommodated primarily through the iron abundance \citep{GarciaDauser2018, Jiang2019, Liu2023}. In the baseline fits, the free reflection parameters are the inner radius $\rin$ (reported in units of $\risco$), the emissivity index $q_1$, the photon index $\Gamma$, the electron temperature $\kte$, the disk-surface ionization $\logxi$, the disk density $\logn$, the iron abundance $\afe$, and the component normalization. In M1, $\Gamma$ and $\kte$ are linked to \texttt{thcomp}; in M2, the reflection fraction is also free.

In 15 of the 18 M1 observations, the relativistic reflection model leaves a narrow residual near the Fe\,K$\alpha$ core. Since \texttt{relxillCp} describes only the relativistically broadened reflection from the inner disk, we model this residual with a narrow \texttt{zgauss} component fixed at a rest-frame energy of 6.4\,keV (neutral Fe\,K$\alpha$), with $\sigma=0.01$\,keV and redshift fixed at zero; only the normalization is left free. This component is interpreted as narrow Fe\,K$\alpha$ emission from distant material, such as the outer disk or the stellar wind, consistent with features resolved by \chandra/HETG \citep{Hanke2009} and \xrism/Resolve \citep{Yamada2025}. The disk-dominated M2 observations do not require this component.

We fix the dimensionless black hole spin to $a_*=0.95$, close to the lower end of the range inferred from continuum-fitting measurements \citep{Gou2011, Zhao2021}. We report $\rin$ in units of $\risco$, but these radii remain conditional on the assumed spin because both the conversion between $\risco$ and $r_g$ and the relativistic line profile depend on $a_*$. We do not fit for spin. The outer disk radius is fixed at $400\,r_g$, the outer emissivity index $q_2$ is linked to the inner emissivity index $q_1$, and the emissivity break radius is fixed at $5\,r_g$; the linked indices make the emissivity profile a single power law, so the break location has no effect. The baseline inclination is fixed at $27\fdg5$ and varied only in the $40\degr$ sensitivity analysis.

\section{Fitting} \label{sec:fitting}

Each observation is fitted in two stages. An initial Levenberg--Marquardt (LM) optimization, repeated from multiple starting values of the disk density (the $\logn$ multi-start separates solution branches differing by $\Delta\chi^2 \sim 10^3$), identifies a high-likelihood solution and determines which model configuration (Section~\ref{sec:model}) is used. Posterior distributions are then sampled using the preconditioned sequential Monte Carlo implementation in \texttt{pocoMC} \citep{Karamanis2022}, which anneals a particle ensemble from the prior to the posterior using normalizing-flow preconditioning while simultaneously estimating the Bayesian evidence. We use this approach because conventional ensemble MCMC methods did not efficiently sample the strongly curved, multi-scale parameter degeneracies present in these fits; the resulting sampling difficulties are described in Appendix~\ref{app:mcmc_rescue}.

We use broad bounded priors. In both model configurations, $\Gamma\in[1.2,3.4]$, $\logxi\in[0,4.7]$, and $\logn\in[15,20]$, covering the full \citet{Ding2024} reflection grid. We adopt $\rin\in[1,30]\,\risco$, $\kte\in[20,300]$\,keV, and $N_{\rm H,wind}\in[0.01,50]\times10^{22}$\,cm$^{-2}$. The baseline implementation uses log-uniform priors for the \texttt{diskbb} and \texttt{relxillCp} component normalizations, with effective bounds constructed around the LM solution and clipped to the \textsc{xspec} hard bounds. The wind column and narrow-line normalization use linear-uniform priors. Other thawed parameters use bounded linear priors over their implemented \textsc{xspec} limits. We fix the inclination at $27\fdg5$ \citep{Orosz2011} and repeat the analysis with $i=40\degr$. The implemented limits, including the data-dependent normalization bounds, are part of the model specification and enter the evidence estimates.

Sampling quality is assessed separately for each observation and recovered mode. The baseline status files report effective sample sizes of approximately 2028--2304, while the saved posterior files contain approximately 2442--2820 weighted samples; 2048 is an algorithmic setting, not the exact returned sample count. No automated prior-boundary accumulation is flagged in the baseline free-abundance posteriors, and the reduced chi-squared values span $\chi^2_\nu=0.87$--1.10. In 24 observations the best sampled $\chi^2$ is no larger than the stored LM value. The two exceptions are ObsIDs 30302019012 and 80902318002, for which the best sampled values are worse by approximately 13 and 31. Table~\ref{tab:results} reports intervals for the baseline recovered modes rather than globally exhaustive posterior intervals.

The population correlations below are descriptive statistics of the 26 posterior medians. They are not hierarchical fits and omit temporal clustering, intrinsic scatter, multimodal weights, and within-observation covariance. The observations are clustered epochs rather than independent draws, and orbital phase is uncontrolled. Since $\Gamma$ defines the state groups and also appears on the horizontal axis, the grouping and the plotted variable are statistically linked; group membership can also change when a sensitivity analysis yields a different $\Gamma$.

Several inferred parameters respond strongly to the assumed iron abundance (Section~\ref{sec:afe}). We use $\afe=1.6$ for the low-abundance sensitivity test and $\afe=4.5$, close to the baseline median, for the high-abundance test. \citet{Ramachandran2025} measure the donor-star iron abundance as $1.33^{+0.19}_{-0.15}$ solar, so 1.6 should be understood as a test value rather than the photospheric Fe measurement. These fixed-abundance fits measure model sensitivity; they do not establish a common physical composition.

For all fits we evaluate the model on an extended energy grid using the directive \texttt{energies 0.01 1000.0 1000 log}, included in every \texttt{fit.xcm} prior to sampling. This follows the recommendation of \citet{Zdziarski2020} for the \texttt{thcomp} convolution model. Because Compton scattering redistributes seed photons over energies well beyond the observed bandpass, the model must be evaluated both below the lower edge of the \nustar{} band, so that up-scattered photons contribute within the observed range, and above the upper edge, so that the high-energy rollover associated with the electron temperature is properly represented. The same energy grid is also used by \texttt{relxillCp}, which computes the reflected spectrum before applying relativistic transfer and folding through the instrumental response. Restricting the calculation to the nominal 3--79\,keV band can bias the inferred reflection fraction and disk ionization.

The LM optimization supplies starting points and is not a search for the global minimum. In several observations, sequential Monte Carlo reaches modes hundreds in $\chi^2$ below the local LM solution; one pathological starting solution differs by more than $\Delta\chi^2\sim5000$. Such values diagnose a failed local start and are not model-comparison statistics. Two baseline sampled results, however, remain above their stored LM values, and the fixed-$q$ tests find lower solutions than several sampled free-$q$ fits. The population summaries and the reduced $\chi^2$ values in Table~\ref{tab:results} refer to the modes reached by the baseline analysis.

\begin{deluxetable*}{lccccccccc}
\tablecaption{Posterior summary for all 26 \nustar{} observations
(baseline free-abundance analysis), grouped by spectral state and sorted
by photon index within each group. Values are posterior medians;
uncertainties are weighted 16th--84th percentile intervals conditional
on the mode recovered in the baseline run.
$\chi^2_\nu$ is evaluated at the best recovered baseline sample.
\label{tab:results}}
\tablewidth{0pt}
\tablehead{
\colhead{ObsID} & \colhead{$\Gamma$} & \colhead{$\log\xi$} &
\colhead{$\log n_e$} & \colhead{$R_{\rm in}/R_{\rm ISCO}$} &
\colhead{$A_{\rm Fe}$} & \colhead{$N_{\rm H,wind}$} &
\colhead{$\log\xi_{\rm wind}$} & \colhead{$f_{\rm cov,wind}$} &
\colhead{$\chi^2_\nu$} \\
& & (erg cm s$^{-1}$) & (cm$^{-3}$) & & ($A_\odot$) &
($10^{22}$ cm$^{-2}$) & (erg cm s$^{-1}$) & &
}
\startdata
\cutinhead{Hard state ($\Gamma < 1.7$)}
90802013002 & $1.591^{+0.012}_{-0.015}$ & $4.062^{+0.064}_{-0.072}$ & $16.82^{+0.74}_{-0.63}$ & $13.5^{+5.3}_{-3.5}$ & $4.32^{+0.81}_{-0.63}$ & $12.9^{+3.3}_{-2.1}$ & $2.186^{+0.094}_{-0.085}$ & $0.342^{+0.053}_{-0.051}$ & 0.90 \\
30002150004 & $1.621^{+0.004}_{-0.003}$ & $3.283^{+0.088}_{-0.103}$ & $16.99^{+0.93}_{-0.82}$ & $4.70^{+1.33}_{-0.85}$ & $4.60^{+0.35}_{-0.35}$ & $3.7^{+5.3}_{-1.7}$ & $2.92^{+0.18}_{-0.24}$ & $0.36^{+0.16}_{-0.15}$ & 0.91 \\
91002320004 & $1.628^{+0.012}_{-0.013}$ & $3.33^{+0.61}_{-0.15}$ & $17.42^{+1.11}_{-0.85}$ & $6.2^{+4.7}_{-1.4}$ & $4.81^{+0.41}_{-0.68}$ & $20.6^{+4.6}_{-6.6}$ & $3.139^{+0.098}_{-0.629}$ & $0.186^{+0.103}_{-0.049}$ & 0.92 \\
30901039002 & $1.628^{+0.021}_{-0.016}$ & $4.05^{+0.10}_{-0.23}$ & $17.96^{+0.56}_{-0.59}$ & $7.4^{+3.5}_{-2.0}$ & $6.0^{+1.9}_{-1.1}$ & $11.9^{+6.2}_{-3.3}$ & $2.46^{+0.29}_{-0.15}$ & $0.221^{+0.068}_{-0.053}$ & 0.91 \\
30002150002 & $1.630^{+0.003}_{-0.003}$ & $2.849^{+0.062}_{-0.072}$ & $18.20^{+0.53}_{-0.34}$ & $7.0^{+3.2}_{-1.9}$ & $4.21^{+0.47}_{-0.40}$ & $10.1^{+3.1}_{-2.1}$ & $3.171^{+0.042}_{-0.043}$ & $0.470^{+0.090}_{-0.079}$ & 0.90 \\
90101020002 & $1.638^{+0.004}_{-0.004}$ & $3.243^{+0.075}_{-0.161}$ & $17.1^{+1.4}_{-1.1}$ & $4.74^{+0.64}_{-0.56}$ & $5.50^{+0.71}_{-0.46}$ & $5.8^{+3.1}_{-2.4}$ & $3.23^{+0.10}_{-0.12}$ & $0.328^{+0.114}_{-0.092}$ & 0.95 \\
30002150008 & $1.640^{+0.004}_{-0.005}$ & $2.879^{+0.138}_{-0.057}$ & $18.09^{+0.41}_{-0.64}$ & $9.3^{+5.1}_{-2.6}$ & $5.31^{+0.75}_{-0.42}$ & $23.1^{+5.5}_{-4.3}$ & $3.303^{+0.046}_{-0.064}$ & $0.247^{+0.039}_{-0.042}$ & 0.93 \\
30202032002 & $1.646^{+0.006}_{-0.006}$ & $3.07^{+0.12}_{-0.15}$ & $17.4^{+1.2}_{-1.1}$ & $12.9^{+11.1}_{-7.3}$ & $5.01^{+0.93}_{-0.75}$ & $21.7^{+7.1}_{-6.0}$ & $3.202^{+0.079}_{-0.095}$ & $0.253^{+0.070}_{-0.051}$ & 0.91 \\
90802013004 & $1.672^{+0.011}_{-0.010}$ & $3.055^{+0.073}_{-0.113}$ & $18.13^{+0.48}_{-0.41}$ & $5.12^{+1.14}_{-0.97}$ & $4.50^{+0.49}_{-0.44}$ & $13.0^{+7.5}_{-4.8}$ & $3.49^{+0.17}_{-0.12}$ & $0.221^{+0.062}_{-0.055}$ & 0.92 \\
30702017006 & $1.674^{+0.016}_{-0.024}$ & $3.174^{+0.131}_{-0.098}$ & $17.90^{+0.56}_{-0.60}$ & $5.15^{+0.91}_{-0.86}$ & $5.17^{+0.85}_{-0.56}$ & $19.6^{+7.3}_{-7.6}$ & $3.83^{+0.26}_{-0.47}$ & $0.230^{+0.053}_{-0.042}$ & 0.95 \\
80502335002 & $1.677^{+0.016}_{-0.018}$ & $3.77^{+0.14}_{-0.17}$ & $15.95^{+0.61}_{-0.43}$ & $9.4^{+3.5}_{-1.7}$ & $4.71^{+0.94}_{-0.72}$ & $8.9^{+6.0}_{-4.8}$ & $2.84^{+0.13}_{-0.33}$ & $0.328^{+0.165}_{-0.086}$ & 0.92 \\
30001011007 & $1.692^{+0.005}_{-0.005}$ & $3.390^{+0.060}_{-0.063}$ & $17.86^{+0.57}_{-0.71}$ & $5.8^{+2.2}_{-1.8}$ & $5.20^{+0.58}_{-0.36}$ & $4.7^{+2.4}_{-1.4}$ & $3.191^{+0.069}_{-0.080}$ & $0.61^{+0.17}_{-0.14}$ & 0.95 \\
\cutinhead{Intermediate state ($1.7 \le \Gamma < 2.1$)}
30302019004 & $1.719^{+0.020}_{-0.017}$ & $3.470^{+0.042}_{-0.045}$ & $16.32^{+0.79}_{-0.73}$ & $4.74^{+0.79}_{-0.58}$ & $5.28^{+0.58}_{-0.44}$ & $4.9^{+1.8}_{-1.1}$ & $3.225^{+0.053}_{-0.055}$ & $0.653^{+0.100}_{-0.104}$ & 0.98 \\
30001011005 & $1.775^{+0.009}_{-0.007}$ & $3.266^{+0.042}_{-0.042}$ & $17.23^{+0.48}_{-0.47}$ & $4.39^{+0.50}_{-0.44}$ & $4.29^{+0.38}_{-0.31}$ & $15.5^{+12.0}_{-8.1}$ & $4.03^{+0.20}_{-0.21}$ & $0.204^{+0.086}_{-0.062}$ & 0.91 \\
30101022002 & $1.843^{+0.009}_{-0.011}$ & $3.465^{+0.041}_{-0.049}$ & $16.67^{+0.81}_{-0.82}$ & $1.93^{+1.97}_{-0.51}$ & $5.55^{+0.77}_{-0.52}$ & $20.6^{+6.7}_{-5.8}$ & $3.99^{+0.14}_{-0.13}$ & $0.287^{+0.071}_{-0.063}$ & 0.90 \\
30302019006 & $1.879^{+0.021}_{-0.019}$ & $3.570^{+0.045}_{-0.041}$ & $18.10^{+0.55}_{-1.22}$ & $5.01^{+0.62}_{-0.50}$ & $5.30^{+0.66}_{-0.35}$ & $4.9^{+5.4}_{-2.1}$ & $3.91^{+0.28}_{-0.27}$ & $0.34^{+0.14}_{-0.10}$ & 0.91 \\
80902318002 & $1.951^{+0.011}_{-0.010}$ & $3.452^{+0.045}_{-0.048}$ & $19.41^{+0.13}_{-0.12}$ & $4.39^{+0.24}_{-0.27}$ & $6.92^{+0.67}_{-0.58}$ & $2.98^{+1.47}_{-0.96}$ & $3.65^{+0.16}_{-0.11}$ & $0.46^{+0.12}_{-0.10}$ & 0.95 \\
30001011011 & $1.999^{+0.005}_{-0.005}$ & $3.514^{+0.063}_{-0.062}$ & $18.63^{+0.27}_{-0.28}$ & $3.61^{+0.35}_{-0.33}$ & $5.63^{+0.61}_{-0.49}$ & $8.9^{+11.4}_{-5.6}$ & $4.11^{+0.37}_{-0.35}$ & $0.186^{+0.124}_{-0.063}$ & 0.89 \\
\cutinhead{Soft state ($\Gamma \ge 2.1$)}
80902318004 & $2.162^{+0.011}_{-0.012}$ & $4.145^{+0.047}_{-0.061}$ & $16.39^{+0.30}_{-0.31}$ & $3.35^{+0.29}_{-0.27}$ & $8.64^{+0.45}_{-0.66}$ & $23.6^{+9.6}_{-11.8}$ & $3.97^{+0.20}_{-0.23}$ & $0.169^{+0.080}_{-0.058}$ & 0.95 \\
80502335006 & $2.216^{+0.021}_{-0.016}$ & $3.956^{+0.039}_{-0.051}$ & $17.70^{+0.13}_{-0.13}$ & $3.79^{+0.30}_{-0.27}$ & $5.92^{+0.56}_{-0.58}$ & $28.7^{+7.6}_{-10.0}$ & $3.64^{+0.16}_{-0.11}$ & $0.191^{+0.045}_{-0.034}$ & 1.04 \\
30302019010 & $2.244^{+0.028}_{-0.025}$ & $4.435^{+0.093}_{-0.091}$ & $16.46^{+0.12}_{-0.13}$ & $4.8^{+1.4}_{-1.0}$ & $4.43^{+1.20}_{-0.80}$ & $31.2^{+2.9}_{-2.3}$ & $3.240^{+0.018}_{-0.018}$ & $0.773^{+0.032}_{-0.034}$ & 0.98 \\
30302019002 & $2.294^{+0.020}_{-0.019}$ & $3.422^{+0.063}_{-0.064}$ & $18.393^{+0.107}_{-0.097}$ & $3.48^{+0.36}_{-0.32}$ & $1.96^{+0.21}_{-0.19}$ & $9.6^{+8.2}_{-4.4}$ & $3.78^{+0.17}_{-0.17}$ & $0.264^{+0.093}_{-0.079}$ & 0.89 \\
30302019012 & $2.354^{+0.020}_{-0.017}$ & $3.708^{+0.040}_{-0.037}$ & $17.705^{+0.090}_{-0.104}$ & $3.97^{+0.68}_{-0.60}$ & $1.63^{+0.17}_{-0.14}$ & $27.9^{+6.6}_{-10.2}$ & $3.84^{+0.14}_{-0.16}$ & $0.277^{+0.065}_{-0.052}$ & 0.91 \\
30001011009 & $2.373^{+0.013}_{-0.014}$ & $3.689^{+0.046}_{-0.050}$ & $17.375^{+0.105}_{-0.099}$ & $4.03^{+0.47}_{-0.60}$ & $1.68^{+0.17}_{-0.13}$ & $27.0^{+2.7}_{-3.2}$ & $3.475^{+0.098}_{-0.051}$ & $0.309^{+0.030}_{-0.030}$ & 0.87 \\
10014001001 & $2.415^{+0.030}_{-0.021}$ & $4.45^{+0.12}_{-0.13}$ & $16.50^{+0.19}_{-0.14}$ & $5.38^{+0.52}_{-0.51}$ & $4.48^{+1.02}_{-0.97}$ & $38.6^{+3.3}_{-4.1}$ & $3.398^{+0.055}_{-0.046}$ & $0.417^{+0.034}_{-0.036}$ & 1.10 \\
30001011002 & $2.477^{+0.023}_{-0.021}$ & $3.884^{+0.062}_{-0.073}$ & $17.28^{+0.10}_{-0.11}$ & $3.95^{+0.42}_{-0.45}$ & $1.86^{+0.19}_{-0.21}$ & $39.2^{+3.8}_{-6.2}$ & $3.416^{+0.045}_{-0.042}$ & $0.339^{+0.033}_{-0.033}$ & 0.95 \\
\enddata
\tablecomments{$\Gamma$ is \texttt{thcomp.Gamma\_tau} in the M1
configuration and \texttt{relxillCp.gamma} in M2. No baseline posterior
median reaches the ISCO ($R_{\rm in}/R_{\rm ISCO}=1$), but this does not
exclude an ISCO mode. Wind parameters are effective quantities from the
phenomenological partial-covering absorber \texttt{zxipcf} and are
sensitive to the continuum, abundance, and orbital-phase sampling.}
\end{deluxetable*}

\section{Results} \label{sec:results}

\subsection{Representative spectra} \label{sec:rep_spectra}

\begin{figure*}[t]
\centering
\includegraphics[width=\textwidth]{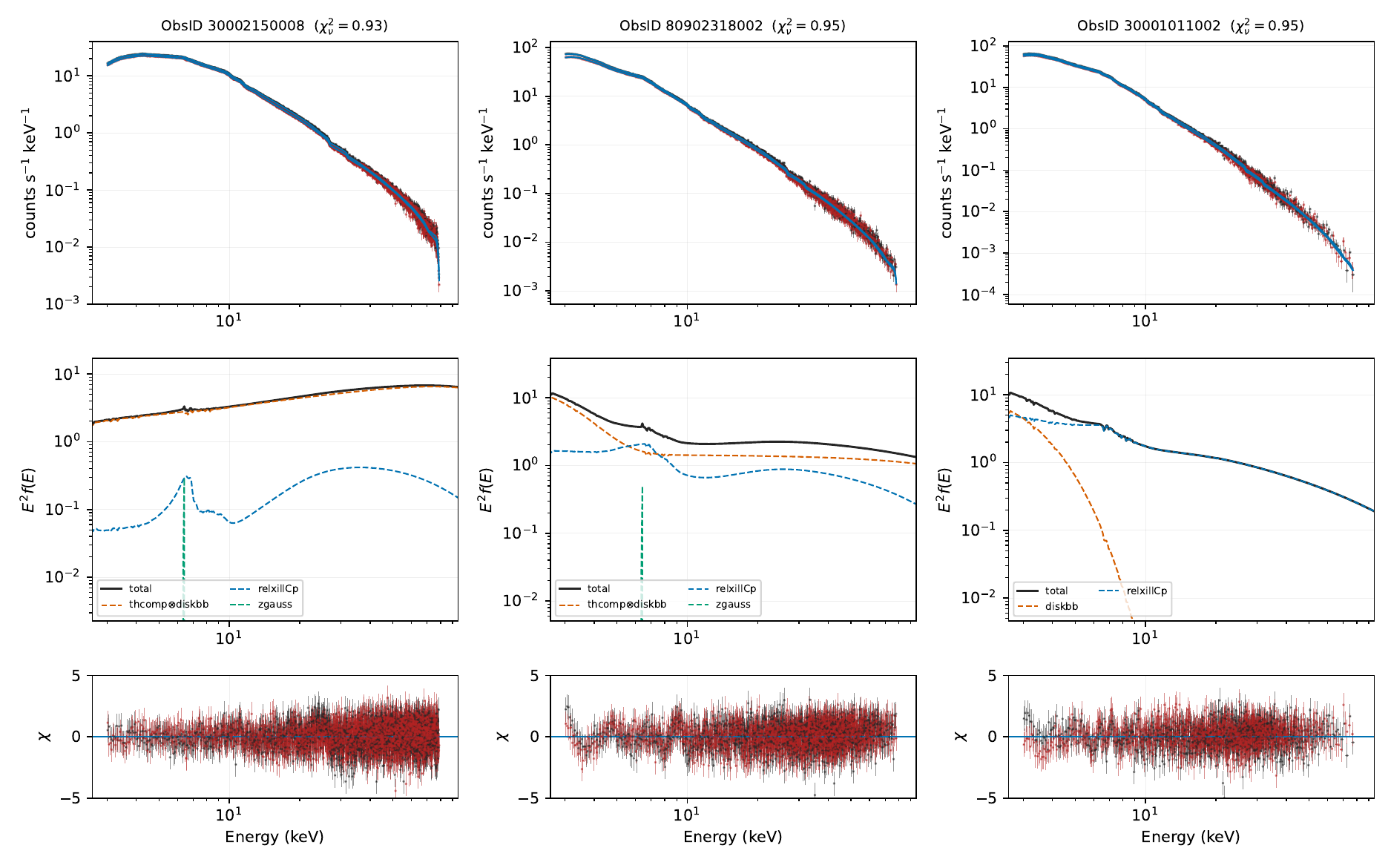}
\caption{Posterior best-fit spectra for three representative observations spanning the state range. \textit{Left column:} hard state, ObsID 30002150008 ($\Gamma = 1.64$). \textit{Middle column:} intermediate state, ObsID 80902318002 ($\Gamma = 1.95$). \textit{Right column:} soft state, ObsID 30001011002 ($\Gamma = 2.48$). \textit{Top row:} folded counts spectrum, FPMA (black) and FPMB (red) with the posterior best-fit model (blue). \textit{Middle row:} unfolded $E^2 f(E)$ decomposition into the model components. \textit{Bottom row:} $\chi$ residuals. The reduced chi-squared values at the posterior best fit are $\chi^2_\nu = 0.93$, $0.95$, and $0.95$. The complete figure set (26 images) is provided in Appendix~\ref{app:perobs}.}
\label{fig:rep_spectra}
\end{figure*}

Figure~\ref{fig:rep_spectra} shows posterior best-fit models for one hard-, intermediate-, and soft-state observation. In the hard-state M1 example, the plotted \texttt{thcomp}$\otimes$\texttt{diskbb} component dominates the 3--30\,keV decomposition, while reflection is subdominant. In M1, this plotted continuum component does not separately identify unscattered disk and Comptonized photons. In the intermediate example, the combined continuum and reflection components both contribute substantially. In the soft-state M2 example, the plotted \texttt{relxillCp} component contains both the direct Comptonized continuum and reflection, so the decomposition does not establish that reflection alone dominates. The reduced chi-squared values are 0.93, 0.95, and 0.95. No large unmodeled broad structure is evident, although mild detector-dependent residual structure remains near 3--5\,keV in the soft example. Appendix~\ref{app:perobs} shows the fits for all 26 observations.

\subsection{Disk-surface ionization and photon index} \label{sec:logxi_gamma}

\begin{figure}[t]
\centering
\includegraphics[width=\columnwidth]{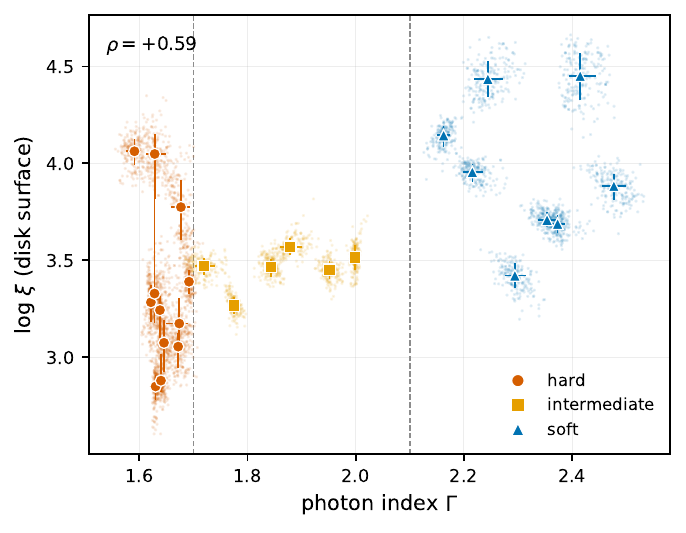}
\caption{Disk-surface ionization $\logxi$ against photon index $\Gamma$ for all 26 observations (free-abundance analysis). Markers show posterior medians with 16th--84th percentile intervals; light points show 200 weighted posterior samples per observation. Colors and shapes denote the $\Gamma$-based state: hard ($\Gamma < 1.7$, vermillion circles), intermediate ($1.7 \le \Gamma < 2.1$, orange squares), and soft ($\Gamma \ge 2.1$, blue triangles). Dashed lines mark the state boundaries. For the 26 baseline medians, Pearson $r=+0.59$ and Spearman $\rho_s=+0.51$; the state median rises from $\logxi\approx3.3$ in the hard state to $\approx3.9$ in the soft state. The statistics describe the recovered modes.}
\label{fig:gamma_logxi}
\end{figure}

Figure~\ref{fig:gamma_logxi} shows a positive cross-observation association between disk-surface ionization and photon index. From the posterior medians, the Pearson coefficients are $r=+0.59$ for the baseline free-abundance analysis, $+0.73$ for $\afe=4.5$, and $+0.78$ for $\afe=1.6$; the corresponding Spearman coefficients are $\rho_s=+0.51$, $+0.55$, and $+0.90$. The baseline state median rises from $\logxi\approx3.3$ in the hard state to $\approx3.9$ in the soft state. The association remains positive in the sampled variants, but is weaker for $i=40\degr$ (Pearson $r\approx+0.46$ and Spearman $\rho_s\approx+0.18$). Changes in incident ionizing illumination, disk density, and coronal geometry could all contribute. The distance in $\xi=4\pi F_{\rm ion}/n$ is the source-to-surface distance, which need not equal the fitted $\rin$, and the observer-directed luminosity does not measure the flux incident on the disk. We do not predict $\xi$ from $\rin$ and an unspecified luminosity. The largest change between the two fixed-abundance fits is 1.88\,dex, or 0.72\,dex after excluding the explicitly multimodal ObsID 30002150002, which motivates an emphasis on state medians rather than individual values.

\subsection{Inner-disk radius across states} \label{sec:rin_gamma}

\begin{figure}[t]
\centering
\includegraphics[width=\columnwidth]{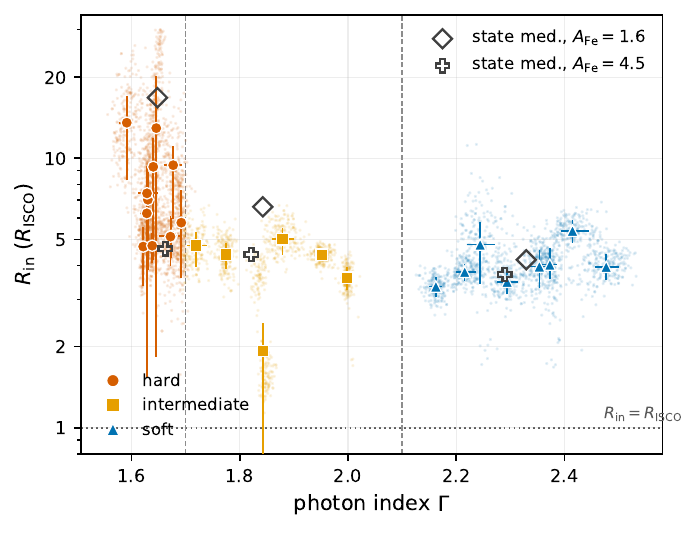}
\caption{Inner-disk radius $\rin/\risco$ against photon index $\Gamma$ for the modes recovered by the free-abundance, free-$q$ fits, on a logarithmic scale. Symbols and posterior samples follow Figure~\ref{fig:gamma_logxi}; the horizontal dotted line marks the ISCO. The median radii are $4.0$, $4.4$, and $6.6\,\risco$ in the soft, intermediate, and hard states, respectively. Open diamonds and crosses show the median for each state in the $\afe=1.6$ and 4.5 fits, plotted at the corresponding median $\Gamma$. The two fixed-abundance hard-state medians differ by a factor of approximately 3.6, compared with factors of approximately 1.5 and 1.1 in the intermediate and soft states; some individual observations shift more. All summaries are conditional on the sampled likelihood modes.}
\label{fig:gamma_rin}
\end{figure}

Figure~\ref{fig:gamma_rin} shows the inferred inner-disk radius as a function of $\Gamma$. Within the modes recovered by the free-abundance, $\afe=1.6$, $\afe=4.5$, and $i=40\degr$ fits, no posterior median is at the ISCO; the smallest baseline median is $\rin=1.9\,\risco$. The baseline state medians decrease from $6.6\,\risco$ in the hard state to $4.4\,\risco$ in the intermediate state and $4.0\,\risco$ in the soft state. The scale changes substantially with the model and recovered mode. The hard-state median ranges from approximately 4.6 to 16.8\,$\risco$ between the fixed-abundance analyses, and the $i=40\degr$ analysis changes some individual radii by factors of several. The soft-state medians are $3.96$, $4.20$, $3.70$, and $2.12\,\risco$ in the baseline, $\afe=1.6$, $\afe=4.5$, and $i=40\degr$ analyses, respectively. Both the ordering and absolute values are therefore conditional on the model and recovered mode. The baseline soft-state median is larger than the near-ISCO soft-state values reported by \citet{Tomsick2014} and \citet{Walton2016}; hard-state truncation measurements and disk models with an inner edge tied to the ISCO are not direct comparisons.

Part of this difference arises from the degeneracy between the emissivity profile and the inner radius. Within each observation, $\rin$ is correlated with the free emissivity index: the $q_1$--$\rin$ posterior correlations range from $+0.64$ to $+0.88$ in the soft state, with median values of $+0.64$ and $+0.77$ in the hard and intermediate states. A steeper emissivity profile concentrates the reflected emission at smaller radii and can compensate for a larger inner radius, so the \nustar{} spectra constrain mainly the combination $(q_1,\rin)$. We define $\Delta\chi^2=\chi^2(q=3)-\chi^2(q\ {\rm free})$. In LM refits of the eight soft-state observations at $q=3$, three move to the ISCO with $\Delta\chi^2=-110$ to $-11$, and a fourth improves by 55 at $\rin=2.4\,\risco$. One moves close to the ISCO at a cost of $+102$, one remains at $3.7\,\risco$ ($+49$), and two give substantially worse fits ($+767$ and $+1296$). Because the $q=3$ model is nested within the free-$q$ model, a fixed-$q$ fit cannot have a lower minimum $\chi^2$ if the global free-$q$ minimum has been found. The four negative values therefore show that the corresponding free-$q$ runs missed a higher-likelihood region. We treat the free-$q$ radii as summaries of the recovered modes, not as globally exhaustive posterior estimates. The fits examined here provide acceptable solutions without requiring either an inner radius at the ISCO or one beyond approximately $20\,\risco$; they do not exclude either geometry.

\subsection{Iron abundance and disk density} \label{sec:afe}
\label{sec:logN}

With disk density free, the \citet{Ding2024} grid gives baseline posterior medians of $\logn=16.0$--19.4, with no baseline median at the upper boundary. The fitted iron abundance is supersolar in most baseline modes: the sample median is $\afe=4.9$ and the range is 1.6--8.6. Across the 26 posterior medians, abundance and density show no clear association: Pearson $r=-0.016$ and Spearman $\rho_s=+0.084$. This population-level result does not exclude a strong density--abundance degeneracy within individual observations. ObsID 30001011002 ended approximately 7.2\,hr before ObsID 10014001001 began; these near-contemporaneous observations give $\afe=1.9\pm0.2$ and $4.5\pm1.0$ when fitted separately. Their bulk composition cannot change over this interval, so the discrepancy shows that fitted $\afe$ is not composition alone. Possible contributors include model inadequacy, calibration, statistical fluctuation, different illumination or absorption conditions, and recovery of different likelihood modes.

In the low-abundance sensitivity analysis, fixing $\afe=1.6$ raises the sample median disk density from $\logn=17.4$ to 18.0, and several posteriors approach the $\logn=20$ grid boundary. Relative to the separate $\afe=4.5$ fits, the median best-fit increase is $\Delta\chi^2=73$ per observation. \citet{Tomsick2018} obtained a solar-abundance fit to one Cygnus\,X-1 epoch with $n_e\sim4\times10^{20}$\,cm$^{-3}$, above the maximum density in the present public grids. This response is consistent with the expected density--abundance trade-off: when the density grid ends too low, a larger fitted $\afe$ can reproduce part of the Fe\,K strength \citep{GarciaDauser2018}. Continuum, illumination, emissivity, calibration, and atomic-model assumptions can produce similar shifts, so these fits do not measure the fraction of the abundance excess attributable to the density ceiling.

\subsection{The stellar wind across states} \label{sec:wind}

\begin{figure*}[t]
\centering
\includegraphics[width=\textwidth]{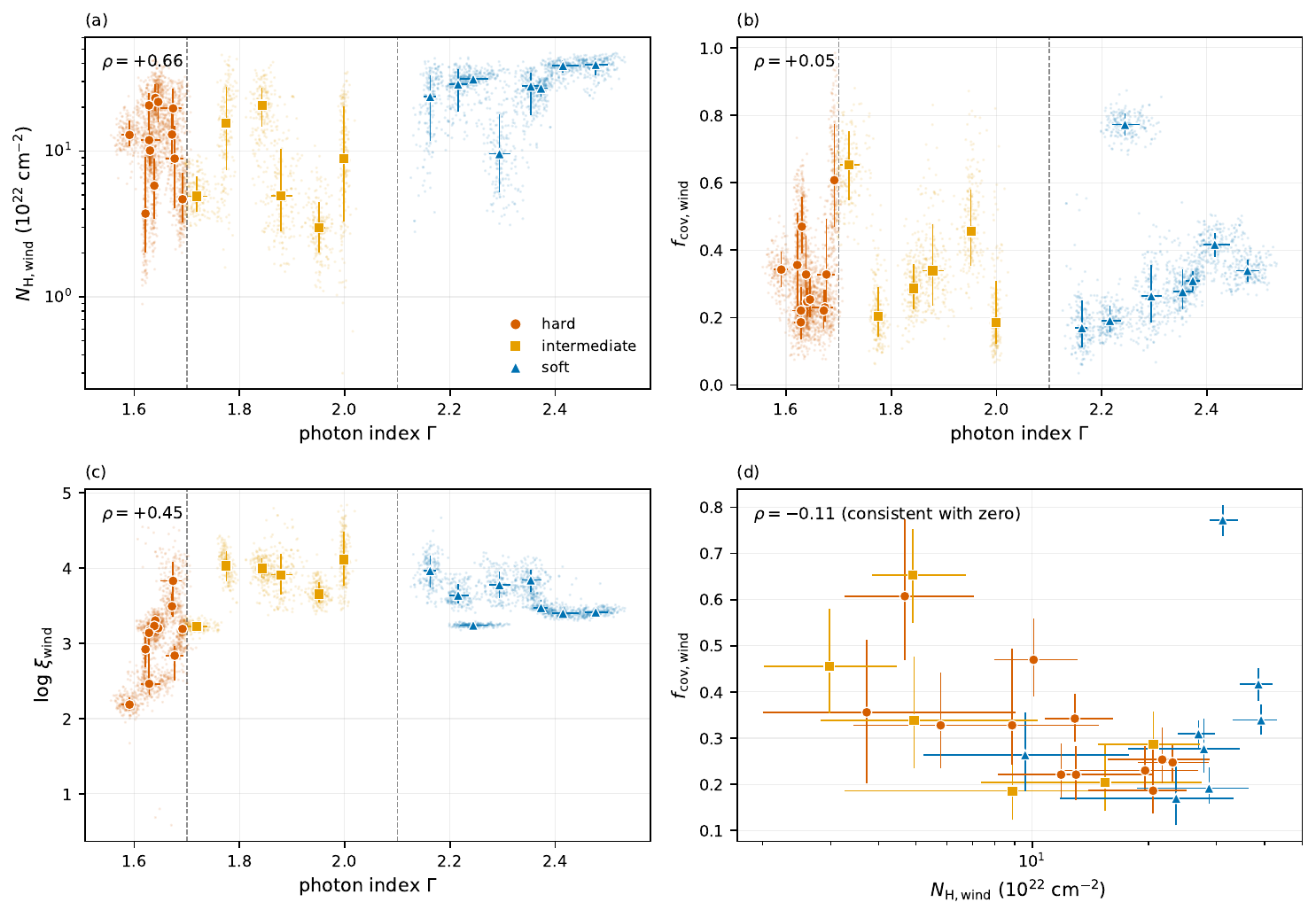}
\caption{Effective wind-absorber parameters from the baseline free-abundance fits. \textit{Panel a:} wind column against $\Gamma$ (Pearson $r=+0.66$, Spearman $\rho_s=+0.49$); the baseline state medians are 12, 7, and $28\times10^{22}$\,cm$^{-2}$ in the hard, intermediate, and soft groups. \textit{Panel b:} covering fraction against $\Gamma$ (median $f_{\rm cov}\approx0.3$; Pearson $r=+0.05$). \textit{Panel c:} wind ionization against $\Gamma$ (Pearson $r=+0.45$); this relation changes between abundance tests. \textit{Panel d:} covering fraction against wind column (Pearson $r=-0.11$). Symbols and posterior samples follow Figure~\ref{fig:gamma_logxi}. These coefficients are descriptive cross-observation statistics of the recovered baseline modes.}
\label{fig:wind}
\end{figure*}

Figure~\ref{fig:wind} summarizes the phenomenological wind-absorber parameters from the recovered baseline modes. The wind column has Pearson $r=+0.659$ and Spearman $\rho_s=+0.485$ with $\Gamma$ in the baseline analysis; for $\afe=4.5$, the corresponding values are $+0.720$ and $+0.419$. The baseline state medians are approximately 12, 7, and $28\times10^{22}$\,cm$^{-2}$ in the hard, intermediate, and soft groups. Since the intermediate median lies below the hard-state value, the three groups do not form a monotonic sequence. The baseline covering fraction has a median of approximately 0.30 and little association with $\Gamma$ (Pearson $r=+0.052$, Spearman $\rho_s=+0.046$). In the low-abundance sensitivity analysis, however, covering fraction becomes strongly state-dependent. Covering fraction and column show little cross-observation association in the baseline analysis (Pearson $r=-0.112$, Spearman $\rho_s=-0.268$), a result that remains qualitatively stable across the archived variants.

The absolute wind parameters are model-sensitive. A normalized parameter-shift comparison finds no general excess over the shifts in the reflection parameters, and the lack of data below 3\,keV leaves partial covering degenerate with the continuum. The baseline column medians span approximately $3$--$39\times10^{22}$\,cm$^{-2}$, with seven of eight soft-state medians above $2\times10^{23}$\,cm$^{-2}$; columns this large may signal continuum or absorber-model systematics. The ordering also disappears in the $\afe=1.6$ analysis, where Spearman $\rho_s\approx0.01$ and the hard/intermediate/soft medians are approximately 16.6, 17.4, and $14.9\times10^{22}$\,cm$^{-2}$. Each spectrum covers a finite orbital-phase interval, and phase is not controlled across states. We treat the \texttt{zxipcf} values as effective absorber parameters, not unique measurements of wind column or clump structure. \citet{Miskovicova2016} establish orbital dependence, but do not test a hard-to-soft increase in wind column.

\section{Discussion} \label{sec:discussion}

\subsection{Where is the disk located?} \label{sec:disc_disk}

In the recovered free-emissivity baseline modes, the median inner radius increases from approximately $4\,\risco$ in the soft state to $6.6\,\risco$ in the hard state. This has the qualitative sign expected in a truncated-disk picture, but it does not establish physical motion of the disk edge because the emissivity prescription, continuum architecture, abundance, inclination, calibration, and recovered likelihood mode all affect $\rin$. The positive $\logxi$--$\Gamma$ association may reflect changes in incident ionizing flux, disk density, and coronal geometry. The present fits neither measure the disk-incident luminosity directly nor identify $\rin$ with the illuminating-source-to-surface distance. The spectra therefore indicate systematic changes in fitted disk and coronal quantities through the state sequence without uniquely determining either geometry.

The timing results support the broad state ranking: $\Gamma$ is strongly anticorrelated with the timing hardness ratio, although the observations are not strictly monotonically ordered. The 8--79\,keV count rate is about an order of magnitude lower in the faintest soft-state bins than in the brightest hard-state bins (Figure~\ref{fig:rate_vs_gamma}), and the 3--8\,keV rate increases over the same range. An inward-moving disk would provide more seed photons, cool the corona by Compton scattering, steepen the spectrum, and shift flux from the hard to the soft band \citep{SunyaevTitarchuk1980, Esin1997, Malzac2001}. The nearly constant 3--79\,keV count rate does not imply constant bolometric luminosity because much of the soft-state disk emission lies below the \nustar{} band \citep{Grinberg2013}. The individual radii nevertheless show considerable scatter. Observations with radii near the ISCO occur mainly in the intermediate and soft states, but truncated solutions occur in all states, and $\rin$ is not monotonic with $\Gamma$ within the hard state.

For $a_*=0.95$, the baseline hard-state median $\rin=6.6\,\risco$ corresponds to approximately $12.8\,r_g$, comparable to the 13--20\,$r_g$ hard-state values reported by \citet{Basak2017}. The baseline soft-state median exceeds the near-ISCO soft-state values reported by \citet{Tomsick2014} and \citet{Walton2016}. \citet{Parker2015} concerns the hard state, while \citet{Zdziarski2024} do not report a directly comparable free measurement of $\rin/\risco$. Fixing $q=3$ moves several observations to or near the ISCO (Section~\ref{sec:rin_gamma}), and changes in abundance or inclination shift some radii by factors of several. Because the fixed-$q$ fits reveal missed regions of the likelihood in several soft-state observations, the individual radii and state summaries remain conditional on the model and recovered mode. \citet{Nosirov2025} likewise find substantial dependence on emissivity and spin assumptions. Differences in state coverage may contribute to the spread among published measurements, although we have not tested that explanation directly.

\ixpe{} polarimetry indicates that the hard-state corona is extended in the disk plane rather than confined to a compact lamppost \citep{Krawczynski2022, Jana2024}; \citet{Steiner2024} draw a similar conclusion in the soft state. An extended corona with a spectrum that softens as the source brightens could contribute to the $\logxi$--$\Gamma$ relation. Ionization alone therefore does not determine the evolution of the disk radius. Simultaneous fits to the spectra, polarization, and timing data are needed to constrain the three-dimensional coronal geometry.

\subsection{Is the iron abundance anomaly a density artifact?} \label{sec:disc_afe}

The upper density limit of current reflection grids may contribute to the fitted iron-abundance anomaly, although these data do not determine how much. Free-density baseline fits with the \citet{Ding2024} grid still give a median $\afe\approx5$. Across the 26 observations, the abundance and density medians have little association even though individual posteriors can retain density--abundance covariance. The near-contemporaneous 2012 observations also give discrepant fitted abundances. At $\afe=1.6$, some density posteriors approach the grid boundary and the fit statistic worsens relative to $\afe=4.5$. A density above the grid maximum could produce this response, as could differences in the continuum, illumination, emissivity, calibration, absorber, atomic model, or likelihood mode. \citet{Tomsick2018} required $n_e\sim4\times10^{20}$\,cm$^{-3}$ for a solar-abundance fit to one observation. Calculations extending to $\logn\gtrsim21$ would test whether higher densities reduce the fitted abundance and its variation among observations. Spin is fixed here, while the emissivity index is free within a fixed single-power-law form.

We treat the fitted $\afe$ as a nuisance parameter of the reflection model, not as a bulk-composition measurement. The $\afe=1.6$ and 4.5 analyses show the sensitivity to two selected values but do not span the full structural uncertainty. We also did not fit a joint model with one abundance tied across all observations.

\subsection{Model-dependent wind absorption across states} \label{sec:disc_wind}

The baseline recovered modes show larger fitted wind columns in the soft group than in the hard group, but this ordering is not robust to the low-abundance sensitivity analysis, and the covering-fraction relation also changes with abundance. The \nustar{} data do not resolve the individual absorption lines or constrain the continuum below 3\,keV, so they cannot distinguish uniquely between physical wind variability and degeneracy between partial covering, continuum curvature, and reflection. Simultaneous soft-band and high-resolution spectra would help separate these possibilities; \chandra{} \citep{Hanke2009} and \xrism{} \citep{Yamada2025} can resolve the relevant absorption structure.

Each fitted spectrum averages over the finite orbital-phase interval covered by that pointing, while the archive samples phase nonuniformly across spectral states \citep{Grinberg2015}. State and orbital variability are therefore confounded in the present descriptive trends. Joint \xmm{} and \nustar{} observations with deliberate phase and state coverage would help separate these effects, although a physical wind model would still be required.

\section{Conclusions} \label{sec:conclusions}

We analyzed a selected sample of 26 archival \nustar{} observations of Cygnus\,X-1 obtained between 2012 and 2024 July using two configurations from one \textsc{xspec} model family. We sampled the recovered posterior modes with preconditioned sequential Monte Carlo and examined completed sensitivity analyses at $\afe=1.6$ and 4.5 and at inclinations of $27\fdg5$ and $40\degr$. The results below are conditional on the adopted spectral architecture, calibration prescription, priors, and likelihood modes reached.

\begin{enumerate}
\item In the recovered baseline modes, disk-surface ionization is positively associated with photon index (Pearson $r=+0.59$, Spearman $\rho_s=+0.51$), with state medians increasing from $\logxi\approx3.3$ in the hard group to approximately 3.9 in the soft group. The positive association also appears in the fixed-abundance analyses but is weaker in the $i=40\degr$ rank test. It is qualitatively compatible with changes in incident ionizing flux, disk density, and coronal geometry; the present fits do not separately establish these contributions.

\item Under the free-emissivity prescription, the recovered baseline modes have median inner radii of 4.0, 4.4, and $6.6\,\risco$ in the soft, intermediate, and hard groups. Similar ordering appears in the completed abundance and inclination variants, although group membership and the absolute scale change. Because $\rin$ and $q_1$ are strongly degenerate, fixing $q=3$ moves several soft-state observations to or near the ISCO. Four fixed-$q$ fits have lower $\chi^2$ than the sampled free-$q$ solutions, proving that those free-$q$ runs did not recover the global minimum. The tested models provide acceptable recovered modes without requiring either an ISCO disk or $\rin\gtrsim20\,\risco$, but they do not exclude either geometry.

\item Baseline free-density fits retain supersolar fitted abundances, spanning $\afe=1.6$--8.6 with a median of 4.9. Across the 26 posterior medians, abundance and density show little association, and two near-contemporaneous observations separated by approximately 7.2\,hr give discrepant fitted abundances. In the low-abundance sensitivity analysis at $\afe=1.6$, some density posteriors approach the upper reflection-grid boundary and the fits worsen relative to $\afe=4.5$. This is consistent with a density--abundance trade-off but does not determine how much of the fitted abundance excess is caused by the grid density limit. Fitted $\afe$ should be treated as a model nuisance parameter, not as a direct composition measurement.

\item The phenomenological wind absorber is not robustly characterized by \nustar{} alone. The baseline modes show a positive fitted column--$\Gamma$ association and little cross-observation association between covering fraction and column, but the column ordering and covering-fraction behavior change in the low-abundance analysis. Continuum degeneracy below 3\,keV, very large effective columns, and uncontrolled orbital-phase sampling prevent a unique physical interpretation of these trends.

\item Sequential Monte Carlo produced well-sampled individual recovered modes but did not establish global posterior exploration. Fixed-$q$ tests exposed additional higher-likelihood regions in four soft-state fits. The table therefore reports baseline mode-conditional intervals, and absolute free-$q$ radii and other mode-sensitive quantities must be interpreted conditionally.
\end{enumerate}

\begin{acknowledgments}
GB acknowledges partial support from a program of the Polish Ministry of Science under the title “Regional Excellence Initiative,” project No. RID/SP/0050/2024/1.
This research has made use of data obtained with \nustar, a project led by
Caltech, funded by NASA and managed by JPL, and of software and data
provided by the High Energy Astrophysics Science Archive Research Center
(HEASARC). We thank the operators of the shared computing cluster on which
the posterior campaigns were run. 
\end{acknowledgments}

\facilities{NuSTAR}

\software{\textsc{heasoft}/\textsc{nustardas} \citep{HEASoft2014},
\textsc{xspec} and PyXspec \citep{Arnaud1996},
\texttt{relxill} v2.4 \citep{Dauser2014, Garcia2014} with the
extended-density tables of \citet{Ding2024},
\texttt{pocoMC} \citep{Karamanis2022},
the \texttt{xspec\_emcee} parallel-evaluation harness (J.~Sanders),
\texttt{emcee} \citep{ForemanMackey2013},
\texttt{numpy}, \texttt{matplotlib}, and \texttt{corner}.}

\appendix

\section{Sampling relativistic-reflection posteriors: failure modes and the adopted scheme} \label{app:mcmc_rescue}

Relativistic-reflection posteriors contain curved degeneracies over several orders of magnitude (for example, density--abundance--normalization and disk--corona continuum trade-offs), hard prior boundaries, and locally flat directions (such as wind ionization at small covering fraction). We tested several sampling methods before selecting the procedure used for the production fits.

\emph{Exploratory ensemble MCMC tests.} In exploratory runs, affine-invariant stretch moves \citep{ForemanMackey2013} and differential-evolution variants often produced acceptance fractions of order 1\%, long autocorrelation estimates, and walkers confined to different regions. Ensemble slice sampling also became inefficient in locally flat directions near boundaries, which led us to use preconditioned sequential Monte Carlo. Since the retained products are not a controlled sampler benchmark, we make no general claim about algorithmic superiority.

\emph{Priors on scale parameters.} Linear-uniform priors on positive scale parameters spanning several orders of magnitude can place substantial prior mass at model fluxes far above the data and slow prior-to-posterior annealing. We use log-uniform priors for the \texttt{diskbb} and \texttt{relxillCp} normalizations, while the narrow-line normalization and wind column remain linear-uniform. These choices are part of the statistical model.

\emph{The adopted scheme.} We first perform multi-start Levenberg--Marquardt optimization over disk density to identify likelihood branches that can differ by $\Delta\chi^2\sim10^3$. We then sample with the preconditioned sequential Monte Carlo algorithm in \texttt{pocoMC} \citep{Karamanis2022}. Baseline effective sample sizes are approximately 2028--2304, and the baseline status files contain no automated boundary-accumulation flags. Two baseline observations nevertheless have best sampled $\chi^2$ values worse than their stored LM solutions, and fixed-$q$ tests expose additional missed likelihood regions. The main table therefore reports intervals conditional on the baseline recovered modes.
\section{Per-observation spectral fits} \label{app:perobs}

The following figures show the posterior best-fit model, the unfolded $E^2 f(E)$ component decomposition, and the $\chi^2$ residuals for each of the 26 \nustar{} observations, ordered by photon index (hardest first), in the same panel layout as Figure~\ref{fig:rep_spectra}.

\begin{figure*}[p]
\centering
\begin{minipage}[t]{0.48\textwidth}\centering
\includegraphics[width=\linewidth,height=0.27\textheight,keepaspectratio]{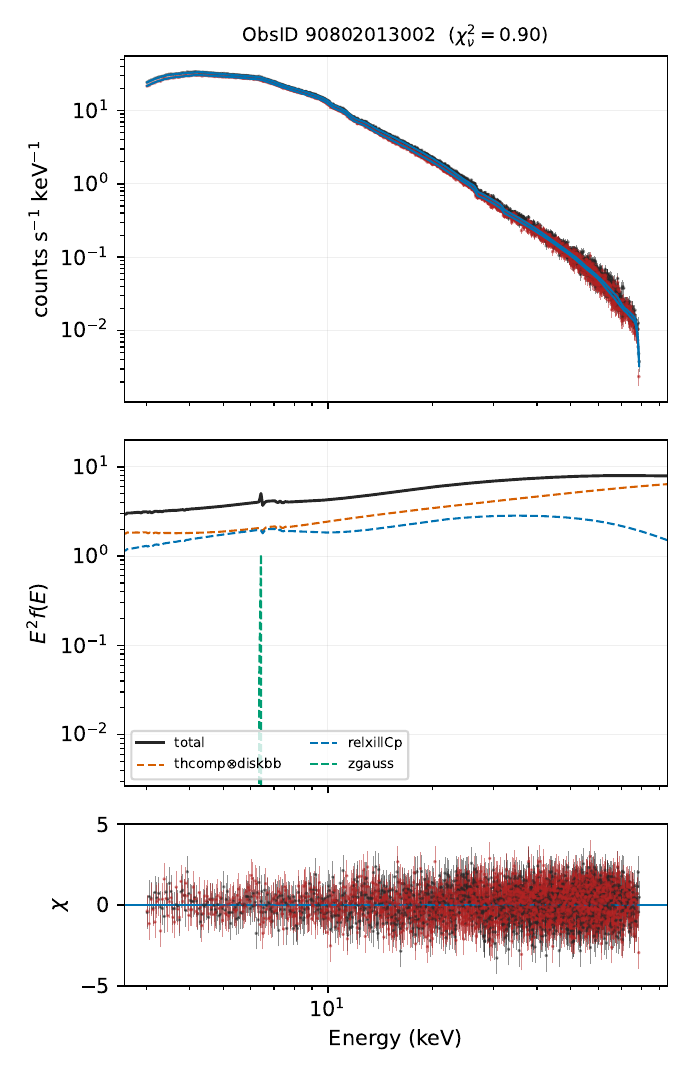}\\
{\footnotesize ObsID 90802013002 --- hard, $\Gamma=1.59$, $\chi^2_\nu=0.90$}
\end{minipage}\hfill\begin{minipage}[t]{0.48\textwidth}\centering
\includegraphics[width=\linewidth,height=0.27\textheight,keepaspectratio]{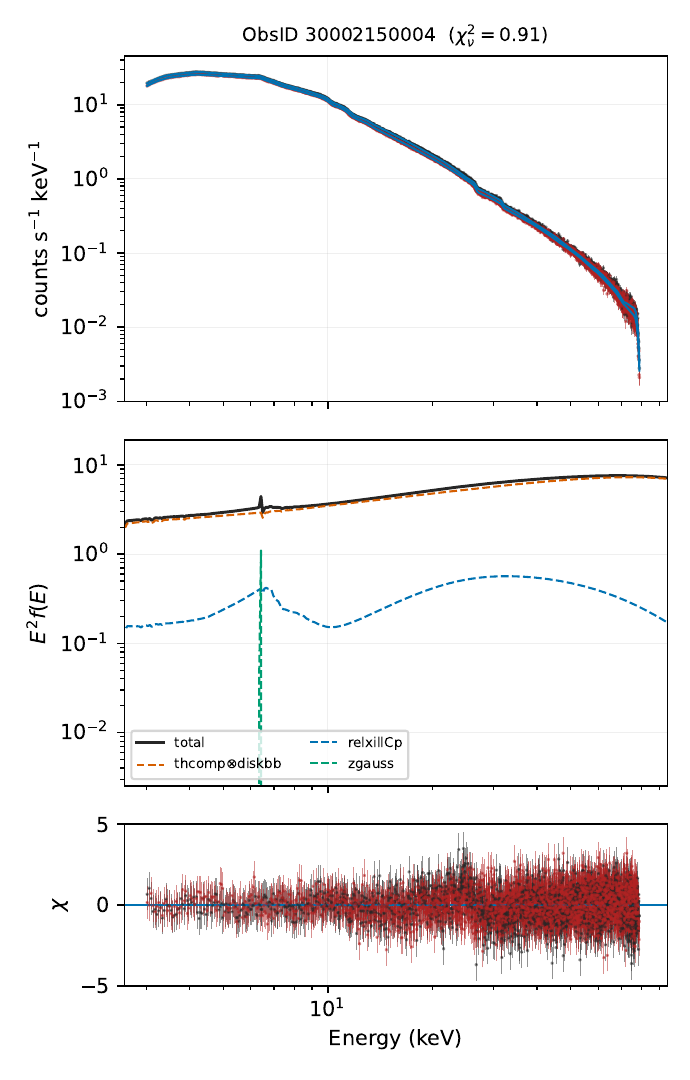}\\
{\footnotesize ObsID 30002150004 --- hard, $\Gamma=1.62$, $\chi^2_\nu=0.91$}
\end{minipage}\\[2mm]
\begin{minipage}[t]{0.48\textwidth}\centering
\includegraphics[width=\linewidth,height=0.27\textheight,keepaspectratio]{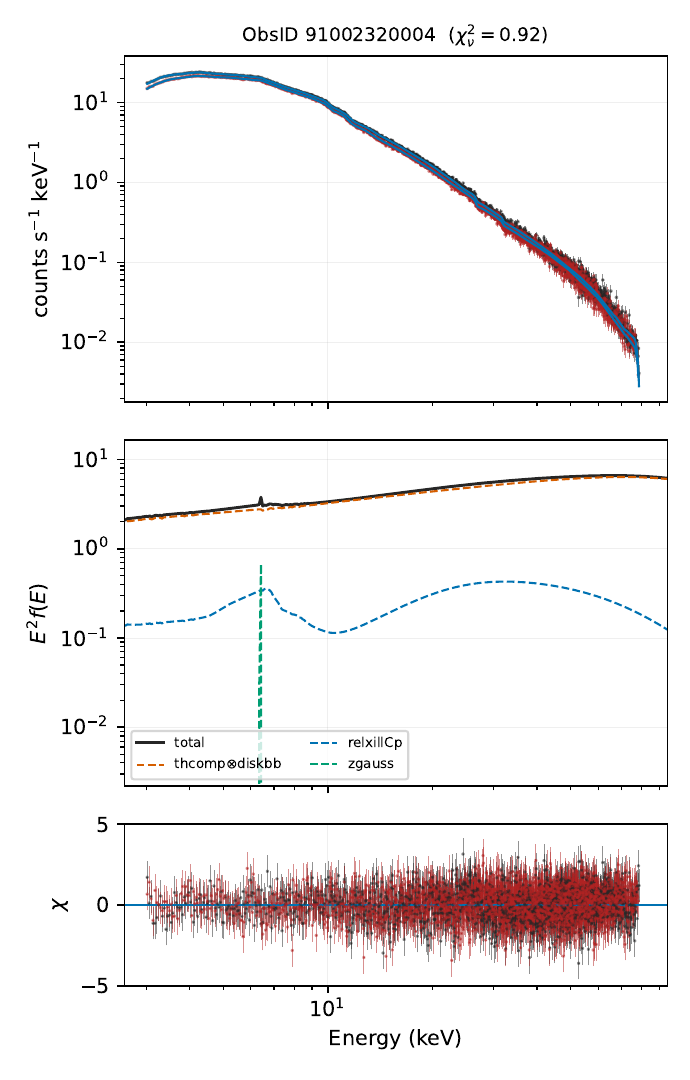}\\
{\footnotesize ObsID 91002320004 --- hard, $\Gamma=1.63$, $\chi^2_\nu=0.92$}
\end{minipage}\hfill\begin{minipage}[t]{0.48\textwidth}\centering
\includegraphics[width=\linewidth,height=0.27\textheight,keepaspectratio]{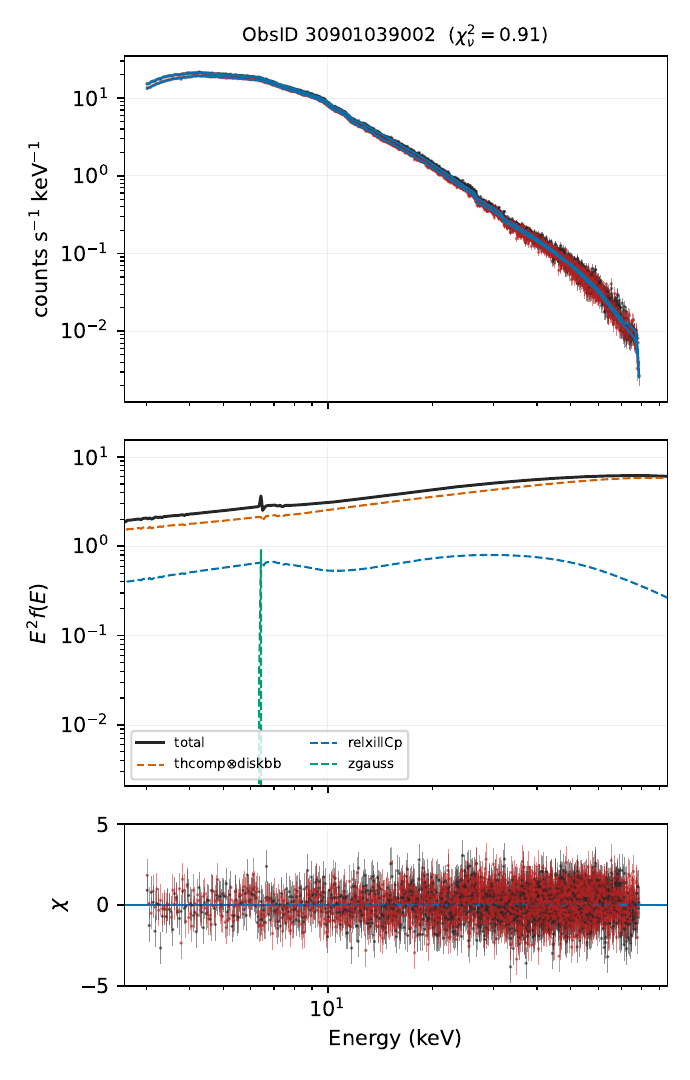}\\
{\footnotesize ObsID 30901039002 --- hard, $\Gamma=1.63$, $\chi^2_\nu=0.91$}
\end{minipage}\\[2mm]
\begin{minipage}[t]{0.48\textwidth}\centering
\includegraphics[width=\linewidth,height=0.27\textheight,keepaspectratio]{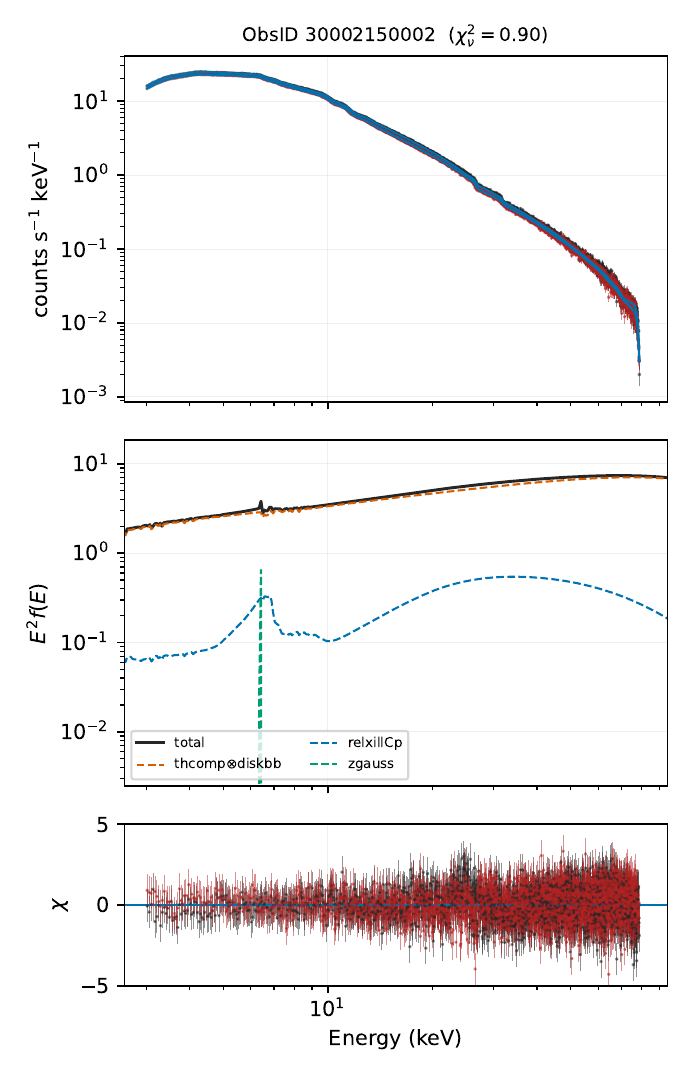}\\
{\footnotesize ObsID 30002150002 --- hard, $\Gamma=1.63$, $\chi^2_\nu=0.90$}
\end{minipage}\hfill\begin{minipage}[t]{0.48\textwidth}\centering
\includegraphics[width=\linewidth,height=0.27\textheight,keepaspectratio]{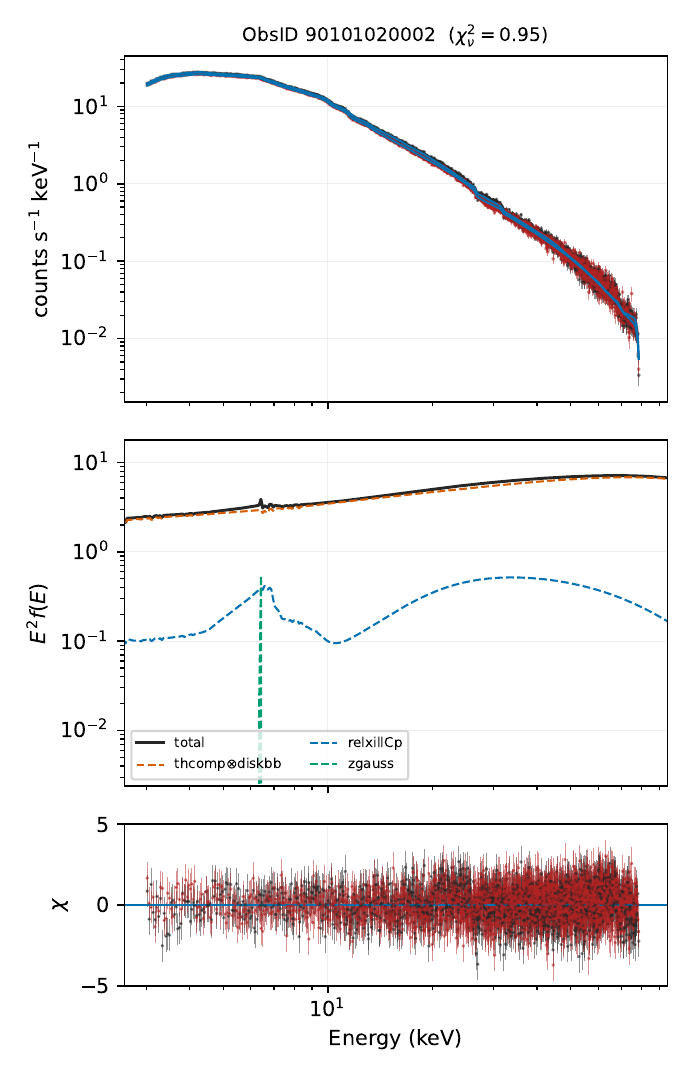}\\
{\footnotesize ObsID 90101020002 --- hard, $\Gamma=1.64$, $\chi^2_\nu=0.95$}
\end{minipage}
\caption{Per-observation posterior best-fit spectra (1 of 5).}
\end{figure*}

\begin{figure*}[p]
\centering
\begin{minipage}[t]{0.48\textwidth}\centering
\includegraphics[width=\linewidth,height=0.27\textheight,keepaspectratio]{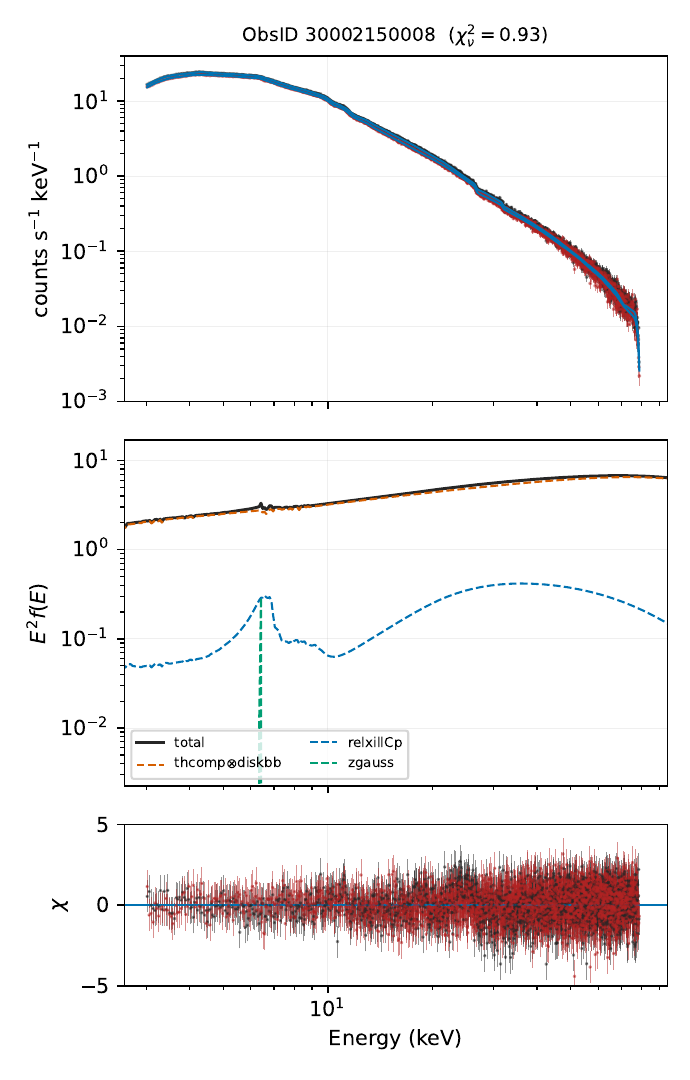}\\
{\footnotesize ObsID 30002150008 --- hard, $\Gamma=1.64$, $\chi^2_\nu=0.93$}
\end{minipage}\hfill\begin{minipage}[t]{0.48\textwidth}\centering
\includegraphics[width=\linewidth,height=0.27\textheight,keepaspectratio]{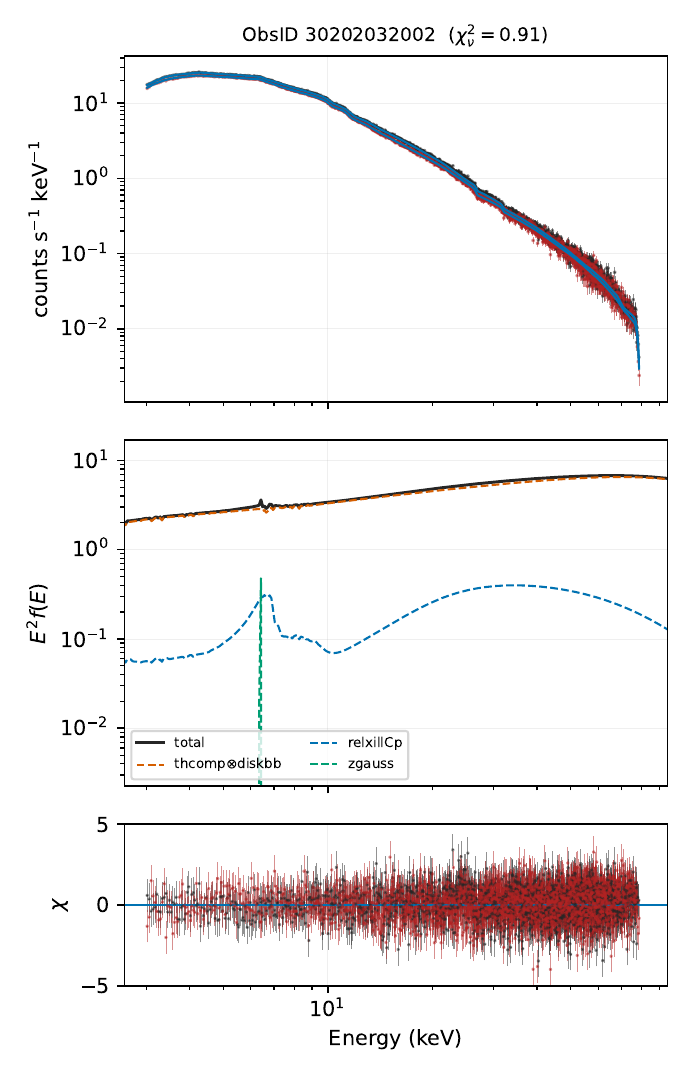}\\
{\footnotesize ObsID 30202032002 --- hard, $\Gamma=1.65$, $\chi^2_\nu=0.91$}
\end{minipage}\\[2mm]
\begin{minipage}[t]{0.48\textwidth}\centering
\includegraphics[width=\linewidth,height=0.27\textheight,keepaspectratio]{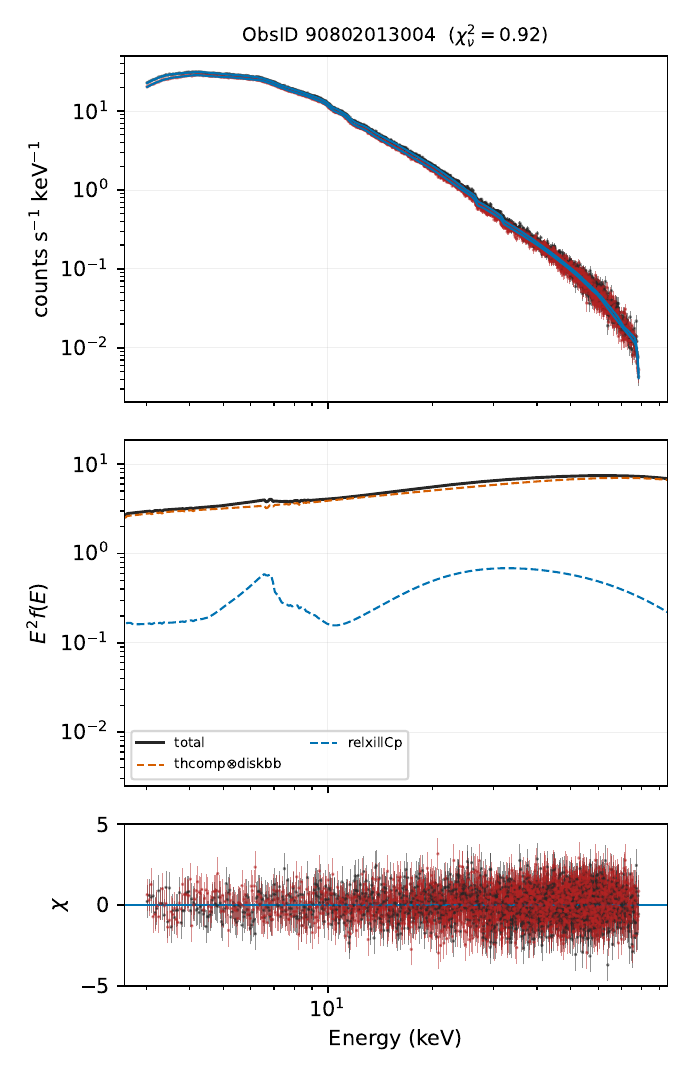}\\
{\footnotesize ObsID 90802013004 --- hard, $\Gamma=1.67$, $\chi^2_\nu=0.92$}
\end{minipage}\hfill\begin{minipage}[t]{0.48\textwidth}\centering
\includegraphics[width=\linewidth,height=0.27\textheight,keepaspectratio]{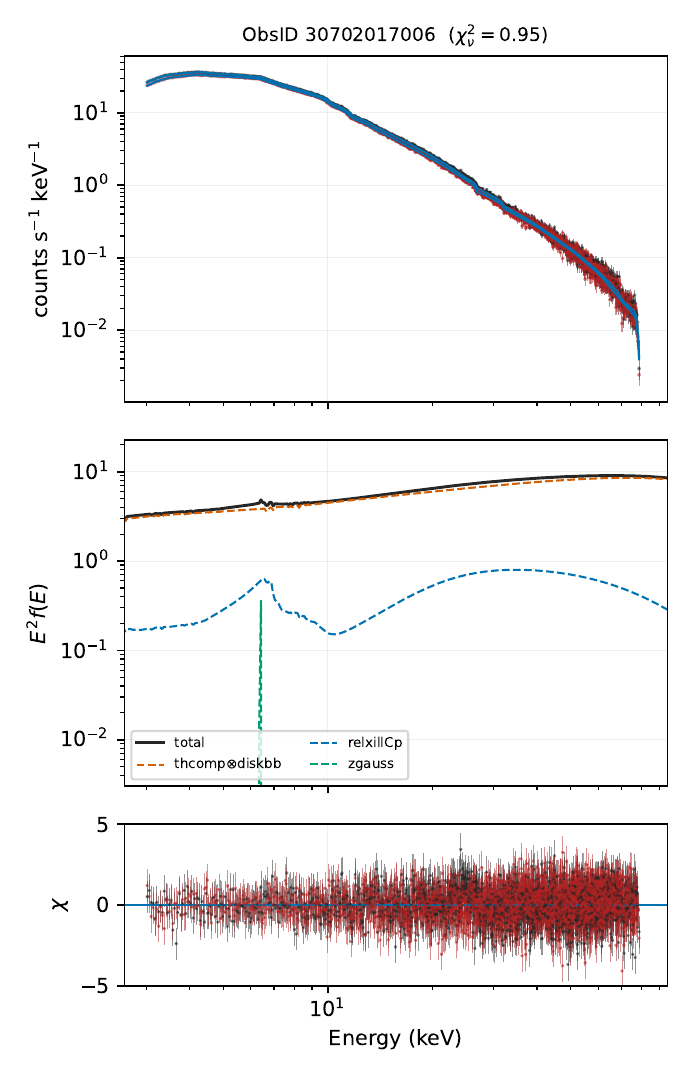}\\
{\footnotesize ObsID 30702017006 --- hard, $\Gamma=1.67$, $\chi^2_\nu=0.95$}
\end{minipage}\\[2mm]
\begin{minipage}[t]{0.48\textwidth}\centering
\includegraphics[width=\linewidth,height=0.27\textheight,keepaspectratio]{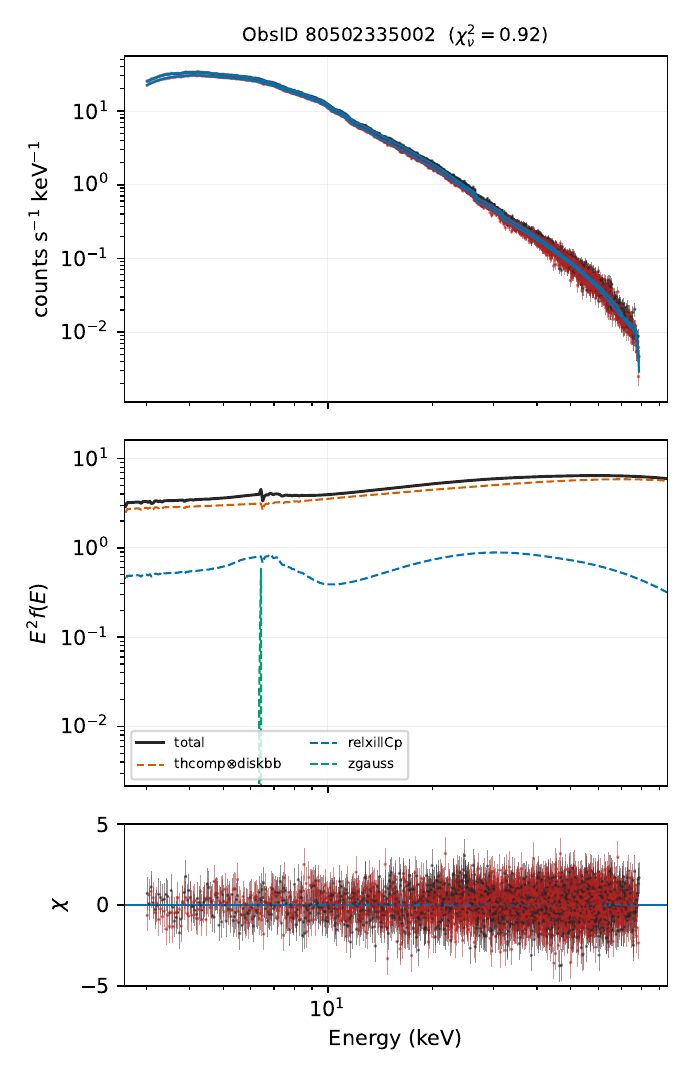}\\
{\footnotesize ObsID 80502335002 --- hard, $\Gamma=1.68$, $\chi^2_\nu=0.92$}
\end{minipage}\hfill\begin{minipage}[t]{0.48\textwidth}\centering
\includegraphics[width=\linewidth,height=0.27\textheight,keepaspectratio]{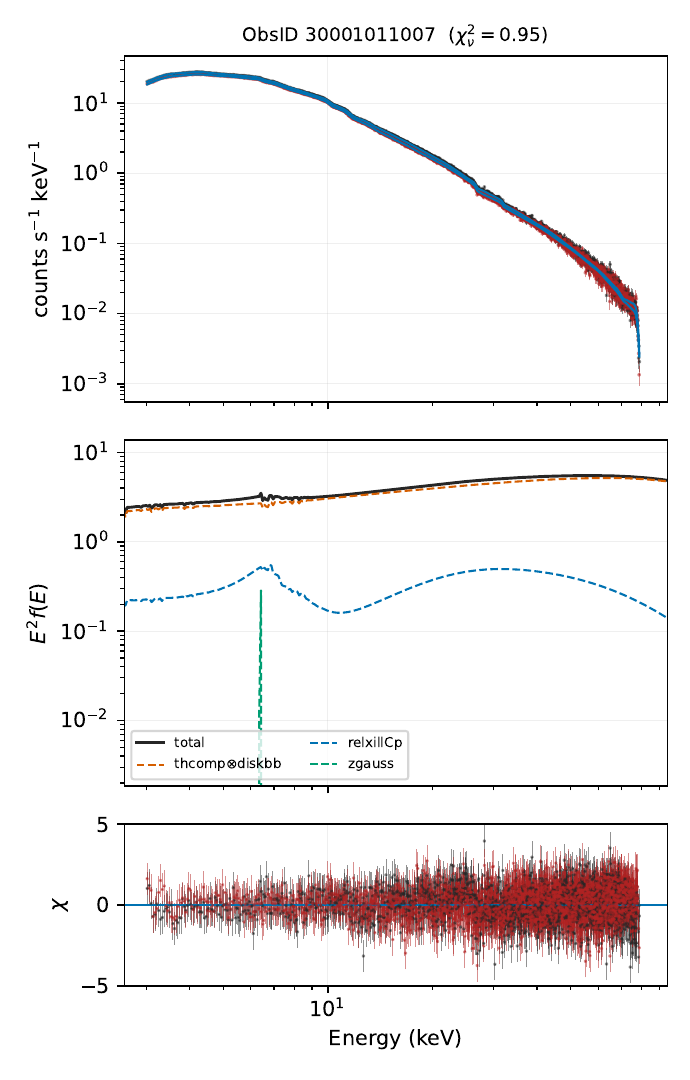}\\
{\footnotesize ObsID 30001011007 --- hard, $\Gamma=1.69$, $\chi^2_\nu=0.95$}
\end{minipage}
\caption{Per-observation posterior best-fit spectra (2 of 5).}
\end{figure*}

\begin{figure*}[p]
\centering
\begin{minipage}[t]{0.48\textwidth}\centering
\includegraphics[width=\linewidth,height=0.27\textheight,keepaspectratio]{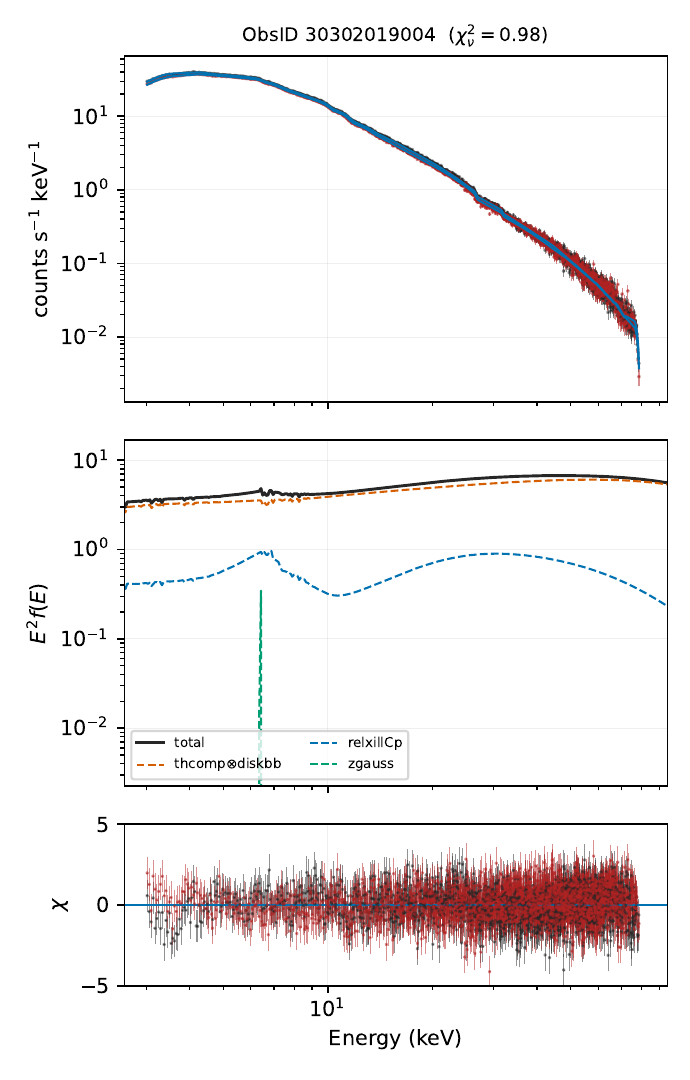}\\
{\footnotesize ObsID 30302019004 --- intermediate, $\Gamma=1.72$, $\chi^2_\nu=0.98$}
\end{minipage}\hfill\begin{minipage}[t]{0.48\textwidth}\centering
\includegraphics[width=\linewidth,height=0.27\textheight,keepaspectratio]{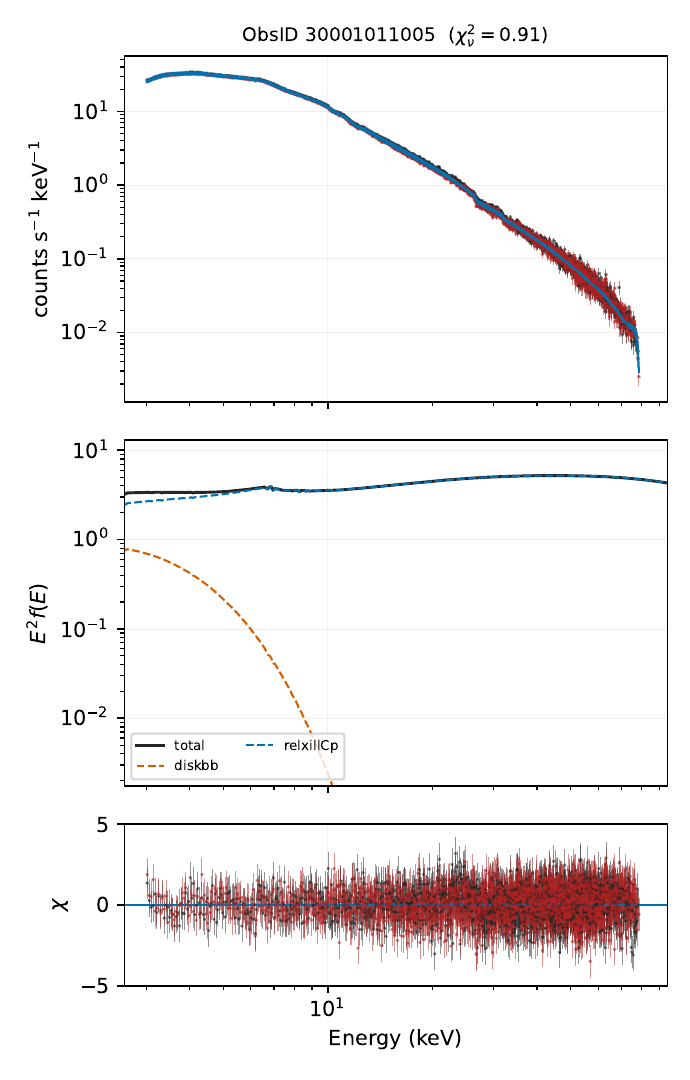}\\
{\footnotesize ObsID 30001011005 --- intermediate, $\Gamma=1.77$, $\chi^2_\nu=0.91$}
\end{minipage}\\[2mm]
\begin{minipage}[t]{0.48\textwidth}\centering
\includegraphics[width=\linewidth,height=0.27\textheight,keepaspectratio]{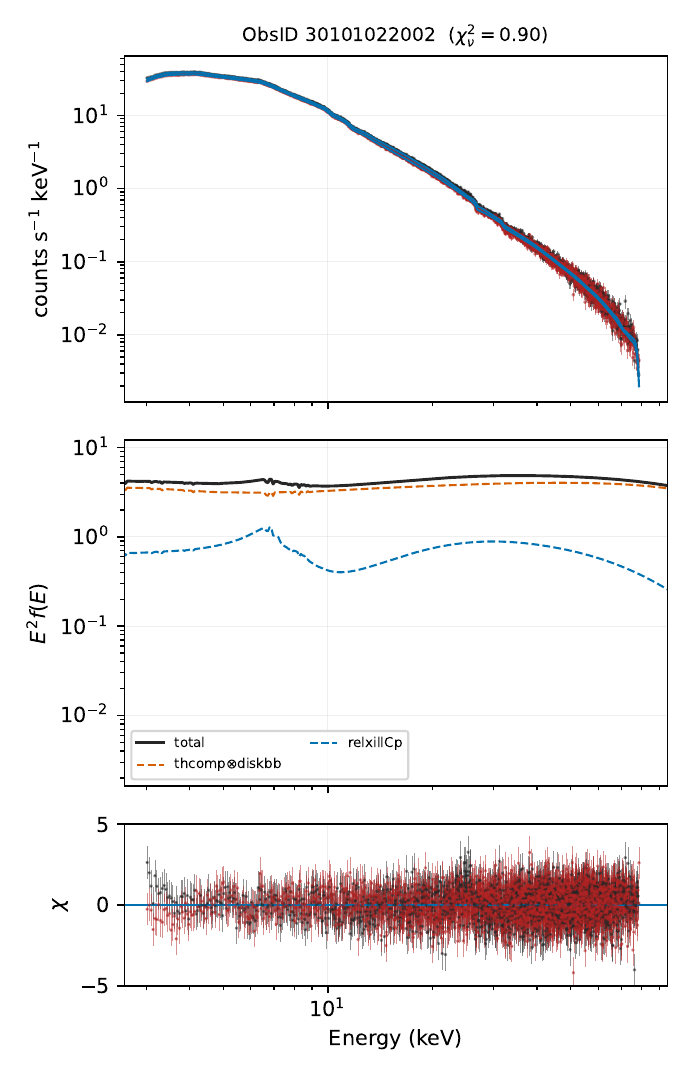}\\
{\footnotesize ObsID 30101022002 --- intermediate, $\Gamma=1.84$, $\chi^2_\nu=0.90$}
\end{minipage}\hfill\begin{minipage}[t]{0.48\textwidth}\centering
\includegraphics[width=\linewidth,height=0.27\textheight,keepaspectratio]{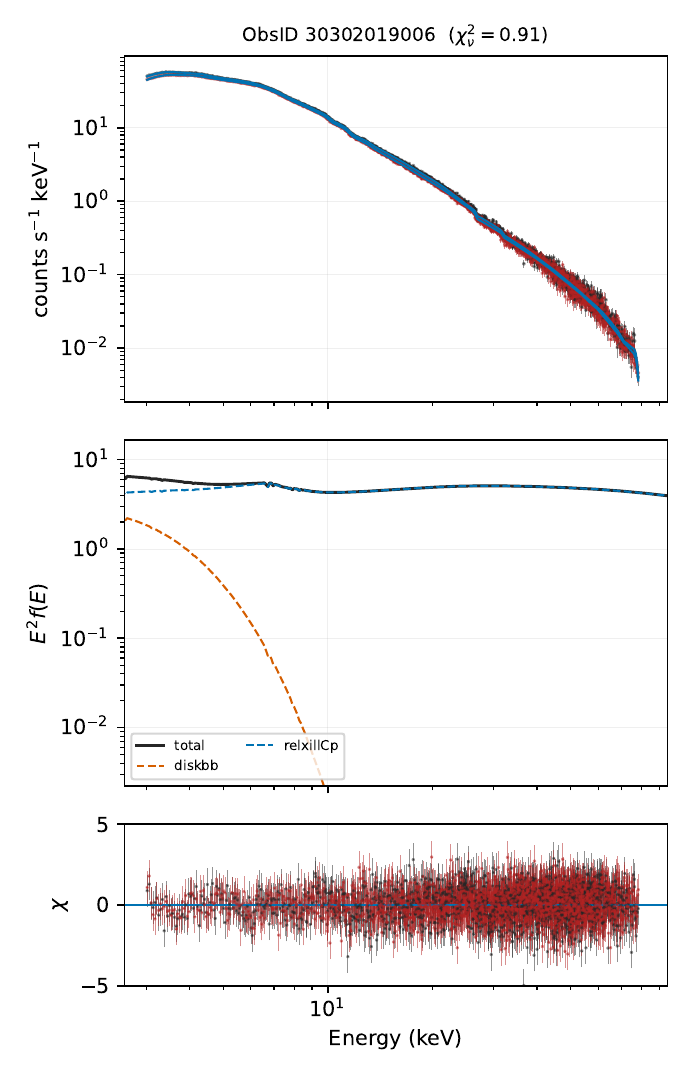}\\
{\footnotesize ObsID 30302019006 --- intermediate, $\Gamma=1.88$, $\chi^2_\nu=0.91$}
\end{minipage}\\[2mm]
\begin{minipage}[t]{0.48\textwidth}\centering
\includegraphics[width=\linewidth,height=0.27\textheight,keepaspectratio]{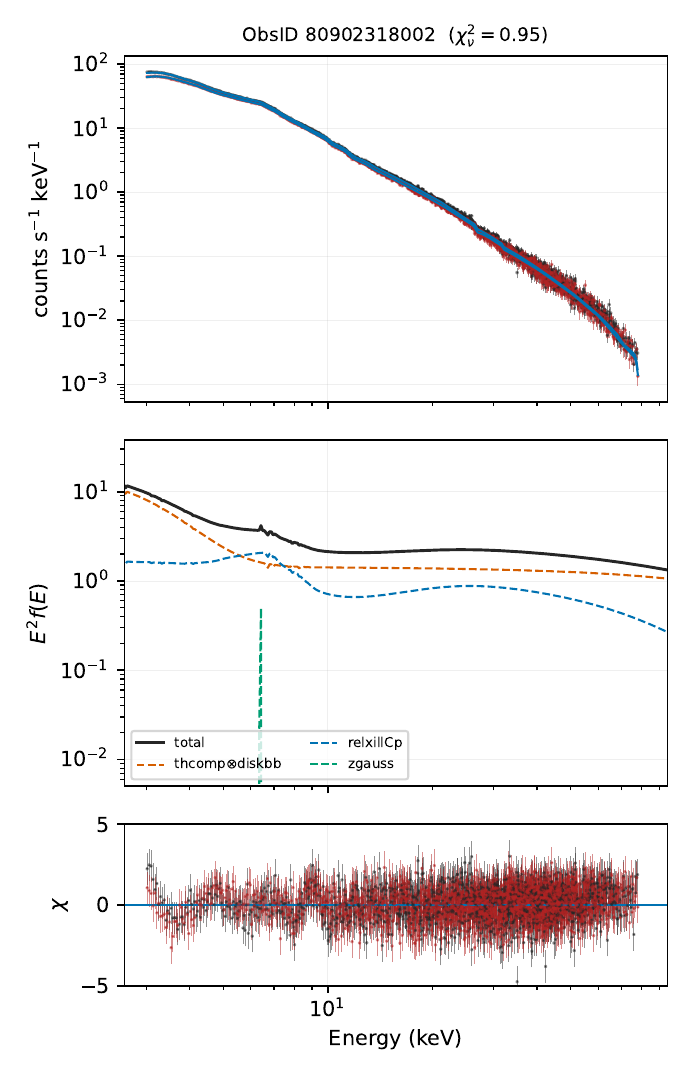}\\
{\footnotesize ObsID 80902318002 --- intermediate, $\Gamma=1.95$, $\chi^2_\nu=0.95$}
\end{minipage}\hfill\begin{minipage}[t]{0.48\textwidth}\centering
\includegraphics[width=\linewidth,height=0.27\textheight,keepaspectratio]{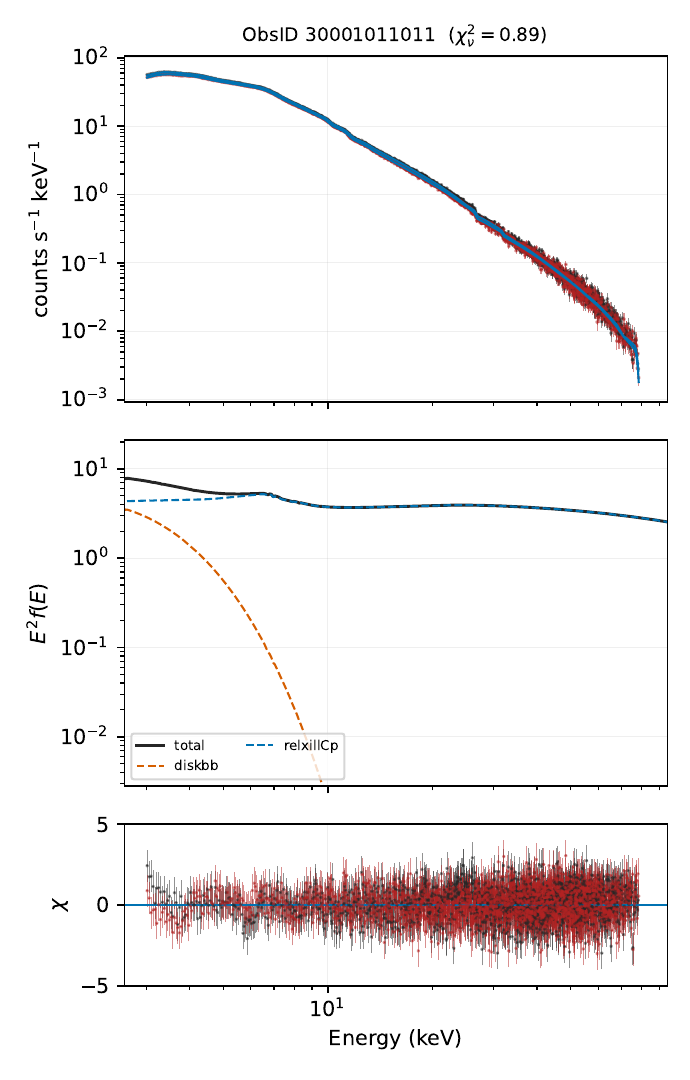}\\
{\footnotesize ObsID 30001011011 --- intermediate, $\Gamma=2.00$, $\chi^2_\nu=0.89$}
\end{minipage}
\caption{Per-observation posterior best-fit spectra (3 of 5).}
\end{figure*}

\begin{figure*}[p]
\centering
\begin{minipage}[t]{0.48\textwidth}\centering
\includegraphics[width=\linewidth,height=0.27\textheight,keepaspectratio]{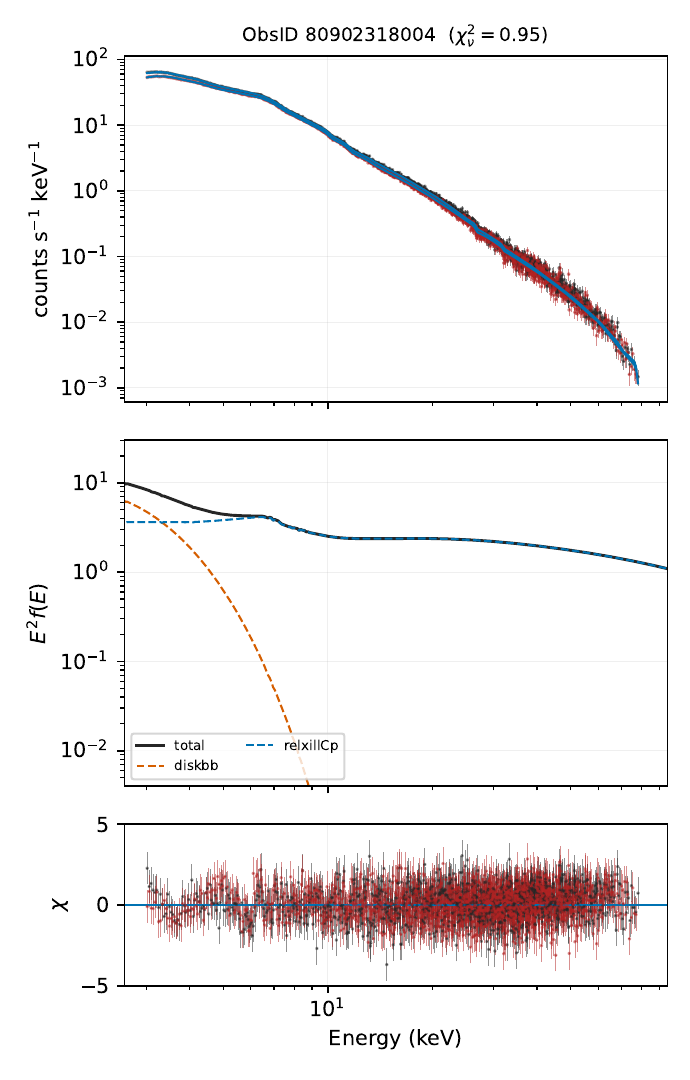}\\
{\footnotesize ObsID 80902318004 --- soft, $\Gamma=2.16$, $\chi^2_\nu=0.95$}
\end{minipage}\hfill\begin{minipage}[t]{0.48\textwidth}\centering
\includegraphics[width=\linewidth,height=0.27\textheight,keepaspectratio]{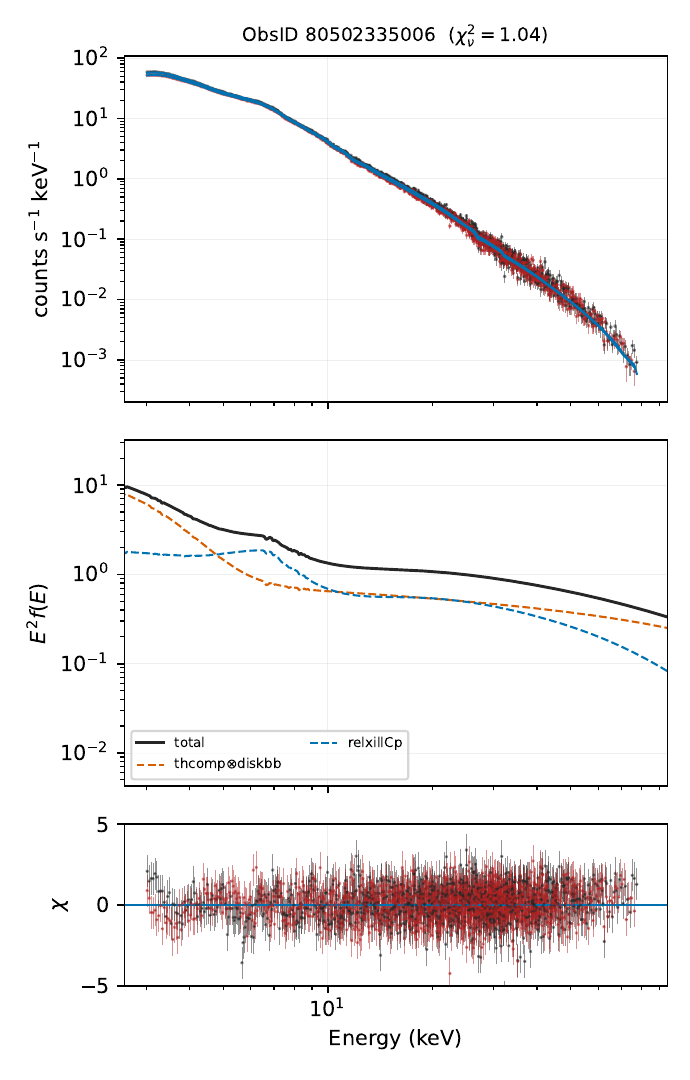}\\
{\footnotesize ObsID 80502335006 --- soft, $\Gamma=2.22$, $\chi^2_\nu=1.04$}
\end{minipage}\\[2mm]
\begin{minipage}[t]{0.48\textwidth}\centering
\includegraphics[width=\linewidth,height=0.27\textheight,keepaspectratio]{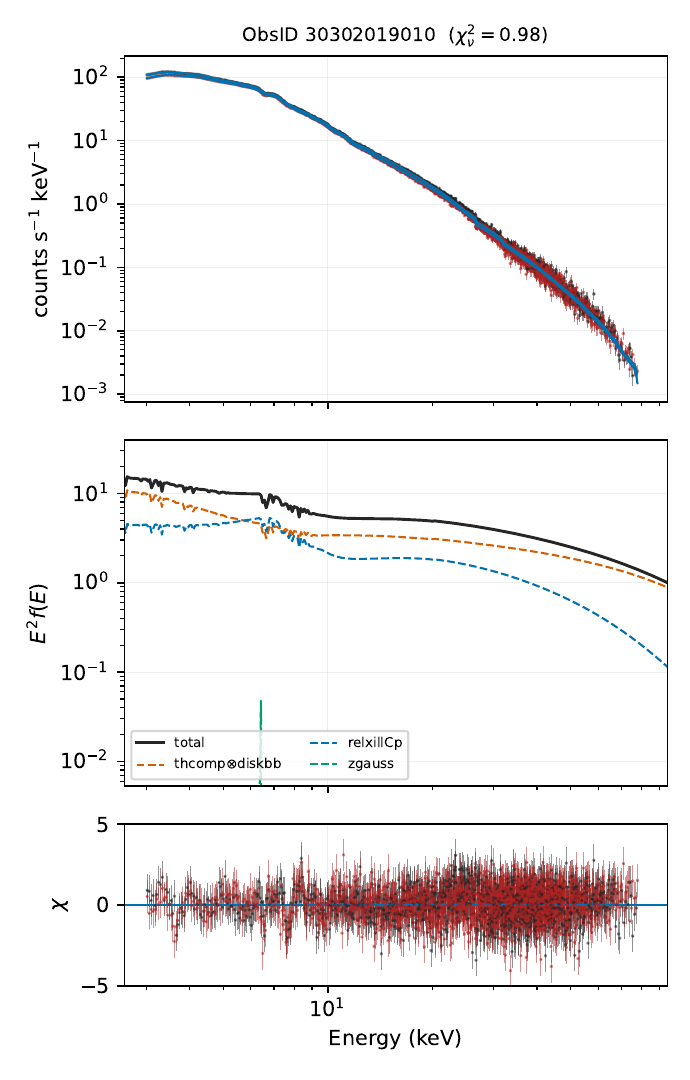}\\
{\footnotesize ObsID 30302019010 --- soft, $\Gamma=2.24$, $\chi^2_\nu=0.98$}
\end{minipage}\hfill\begin{minipage}[t]{0.48\textwidth}\centering
\includegraphics[width=\linewidth,height=0.27\textheight,keepaspectratio]{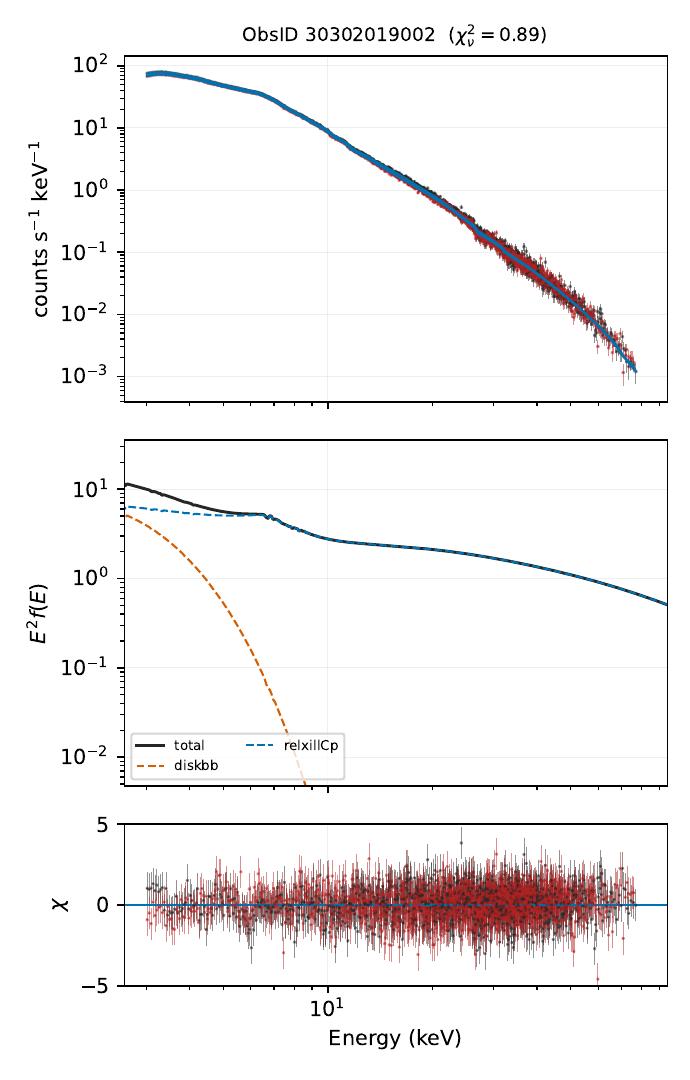}\\
{\footnotesize ObsID 30302019002 --- soft, $\Gamma=2.29$, $\chi^2_\nu=0.89$}
\end{minipage}\\[2mm]
\begin{minipage}[t]{0.48\textwidth}\centering
\includegraphics[width=\linewidth,height=0.27\textheight,keepaspectratio]{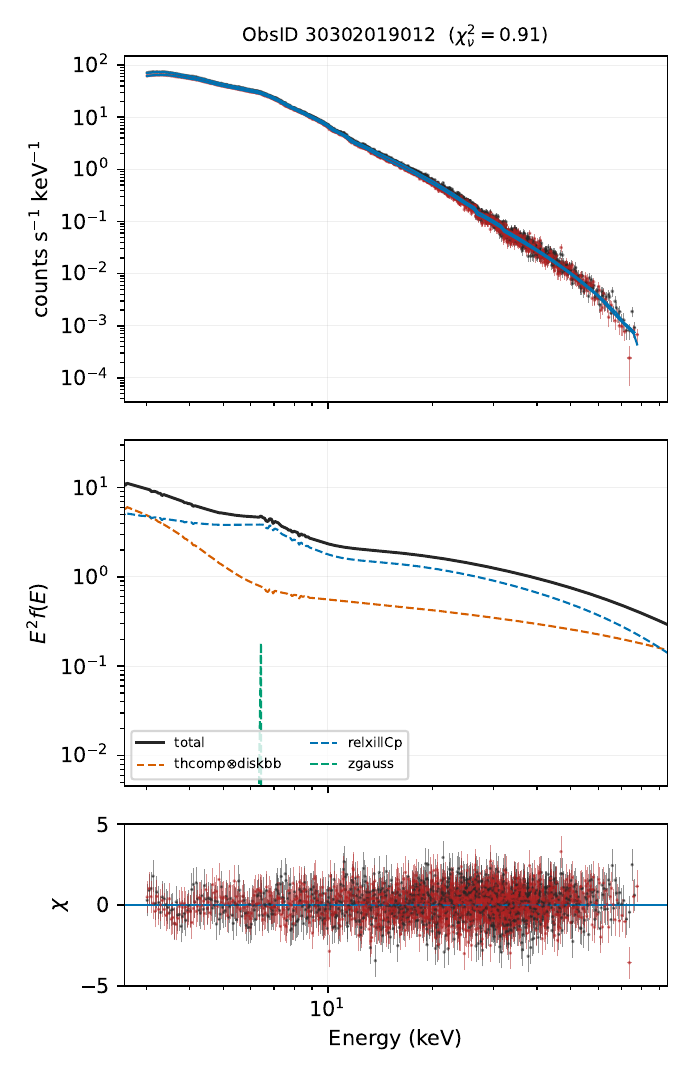}\\
{\footnotesize ObsID 30302019012 --- soft, $\Gamma=2.35$, $\chi^2_\nu=0.91$}
\end{minipage}\hfill\begin{minipage}[t]{0.48\textwidth}\centering
\includegraphics[width=\linewidth,height=0.27\textheight,keepaspectratio]{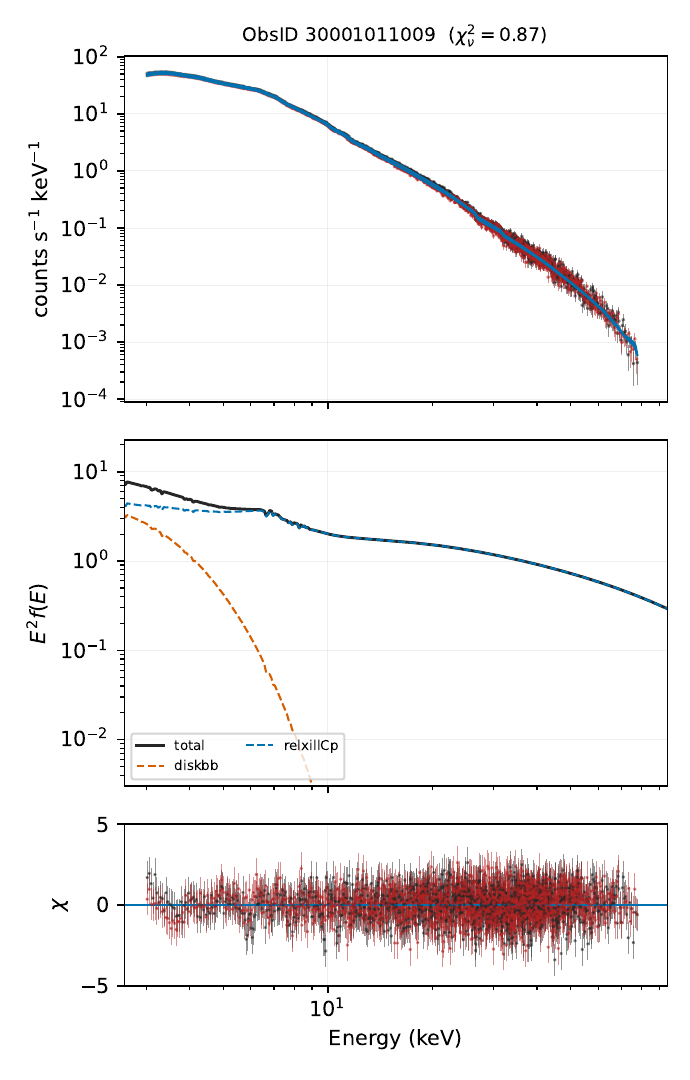}\\
{\footnotesize ObsID 30001011009 --- soft, $\Gamma=2.37$, $\chi^2_\nu=0.87$}
\end{minipage}
\caption{Per-observation posterior best-fit spectra (4 of 5).}
\end{figure*}

\begin{figure*}[p]
\centering
\begin{minipage}[t]{0.48\textwidth}\centering
\includegraphics[width=\linewidth,height=0.27\textheight,keepaspectratio]{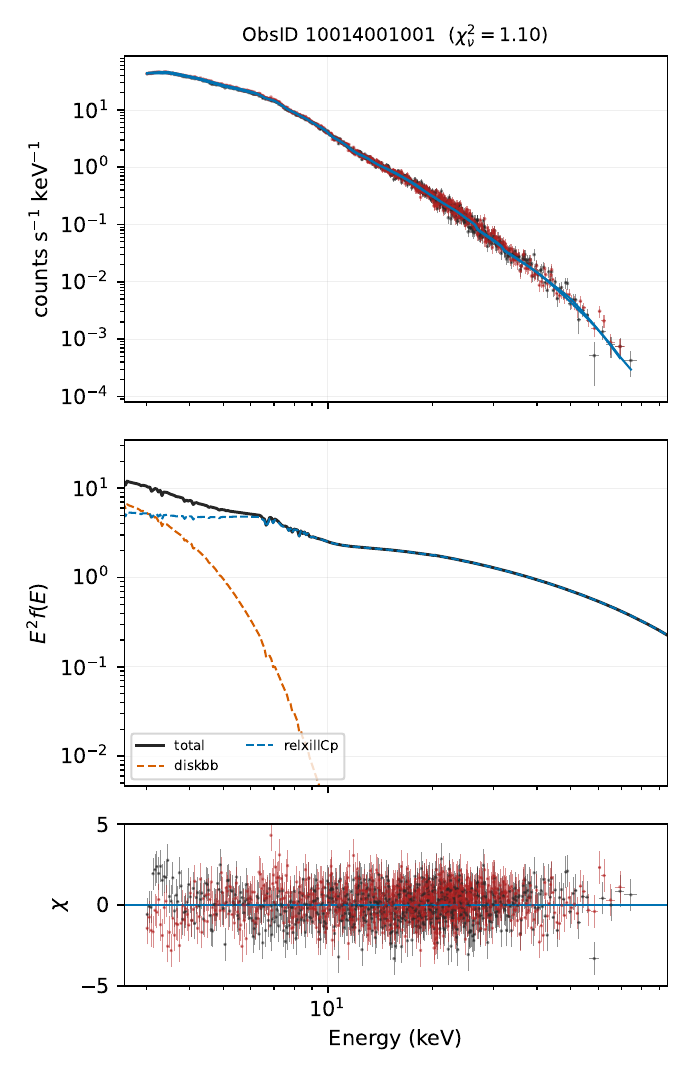}\\
{\footnotesize ObsID 10014001001 --- soft, $\Gamma=2.41$, $\chi^2_\nu=1.10$}
\end{minipage}\hfill\begin{minipage}[t]{0.48\textwidth}\centering
\includegraphics[width=\linewidth,height=0.27\textheight,keepaspectratio]{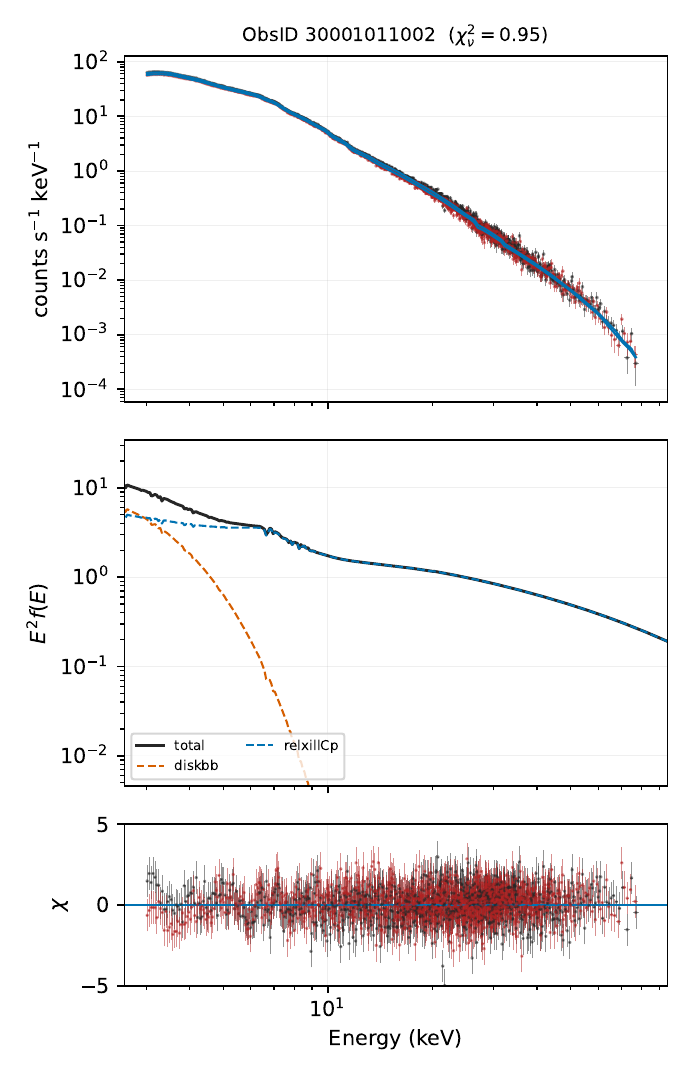}\\
{\footnotesize ObsID 30001011002 --- soft, $\Gamma=2.48$, $\chi^2_\nu=0.95$}
\end{minipage}
\caption{Per-observation posterior best-fit spectra (5 of 5).}
\end{figure*}

\section{Posterior structure of the reflection parameters} \label{app:corner}

\begin{figure*}[p]
\centering
\includegraphics[width=0.85\textwidth]{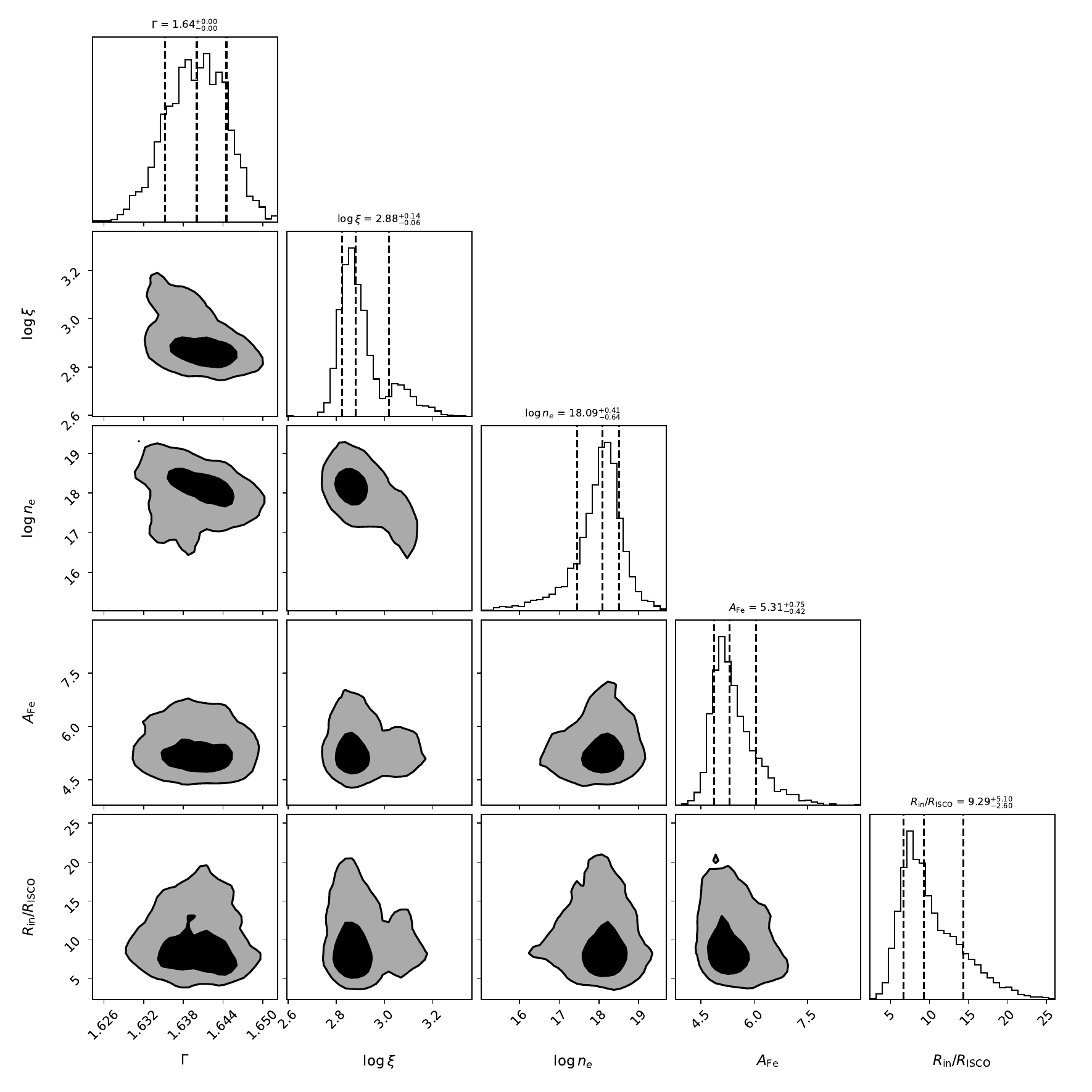}
\caption{Posterior corner plot of five selected reflection quantities for the hard-state ObsID 30002150008 ($\Gamma = 1.64$). Posterior medians and 16th--84th intervals are given above each diagonal panel. The plot shows the mode recovered in the baseline run.}
\label{fig:corner_hard}
\end{figure*}

\begin{figure*}[p]
\centering
\includegraphics[width=0.85\textwidth]{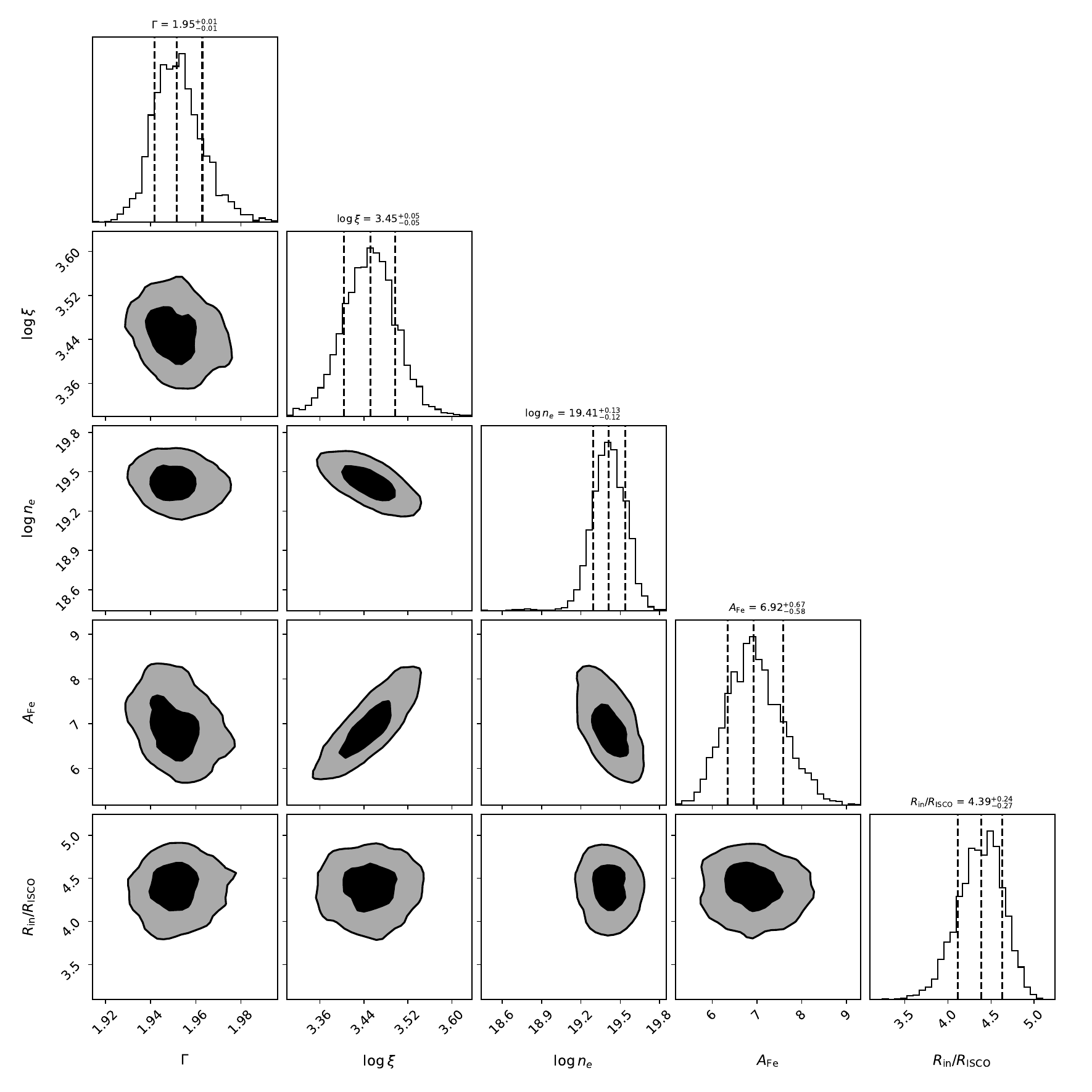}
\caption{Posterior corner plot for the intermediate-state representative ObsID 80902318002 ($\Gamma = 1.95$). Format follows Figure~\ref{fig:corner_hard}.}
\label{fig:corner_intermediate}
\end{figure*}

\begin{figure*}[p]
\centering
\includegraphics[width=0.85\textwidth]{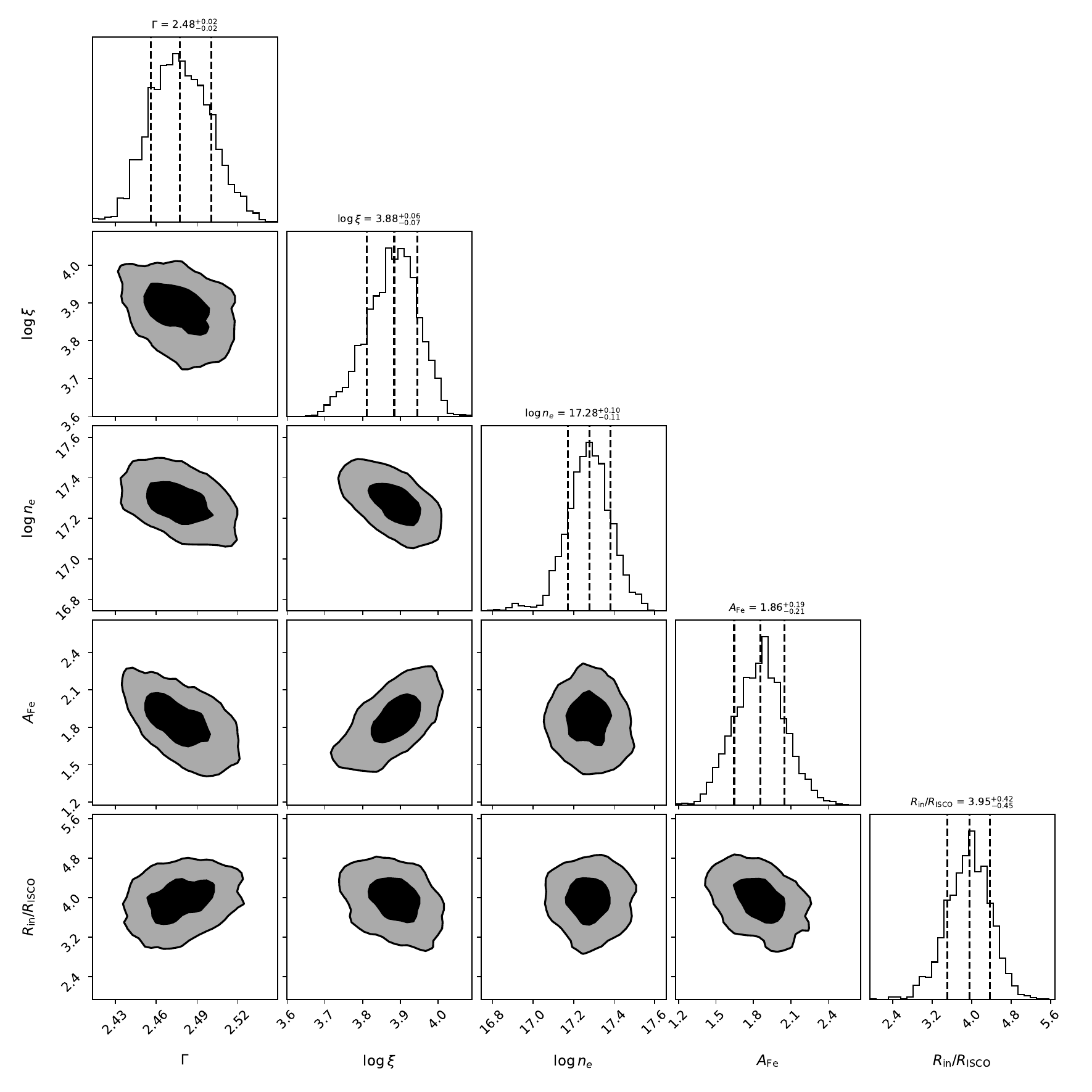}
\caption{Posterior corner plot for the soft-state representative ObsID 30001011002 ($\Gamma = 2.48$). Format follows Figure~\ref{fig:corner_hard}.}
\label{fig:corner_soft}
\end{figure*}

Figures~\ref{fig:corner_hard}, \ref{fig:corner_intermediate}, and \ref{fig:corner_soft} show five selected quantities from the importance-weighted baseline \texttt{pocoMC} posteriors: $\Gamma$, $\logxi$, $\logn$, $\afe$, and $\rin/\risco$. Other free reflection parameters include $q_1$, $\kte$, the component normalization, and, in M2, reflection fraction. The diagonal panels mark the 16th, 50th, and 84th percentiles of the one-dimensional marginals; the off-diagonal panels show approximate enclosed posterior-mass contours. Density--abundance covariance is strongest in the intermediate-state example. The plots omit $q_1$ and therefore do not show the $q_1$--$\rin$ relation discussed in Section~\ref{sec:rin_gamma}. The hard-state figure shows the baseline recovered mode, with no claim about its weight relative to other modes.

The corner plots for the complete set of 26 observations follow, ordered by photon index (hardest first); the format matches Figure~\ref{fig:corner_hard}.

\begin{figure*}[p]
\centering
\begin{minipage}[t]{0.48\textwidth}\centering
\includegraphics[width=\linewidth,height=0.27\textheight,keepaspectratio]{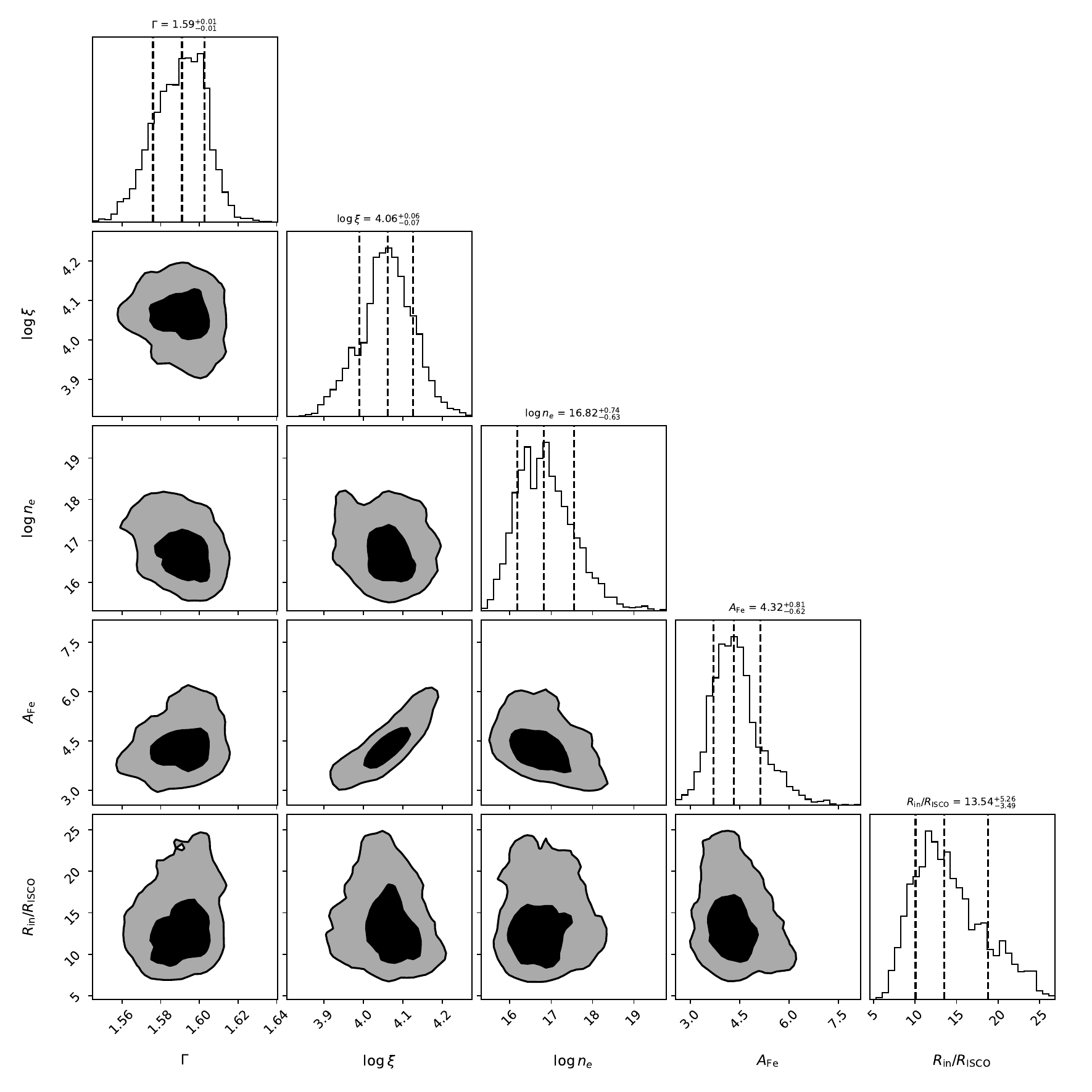}\\
{\footnotesize ObsID 90802013002 --- hard, $\Gamma=1.59$, $\chi^2_\nu=0.90$}
\end{minipage}\hfill\begin{minipage}[t]{0.48\textwidth}\centering
\includegraphics[width=\linewidth,height=0.27\textheight,keepaspectratio]{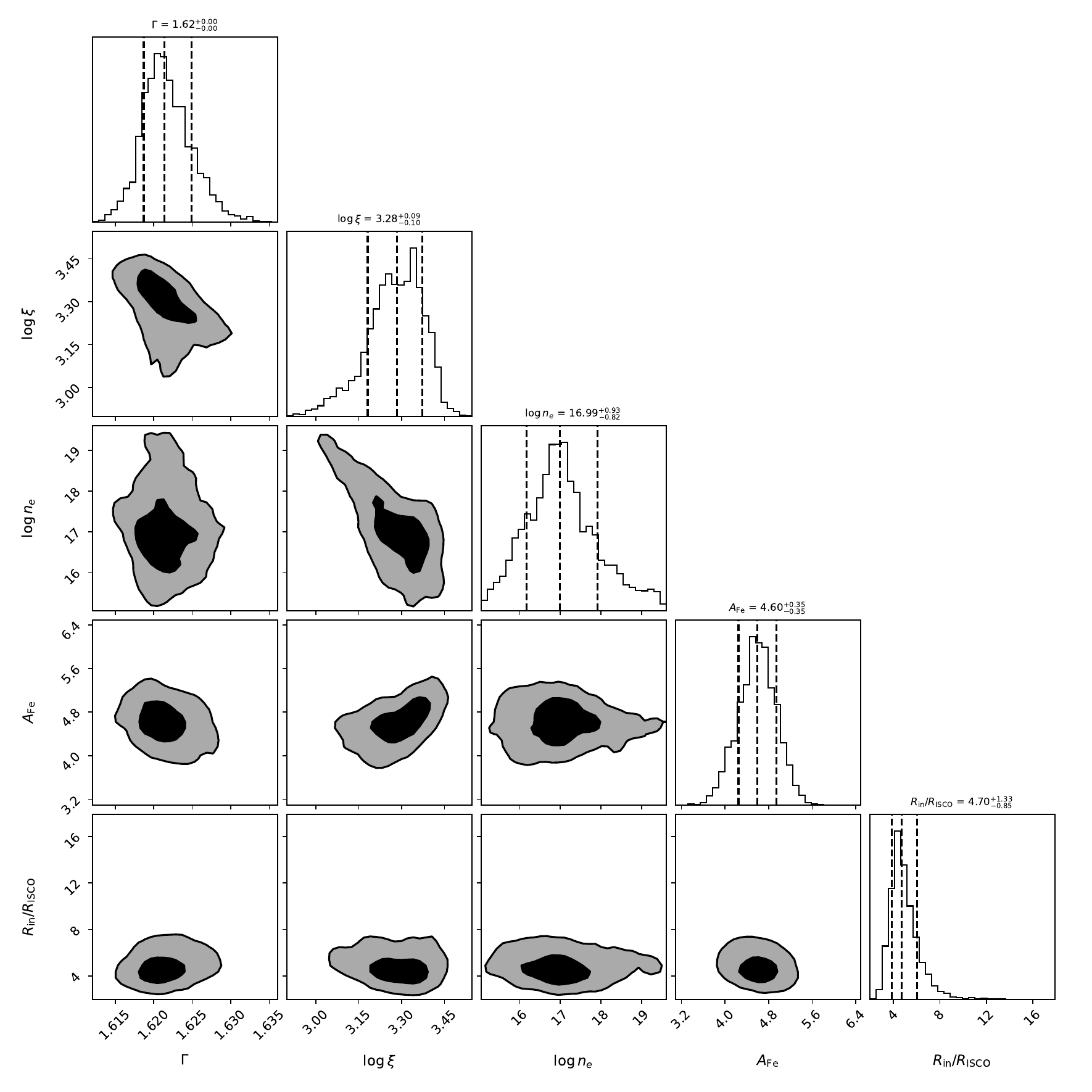}\\
{\footnotesize ObsID 30002150004 --- hard, $\Gamma=1.62$, $\chi^2_\nu=0.91$}
\end{minipage}\\[2mm]
\begin{minipage}[t]{0.48\textwidth}\centering
\includegraphics[width=\linewidth,height=0.27\textheight,keepaspectratio]{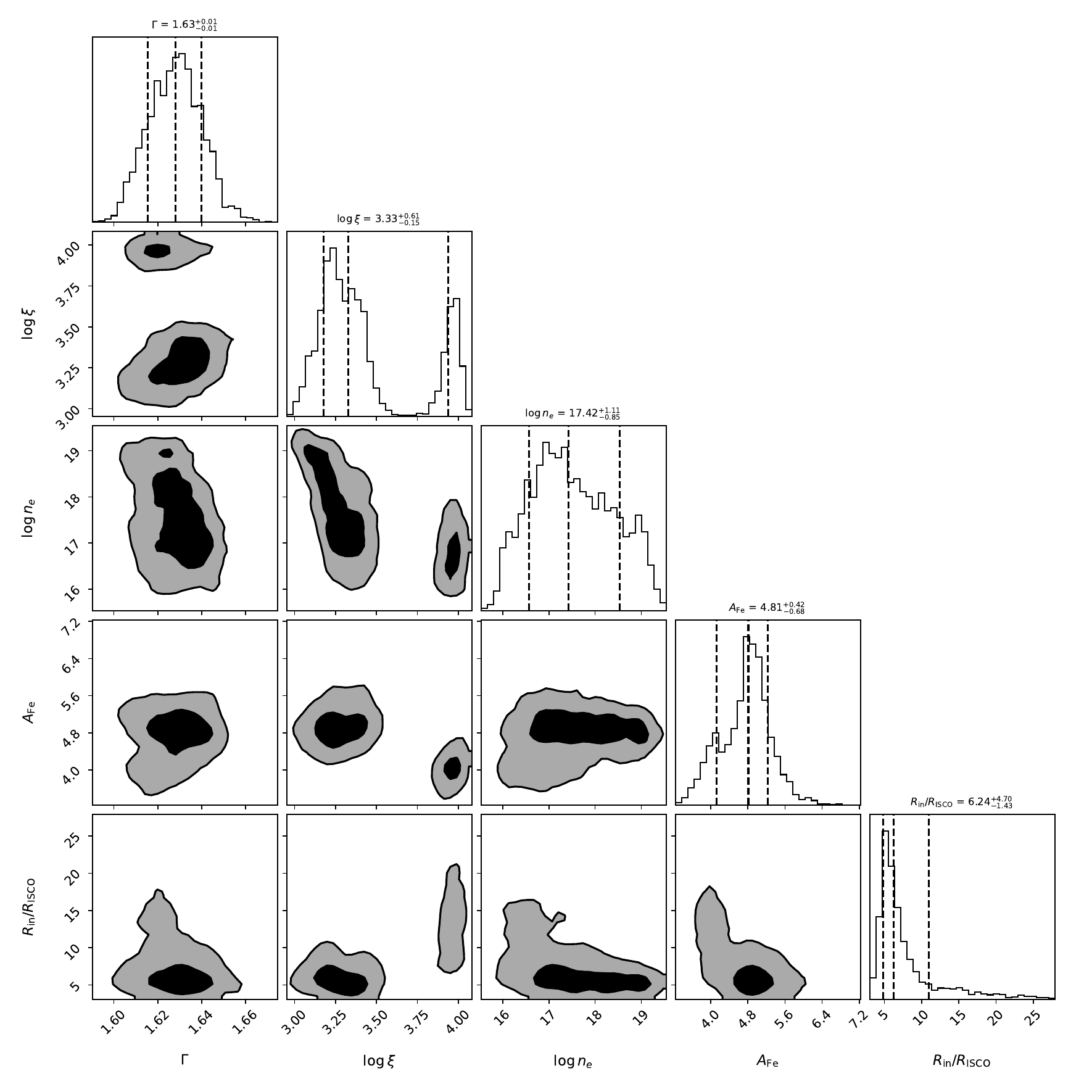}\\
{\footnotesize ObsID 91002320004 --- hard, $\Gamma=1.63$, $\chi^2_\nu=0.92$}
\end{minipage}\hfill\begin{minipage}[t]{0.48\textwidth}\centering
\includegraphics[width=\linewidth,height=0.27\textheight,keepaspectratio]{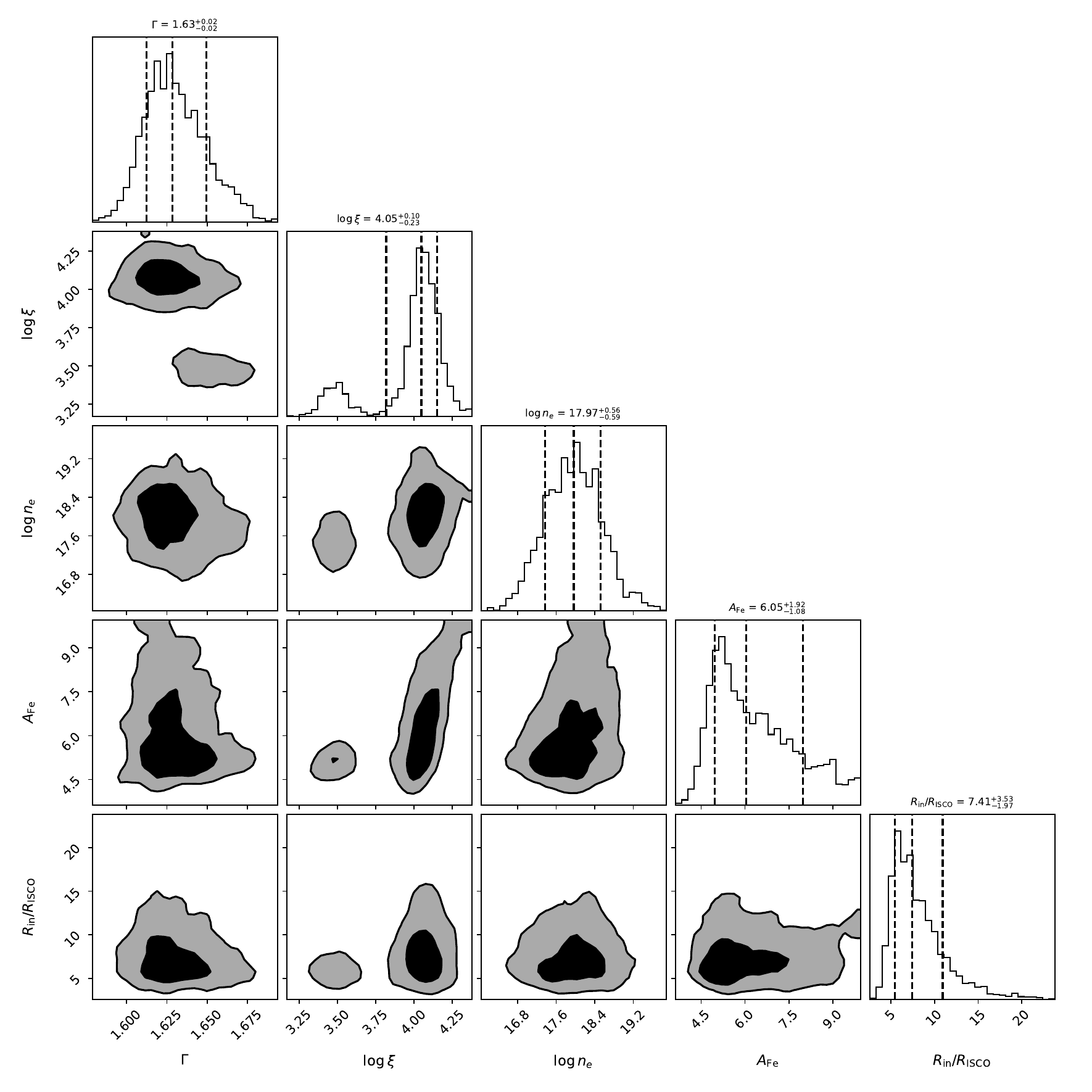}\\
{\footnotesize ObsID 30901039002 --- hard, $\Gamma=1.63$, $\chi^2_\nu=0.91$}
\end{minipage}
\caption{Posterior corner plots for the complete sample (1 of 7).}
\end{figure*}

\begin{figure*}[p]
\centering
\begin{minipage}[t]{0.48\textwidth}\centering
\includegraphics[width=\linewidth,height=0.27\textheight,keepaspectratio]{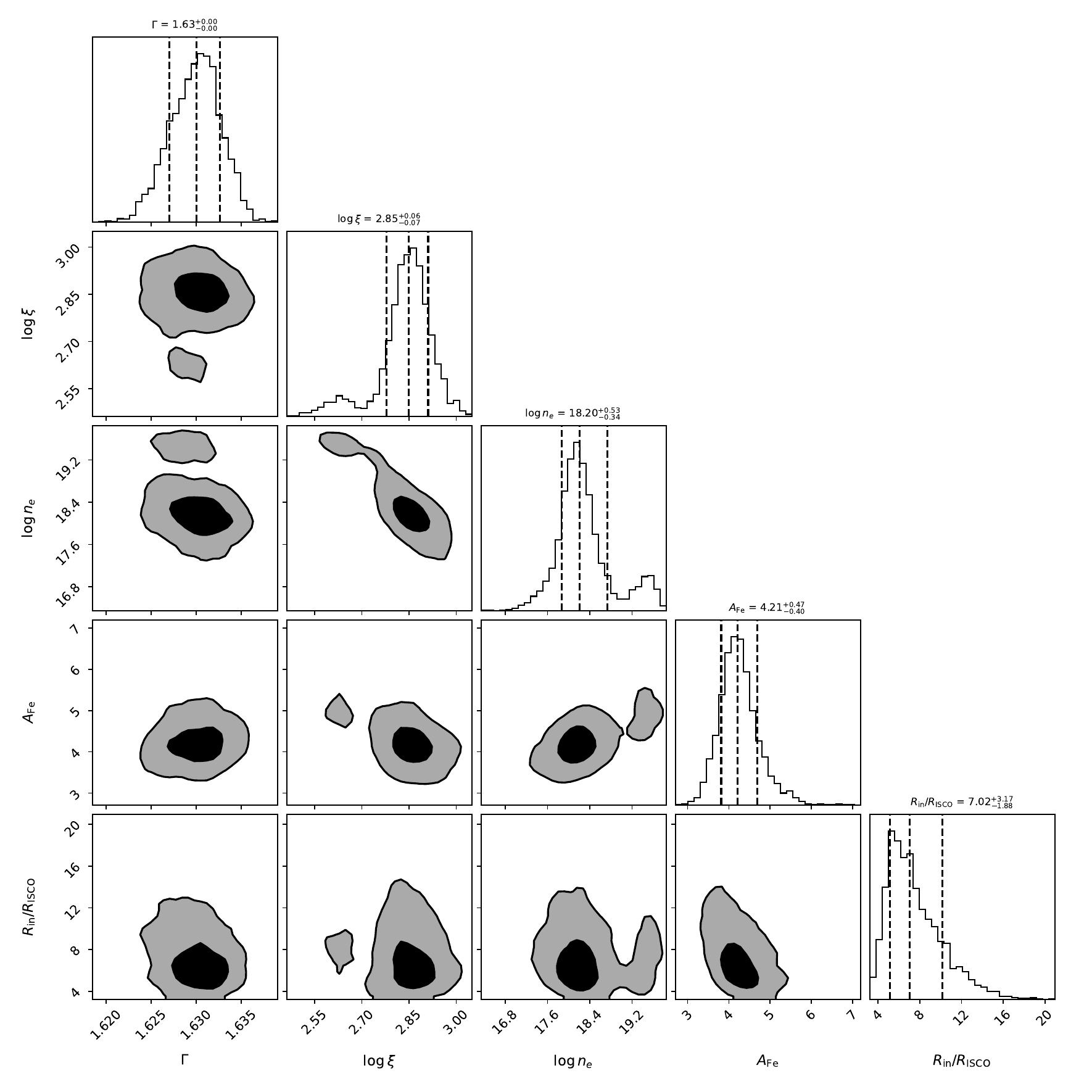}\\
{\footnotesize ObsID 30002150002 --- hard, $\Gamma=1.63$, $\chi^2_\nu=0.90$}
\end{minipage}\hfill\begin{minipage}[t]{0.48\textwidth}\centering
\includegraphics[width=\linewidth,height=0.27\textheight,keepaspectratio]{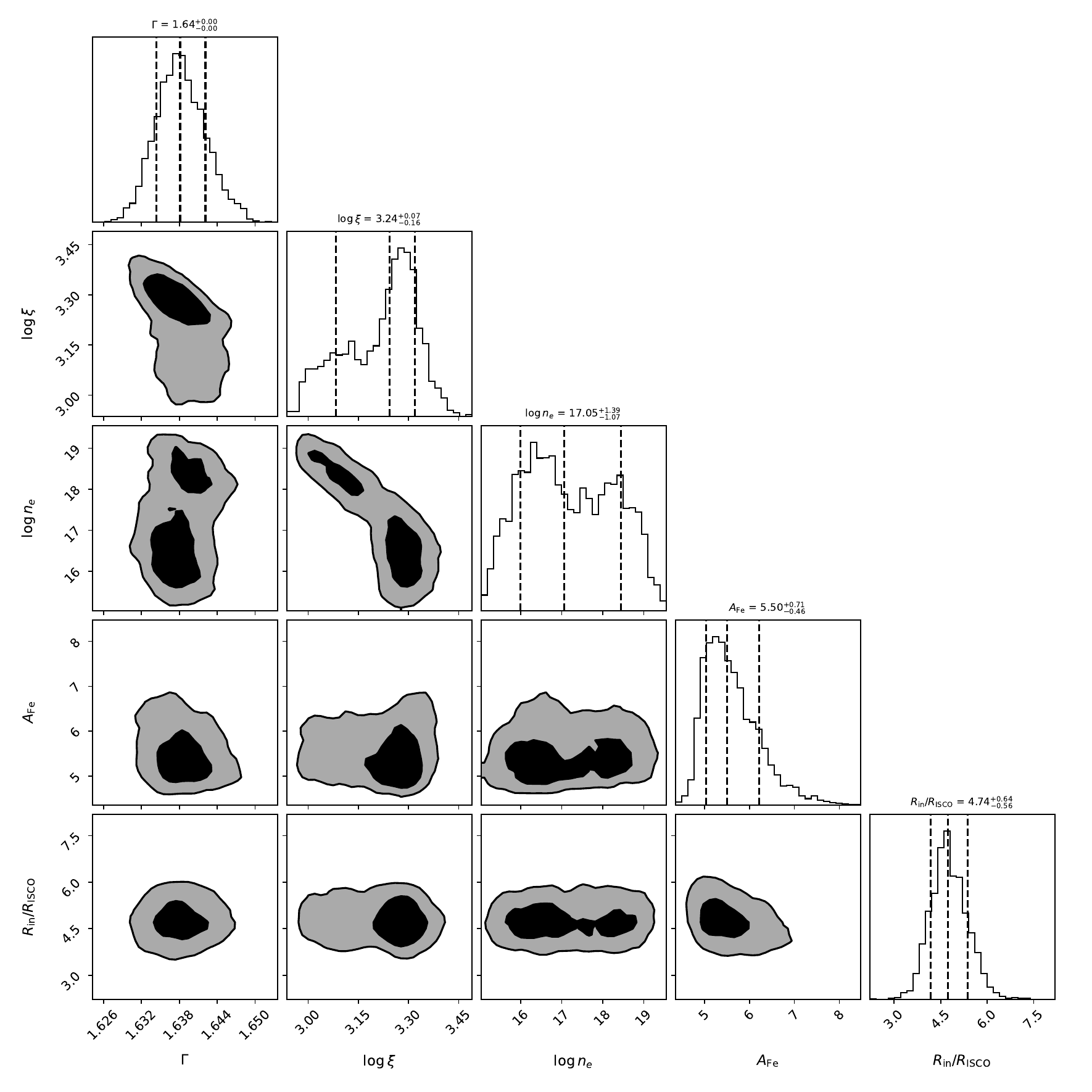}\\
{\footnotesize ObsID 90101020002 --- hard, $\Gamma=1.64$, $\chi^2_\nu=0.95$}
\end{minipage}\\[2mm]
\begin{minipage}[t]{0.48\textwidth}\centering
\includegraphics[width=\linewidth,height=0.27\textheight,keepaspectratio]{corners/30002150008}\\
{\footnotesize ObsID 30002150008 --- hard, $\Gamma=1.64$, $\chi^2_\nu=0.93$}
\end{minipage}\hfill\begin{minipage}[t]{0.48\textwidth}\centering
\includegraphics[width=\linewidth,height=0.27\textheight,keepaspectratio]{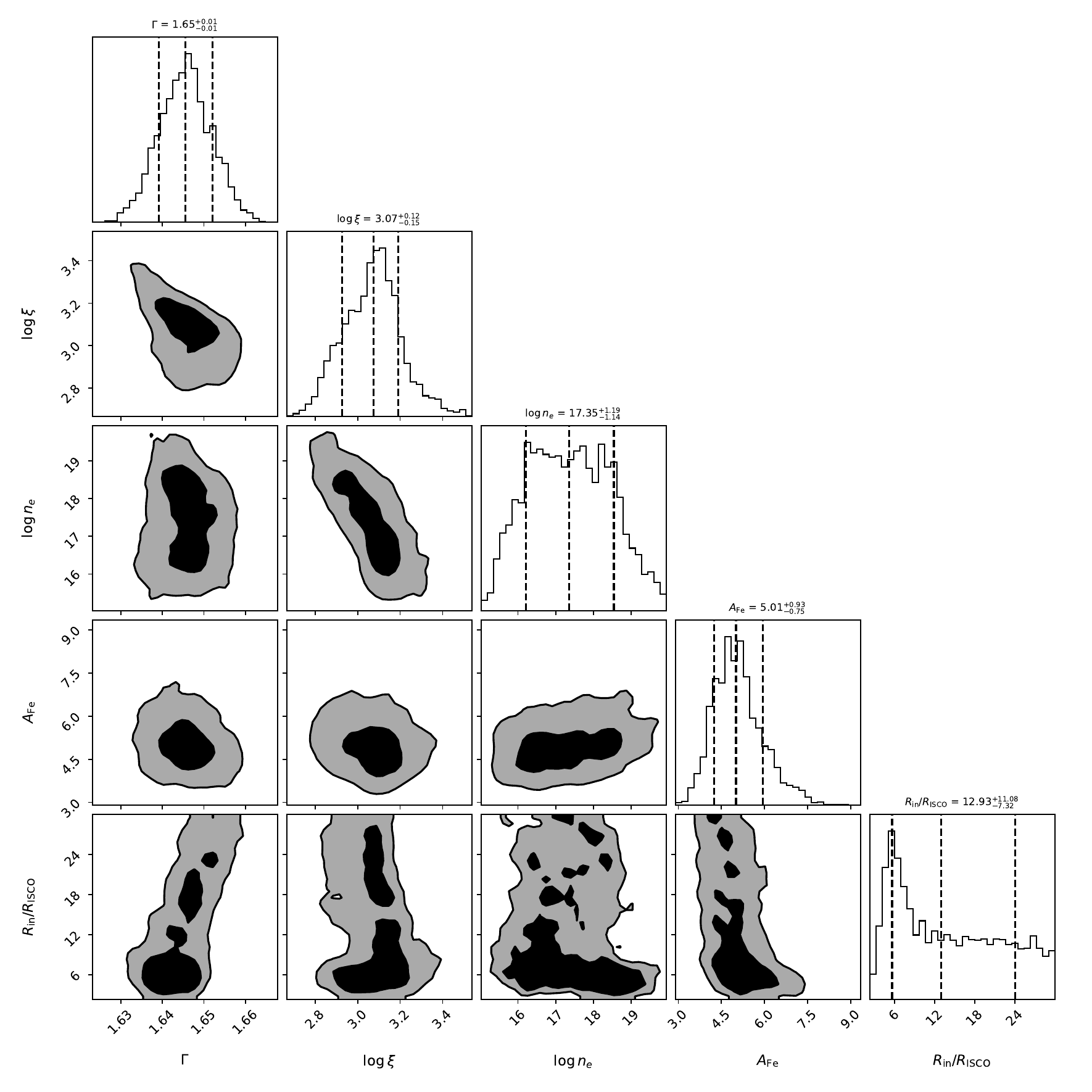}\\
{\footnotesize ObsID 30202032002 --- hard, $\Gamma=1.65$, $\chi^2_\nu=0.91$}
\end{minipage}
\caption{Posterior corner plots for the complete sample (2 of 7).}
\end{figure*}

\begin{figure*}[p]
\centering
\begin{minipage}[t]{0.48\textwidth}\centering
\includegraphics[width=\linewidth,height=0.27\textheight,keepaspectratio]{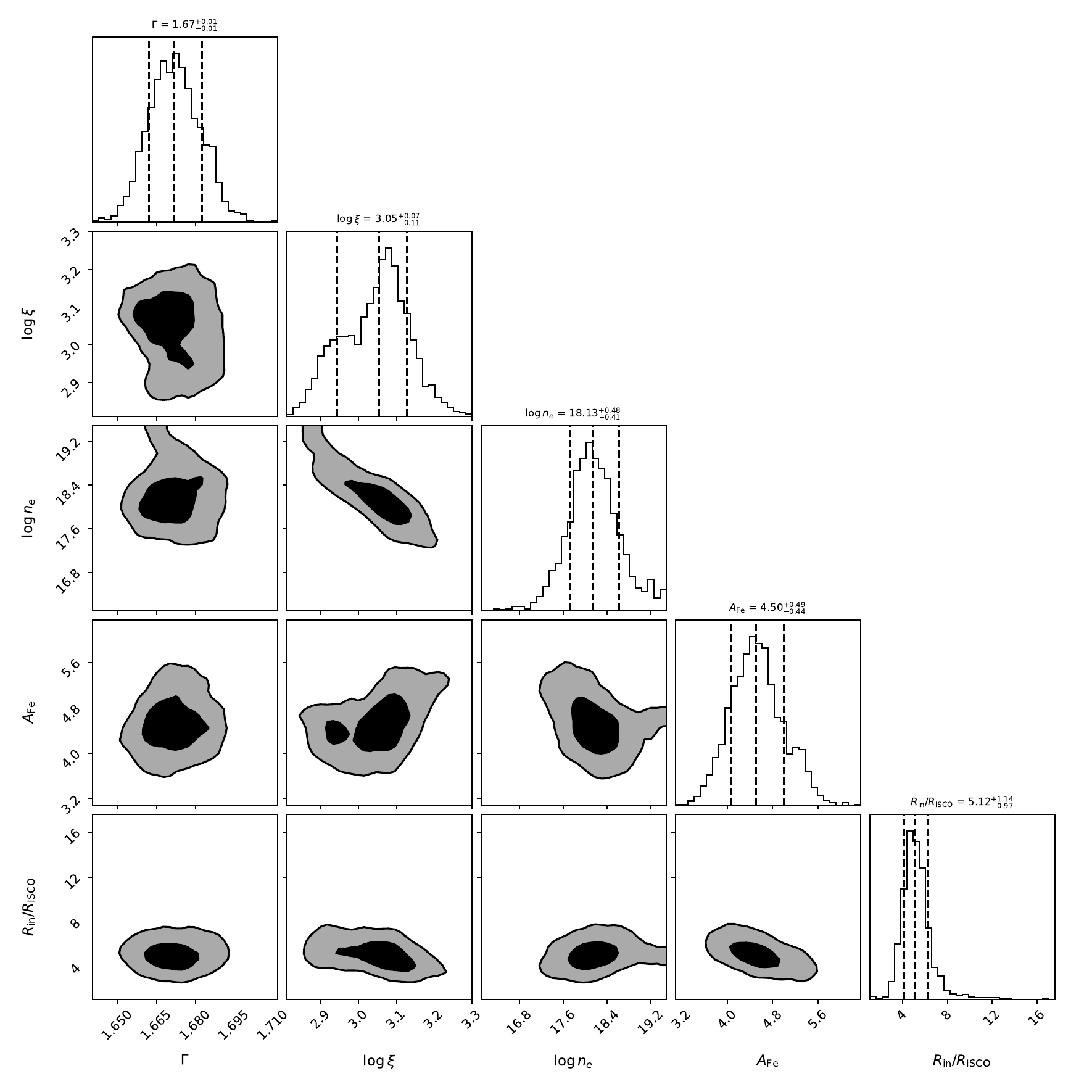}\\
{\footnotesize ObsID 90802013004 --- hard, $\Gamma=1.67$, $\chi^2_\nu=0.92$}
\end{minipage}\hfill\begin{minipage}[t]{0.48\textwidth}\centering
\includegraphics[width=\linewidth,height=0.27\textheight,keepaspectratio]{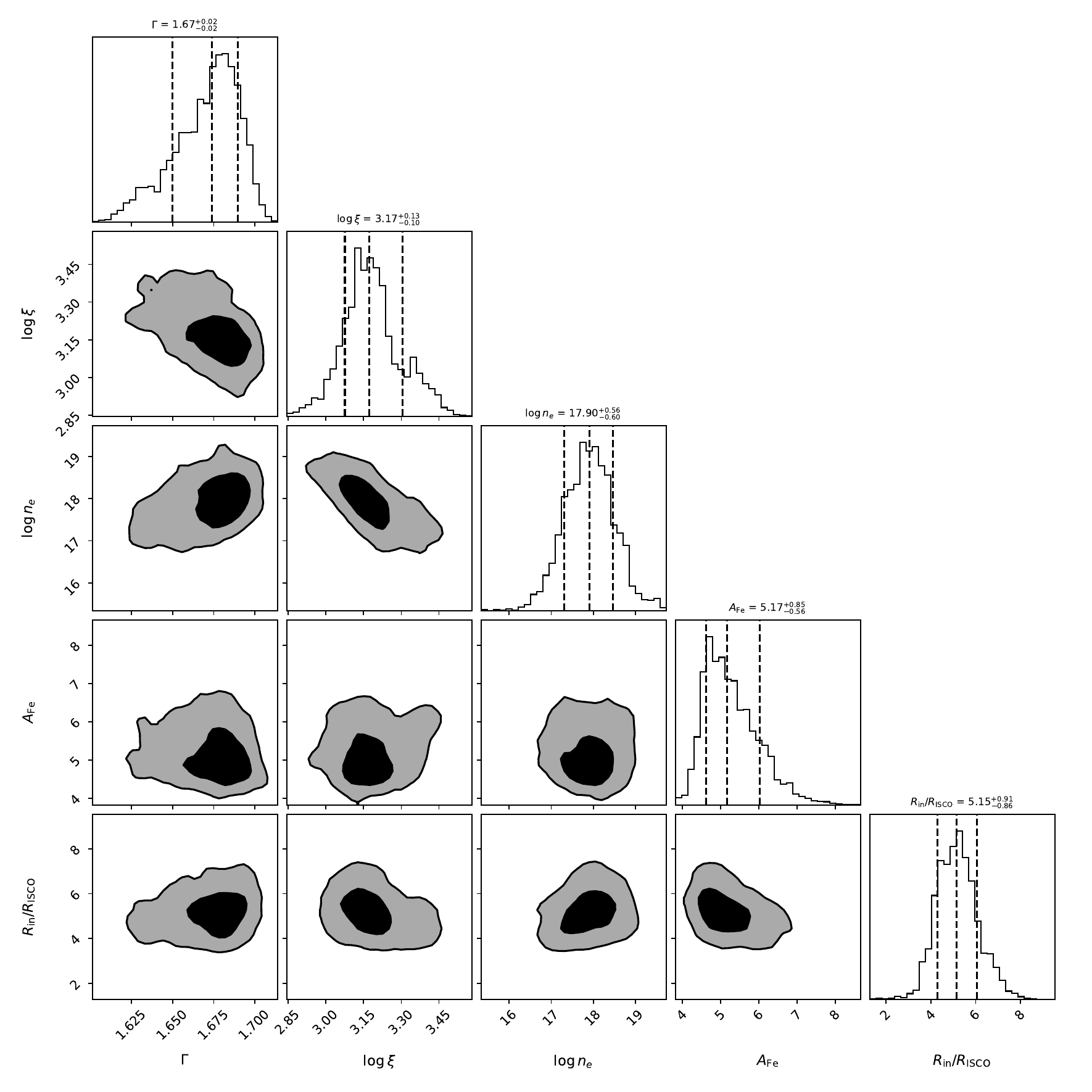}\\
{\footnotesize ObsID 30702017006 --- hard, $\Gamma=1.67$, $\chi^2_\nu=0.95$}
\end{minipage}\\[2mm]
\begin{minipage}[t]{0.48\textwidth}\centering
\includegraphics[width=\linewidth,height=0.27\textheight,keepaspectratio]{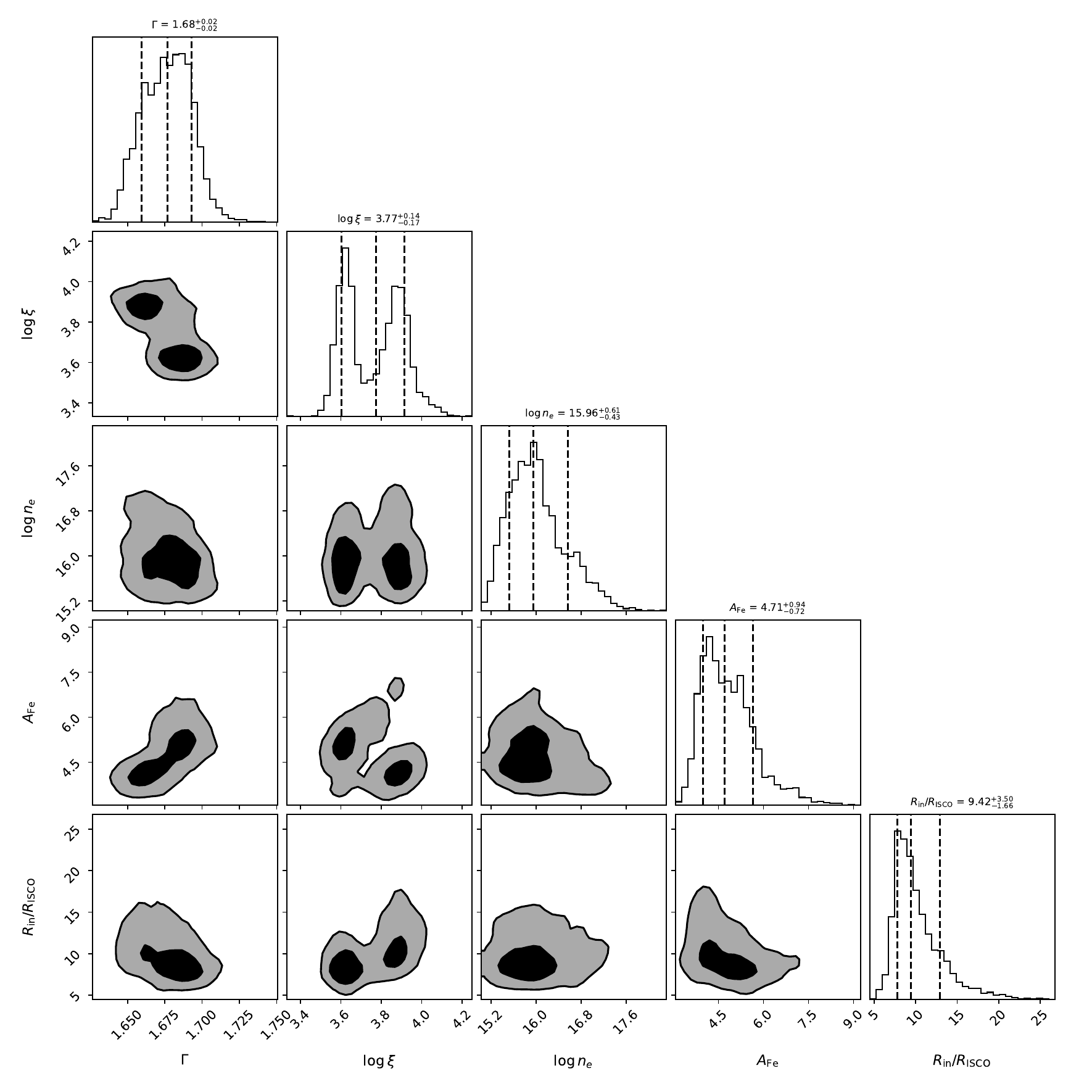}\\
{\footnotesize ObsID 80502335002 --- hard, $\Gamma=1.68$, $\chi^2_\nu=0.92$}
\end{minipage}\hfill\begin{minipage}[t]{0.48\textwidth}\centering
\includegraphics[width=\linewidth,height=0.27\textheight,keepaspectratio]{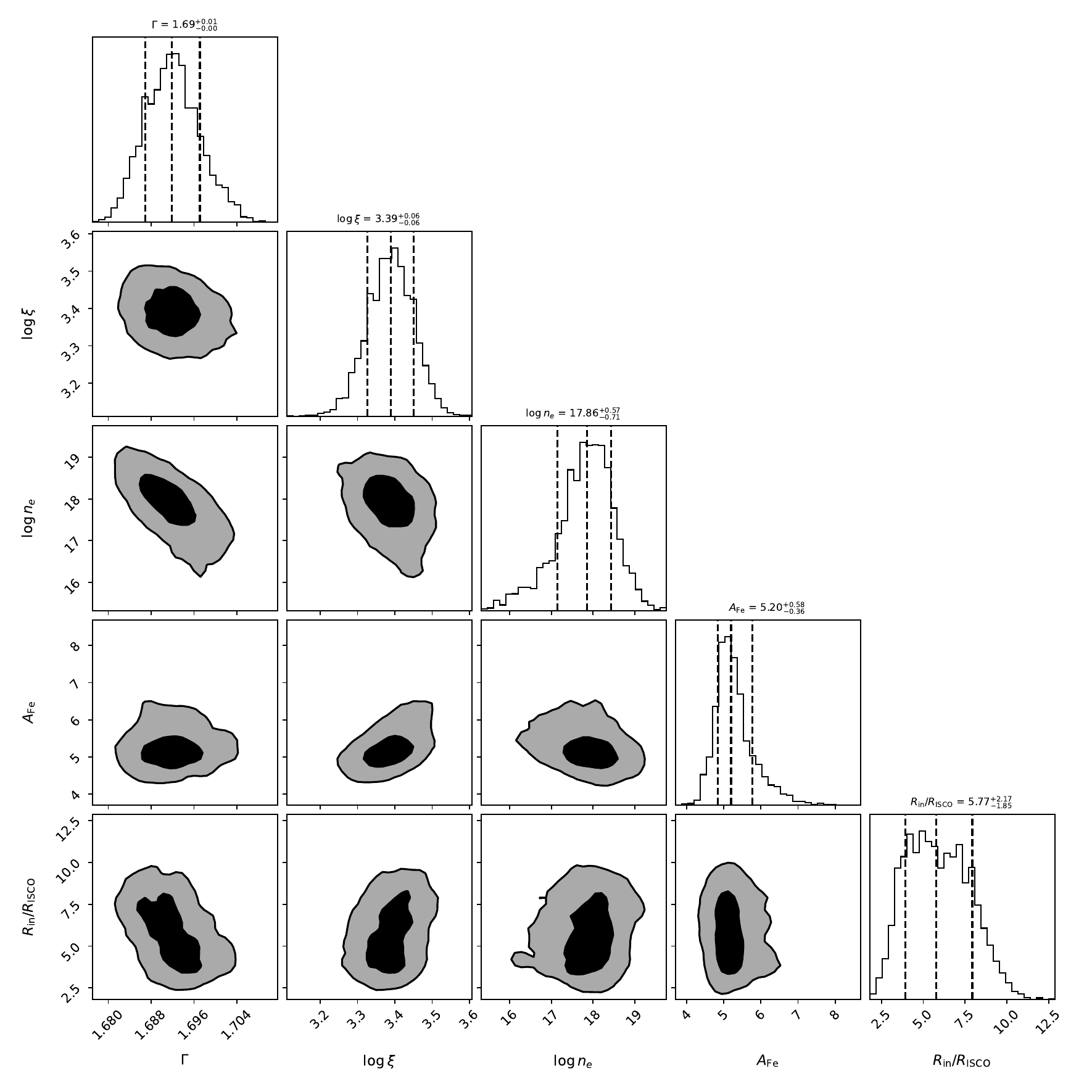}\\
{\footnotesize ObsID 30001011007 --- hard, $\Gamma=1.69$, $\chi^2_\nu=0.95$}
\end{minipage}
\caption{Posterior corner plots for the complete sample (3 of 7).}
\end{figure*}

\begin{figure*}[p]
\centering
\begin{minipage}[t]{0.48\textwidth}\centering
\includegraphics[width=\linewidth,height=0.27\textheight,keepaspectratio]{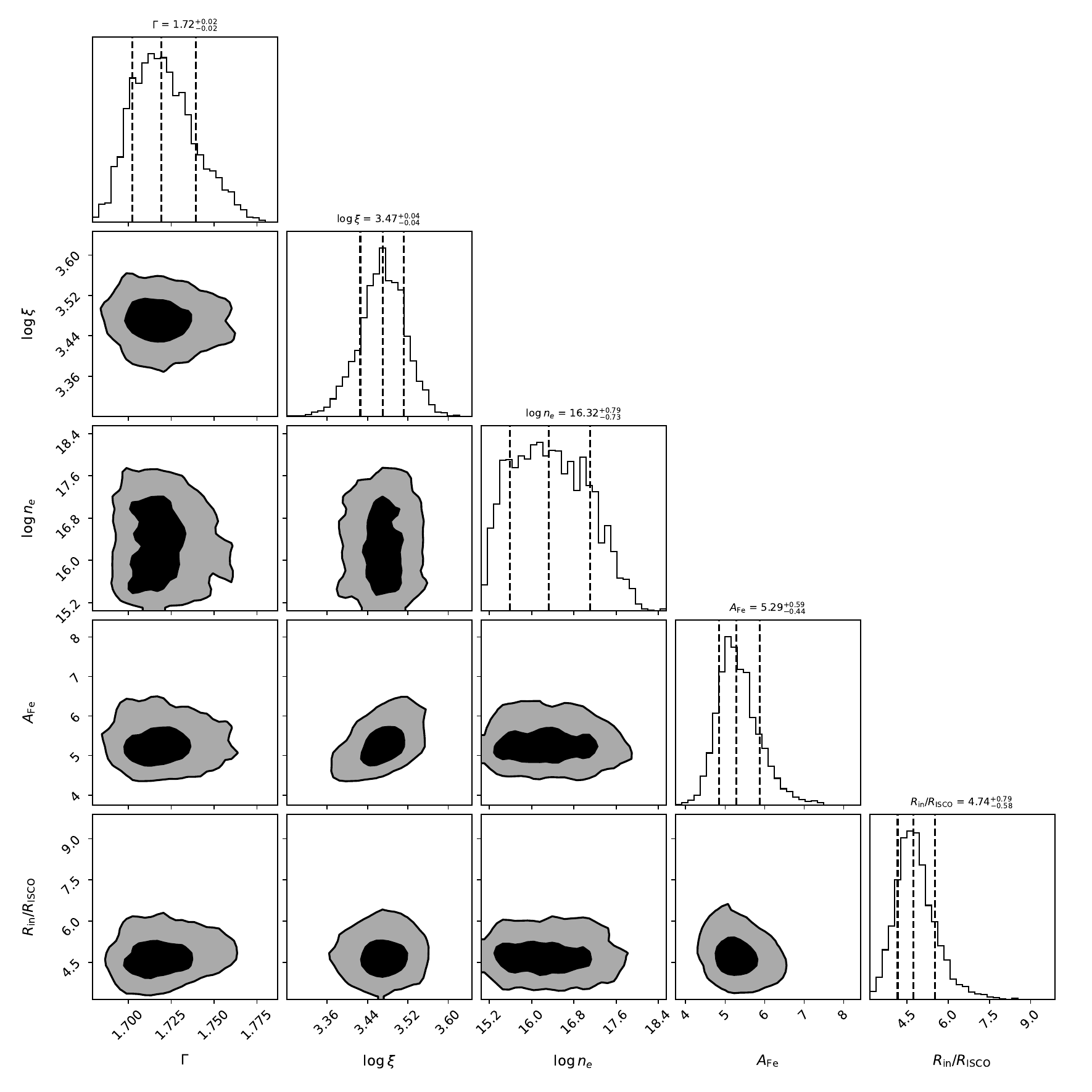}\\
{\footnotesize ObsID 30302019004 --- intermediate, $\Gamma=1.72$, $\chi^2_\nu=0.98$}
\end{minipage}\hfill\begin{minipage}[t]{0.48\textwidth}\centering
\includegraphics[width=\linewidth,height=0.27\textheight,keepaspectratio]{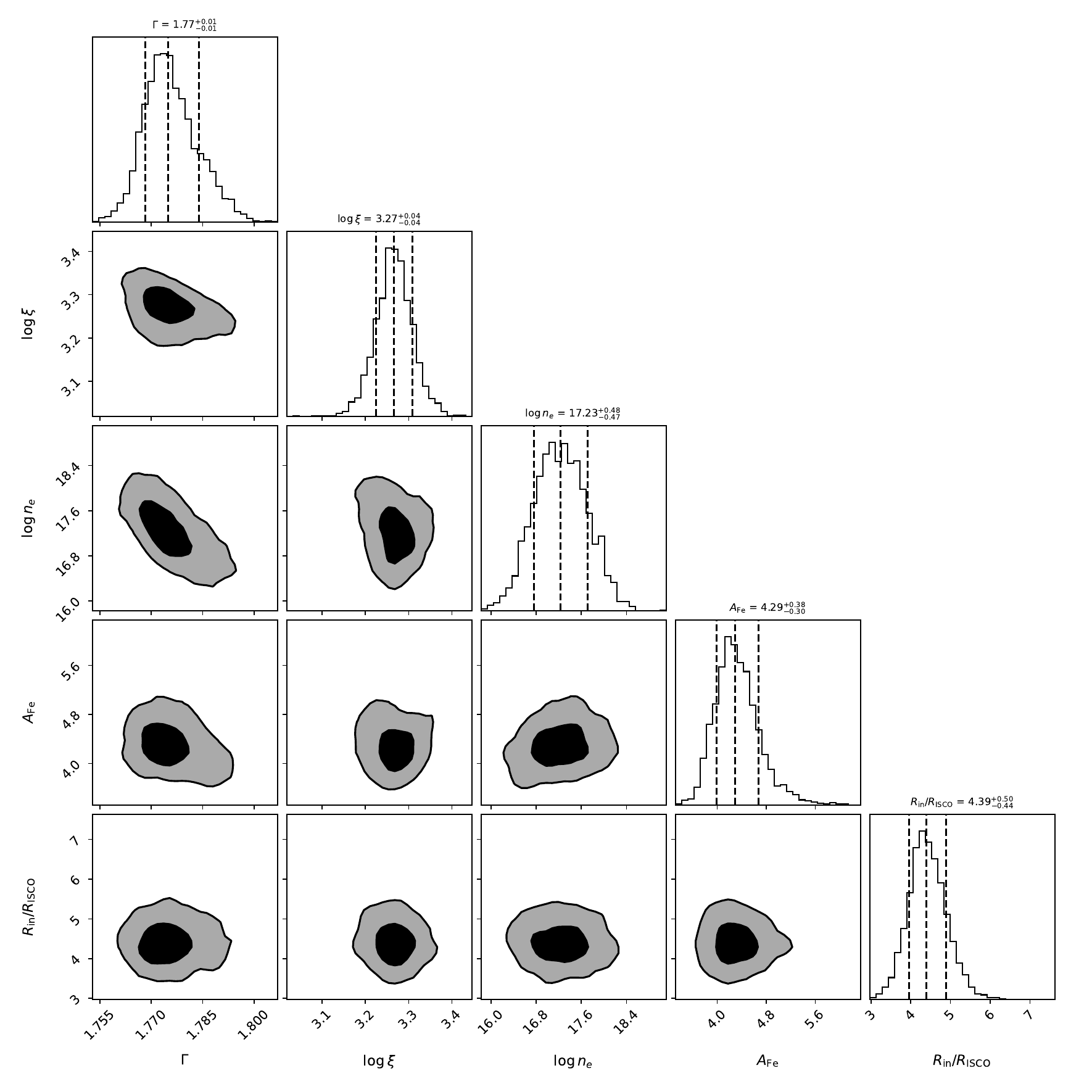}\\
{\footnotesize ObsID 30001011005 --- intermediate, $\Gamma=1.77$, $\chi^2_\nu=0.91$}
\end{minipage}\\[2mm]
\begin{minipage}[t]{0.48\textwidth}\centering
\includegraphics[width=\linewidth,height=0.27\textheight,keepaspectratio]{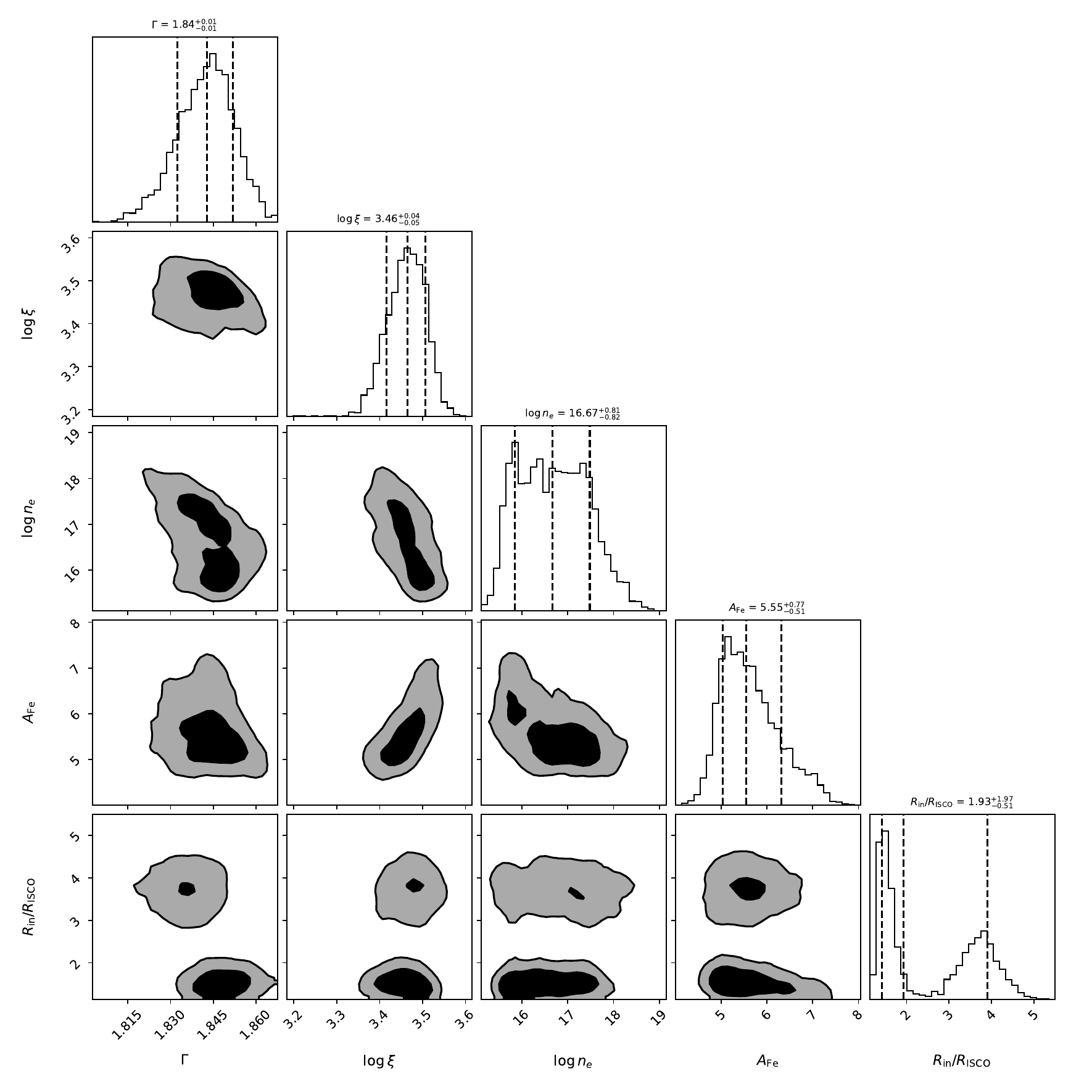}\\
{\footnotesize ObsID 30101022002 --- intermediate, $\Gamma=1.84$, $\chi^2_\nu=0.90$}
\end{minipage}\hfill\begin{minipage}[t]{0.48\textwidth}\centering
\includegraphics[width=\linewidth,height=0.27\textheight,keepaspectratio]{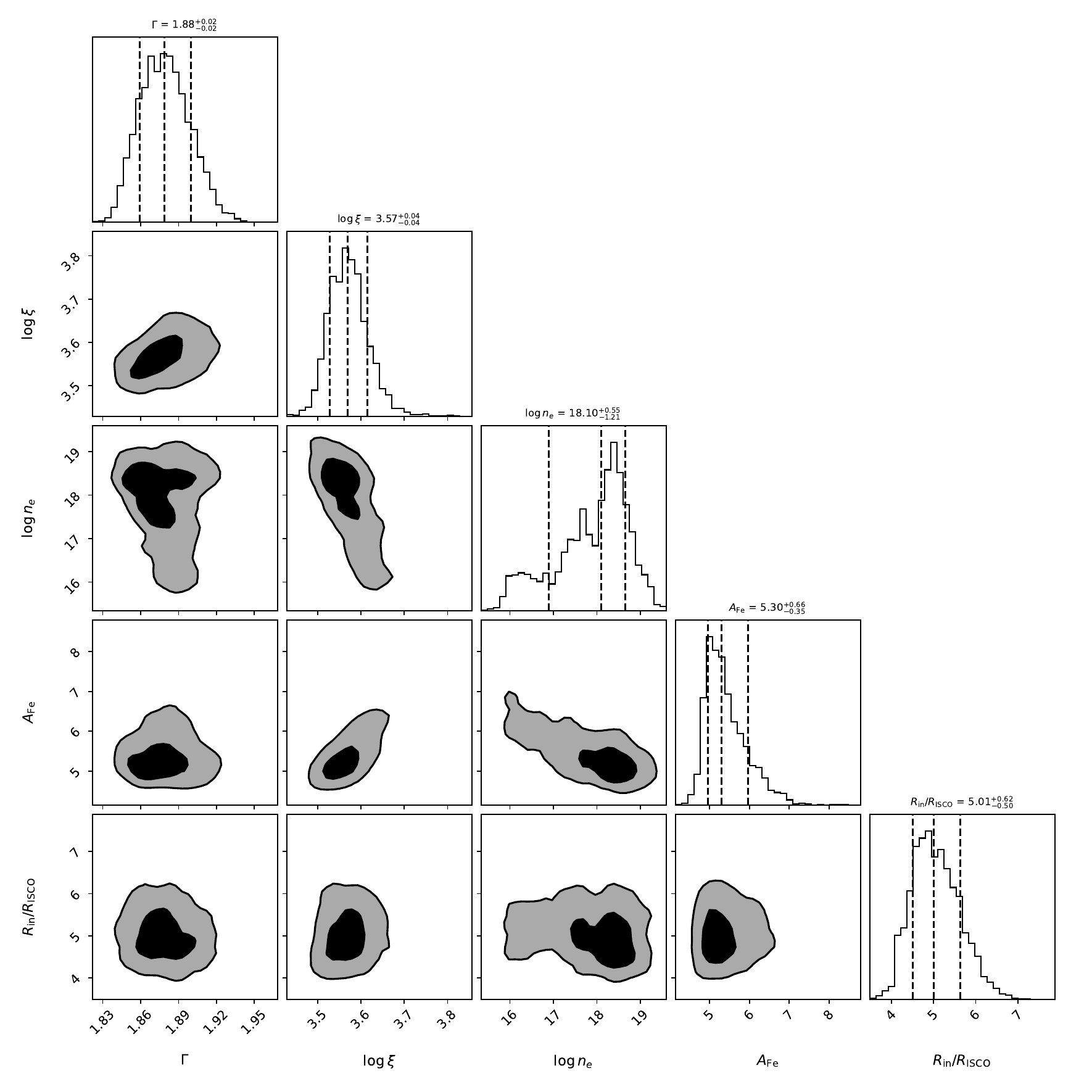}\\
{\footnotesize ObsID 30302019006 --- intermediate, $\Gamma=1.88$, $\chi^2_\nu=0.91$}
\end{minipage}
\caption{Posterior corner plots for the complete sample (4 of 7).}
\end{figure*}

\begin{figure*}[p]
\centering
\begin{minipage}[t]{0.48\textwidth}\centering
\includegraphics[width=\linewidth,height=0.27\textheight,keepaspectratio]{corners/80902318002}\\
{\footnotesize ObsID 80902318002 --- intermediate, $\Gamma=1.95$, $\chi^2_\nu=0.95$}
\end{minipage}\hfill\begin{minipage}[t]{0.48\textwidth}\centering
\includegraphics[width=\linewidth,height=0.27\textheight,keepaspectratio]{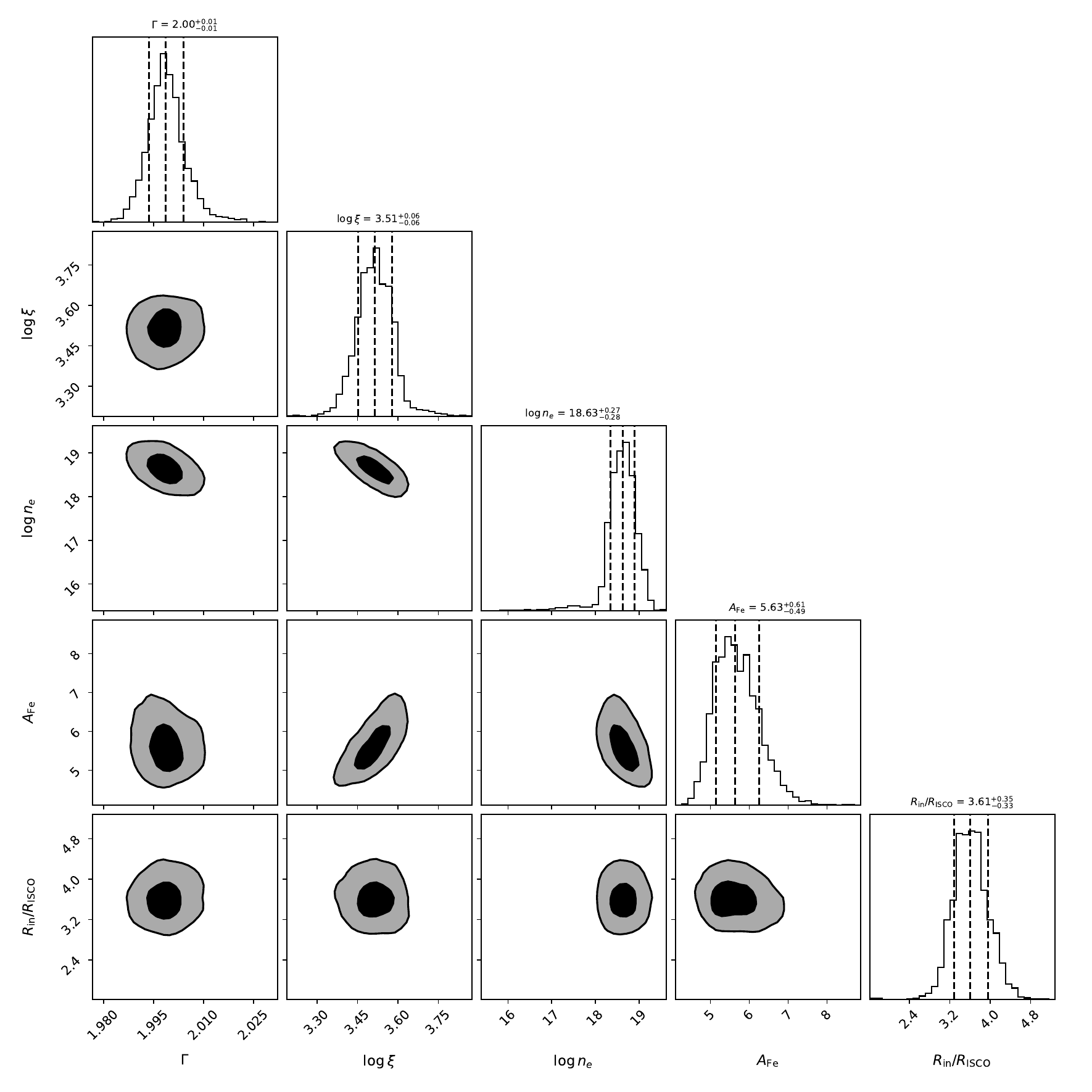}\\
{\footnotesize ObsID 30001011011 --- intermediate, $\Gamma=2.00$, $\chi^2_\nu=0.89$}
\end{minipage}\\[2mm]
\begin{minipage}[t]{0.48\textwidth}\centering
\includegraphics[width=\linewidth,height=0.27\textheight,keepaspectratio]{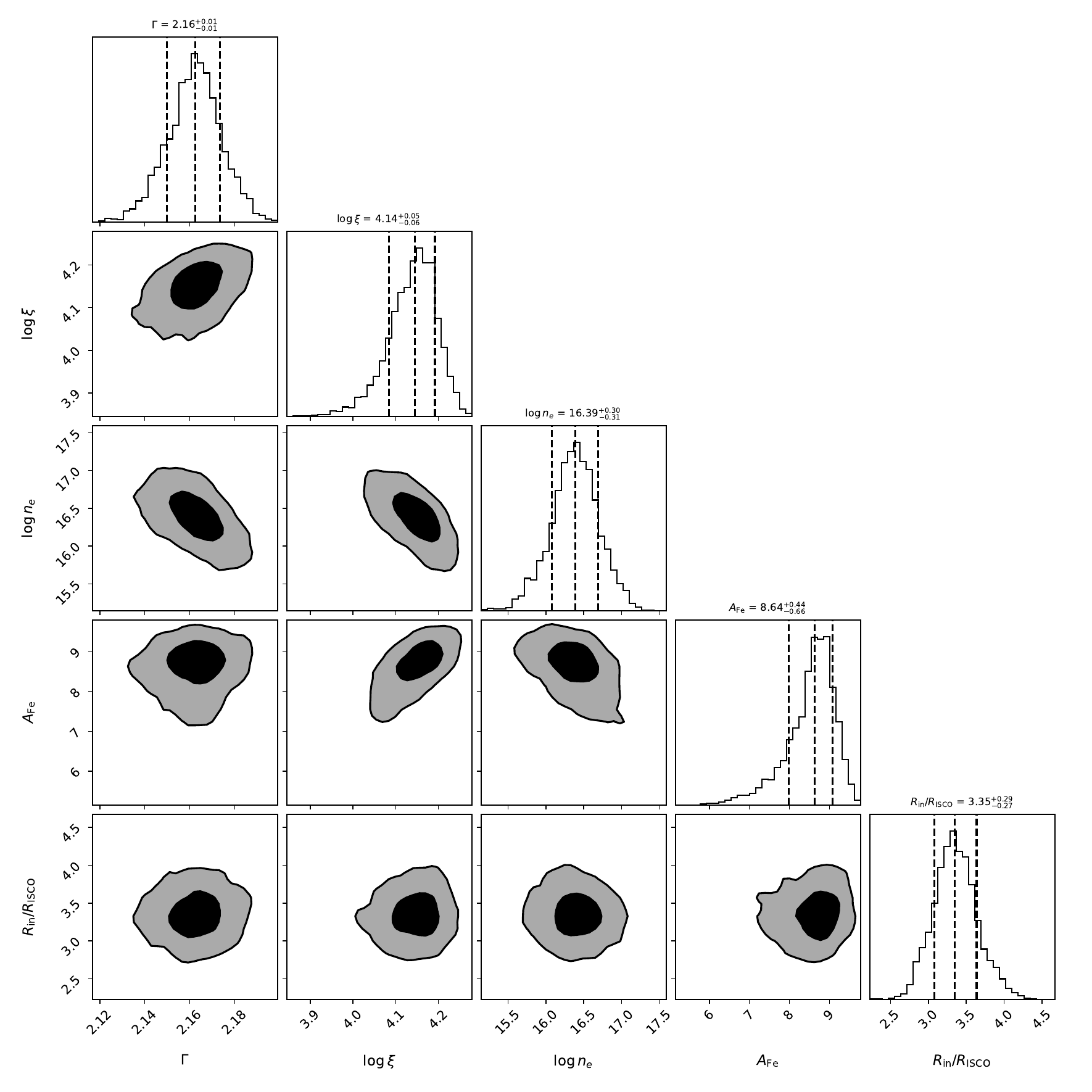}\\
{\footnotesize ObsID 80902318004 --- soft, $\Gamma=2.16$, $\chi^2_\nu=0.95$}
\end{minipage}\hfill\begin{minipage}[t]{0.48\textwidth}\centering
\includegraphics[width=\linewidth,height=0.27\textheight,keepaspectratio]{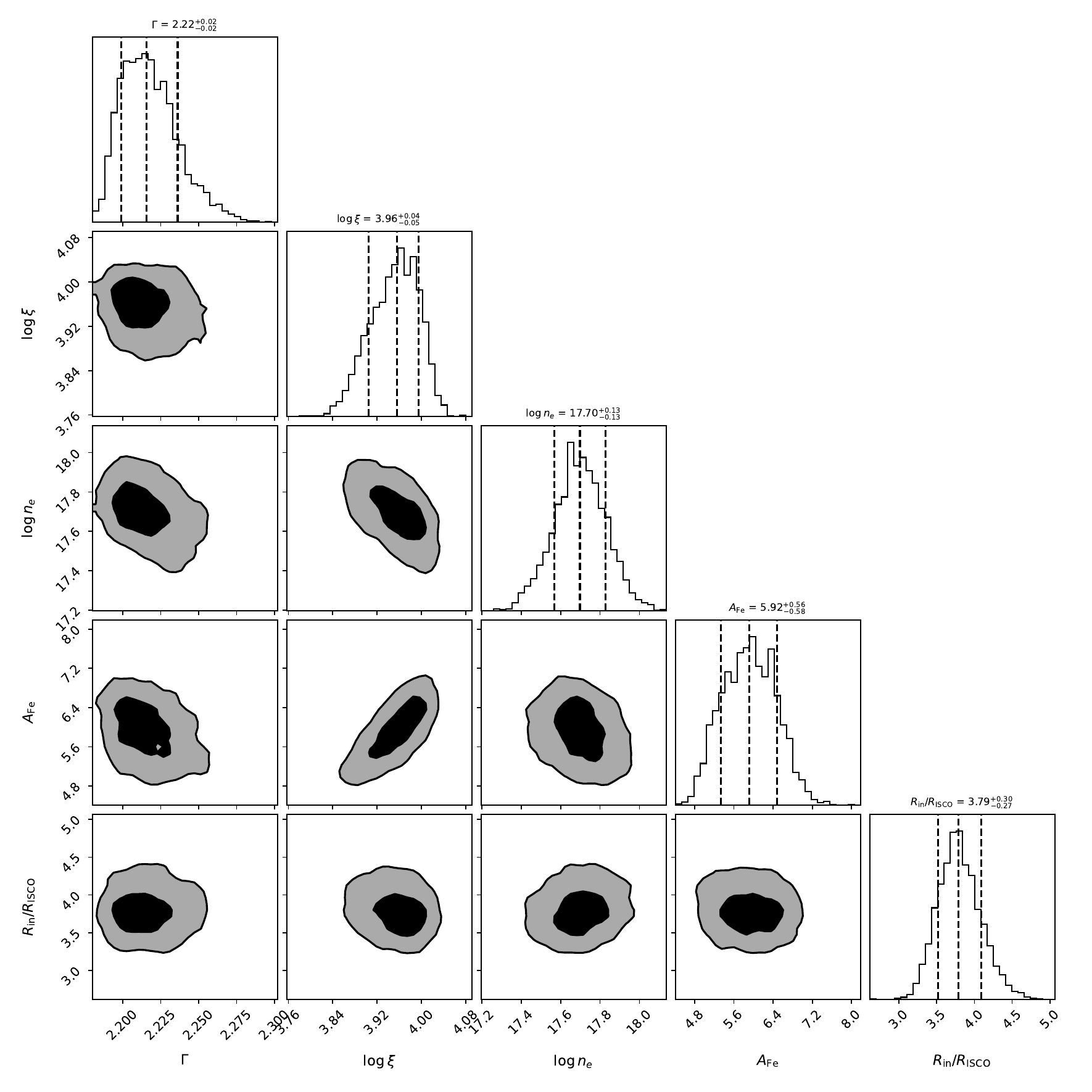}\\
{\footnotesize ObsID 80502335006 --- soft, $\Gamma=2.22$, $\chi^2_\nu=1.04$}
\end{minipage}
\caption{Posterior corner plots for the complete sample (5 of 7).}
\end{figure*}

\begin{figure*}[p]
\centering
\begin{minipage}[t]{0.48\textwidth}\centering
\includegraphics[width=\linewidth,height=0.27\textheight,keepaspectratio]{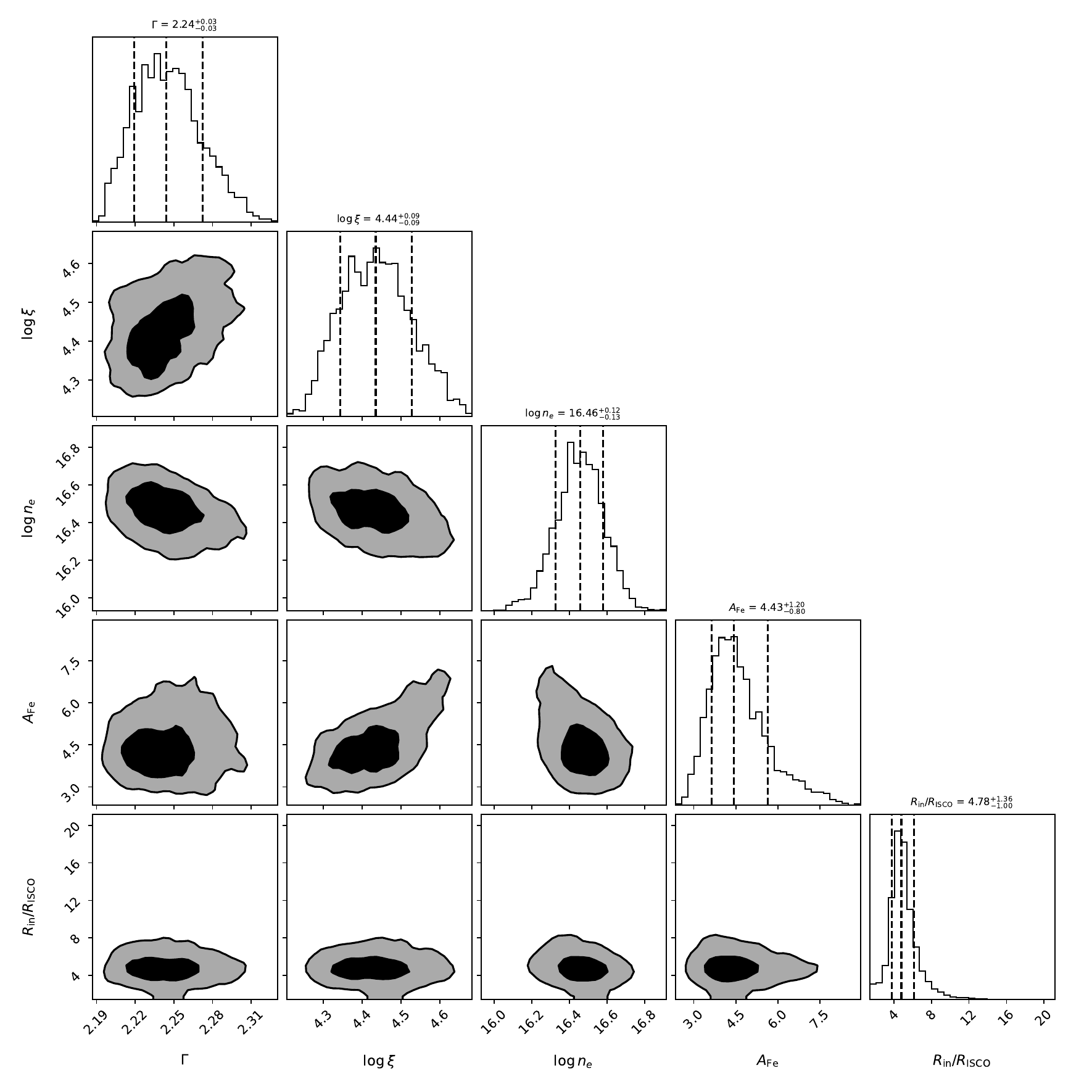}\\
{\footnotesize ObsID 30302019010 --- soft, $\Gamma=2.24$, $\chi^2_\nu=0.98$}
\end{minipage}\hfill\begin{minipage}[t]{0.48\textwidth}\centering
\includegraphics[width=\linewidth,height=0.27\textheight,keepaspectratio]{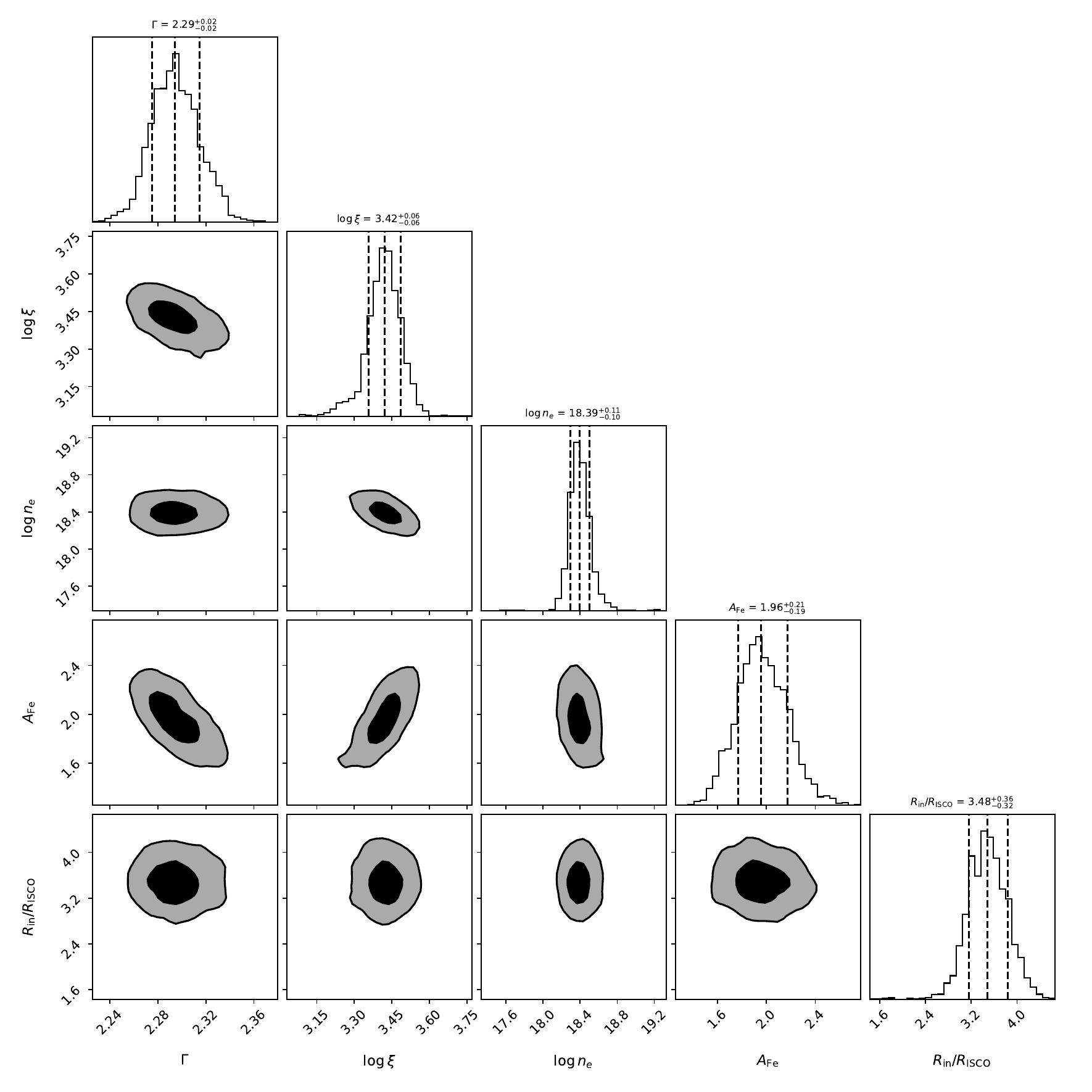}\\
{\footnotesize ObsID 30302019002 --- soft, $\Gamma=2.29$, $\chi^2_\nu=0.89$}
\end{minipage}\\[2mm]
\begin{minipage}[t]{0.48\textwidth}\centering
\includegraphics[width=\linewidth,height=0.27\textheight,keepaspectratio]{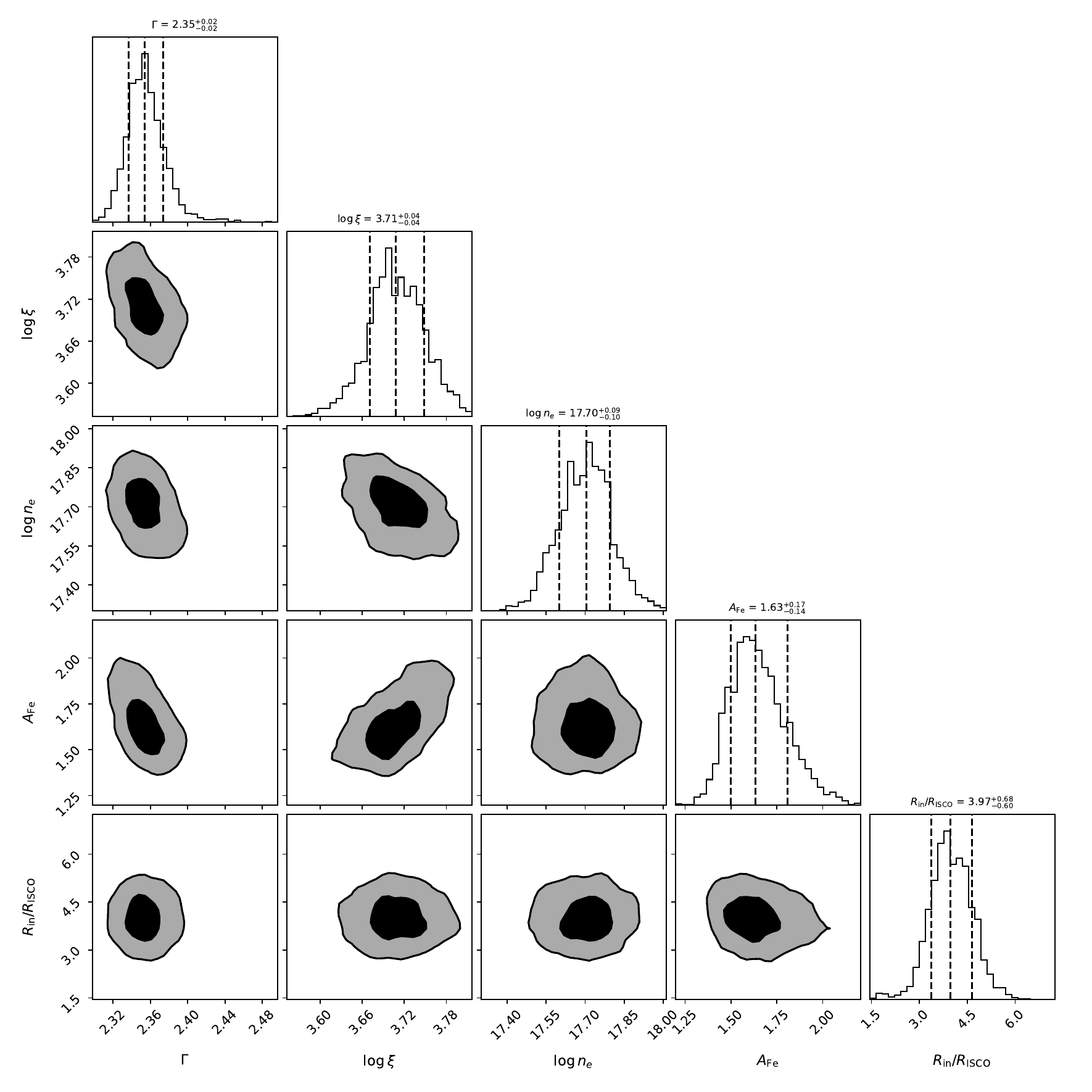}\\
{\footnotesize ObsID 30302019012 --- soft, $\Gamma=2.35$, $\chi^2_\nu=0.91$}
\end{minipage}\hfill\begin{minipage}[t]{0.48\textwidth}\centering
\includegraphics[width=\linewidth,height=0.27\textheight,keepaspectratio]{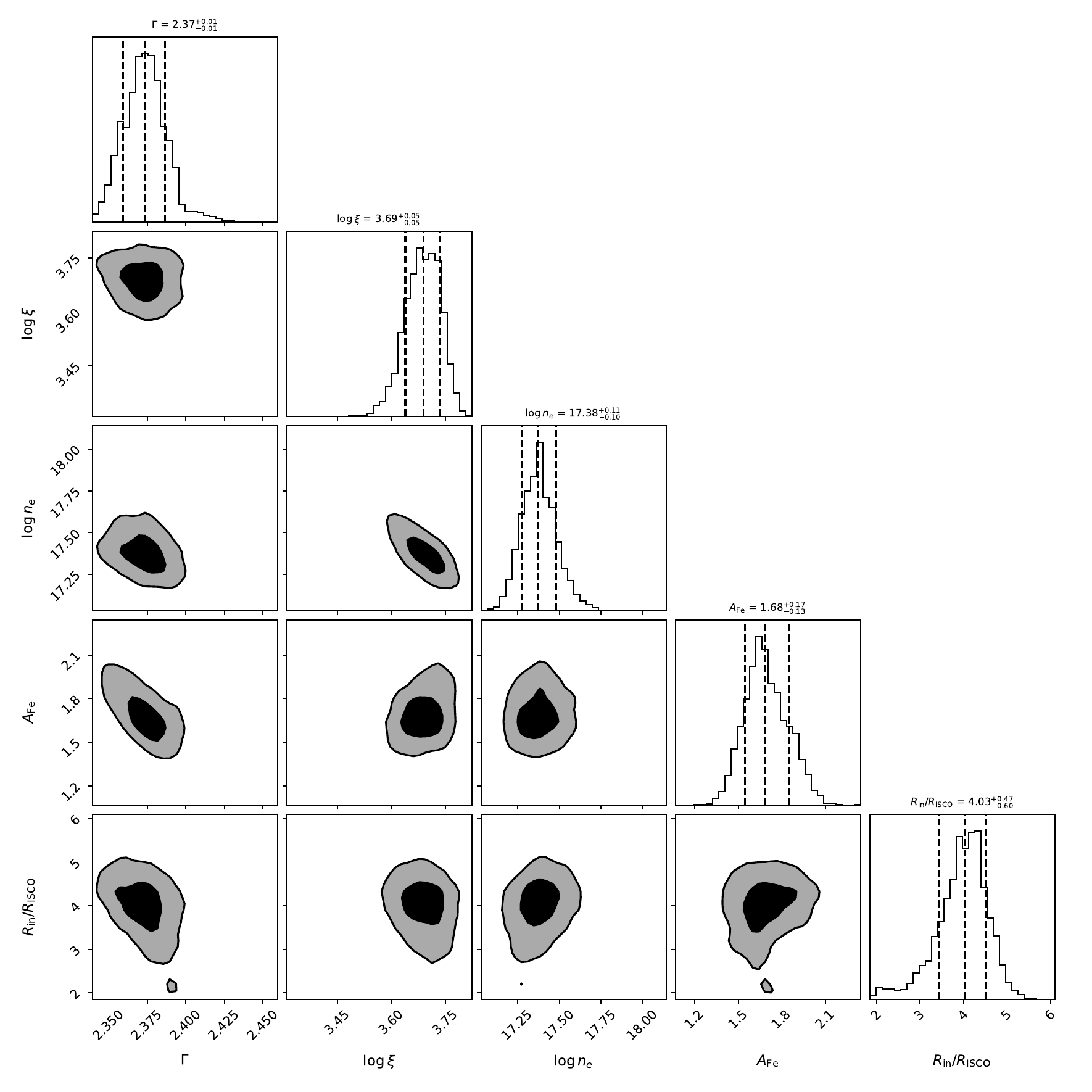}\\
{\footnotesize ObsID 30001011009 --- soft, $\Gamma=2.37$, $\chi^2_\nu=0.87$}
\end{minipage}
\caption{Posterior corner plots for the complete sample (6 of 7).}
\end{figure*}

\begin{figure*}[p]
\centering
\begin{minipage}[t]{0.48\textwidth}\centering
\includegraphics[width=\linewidth,height=0.27\textheight,keepaspectratio]{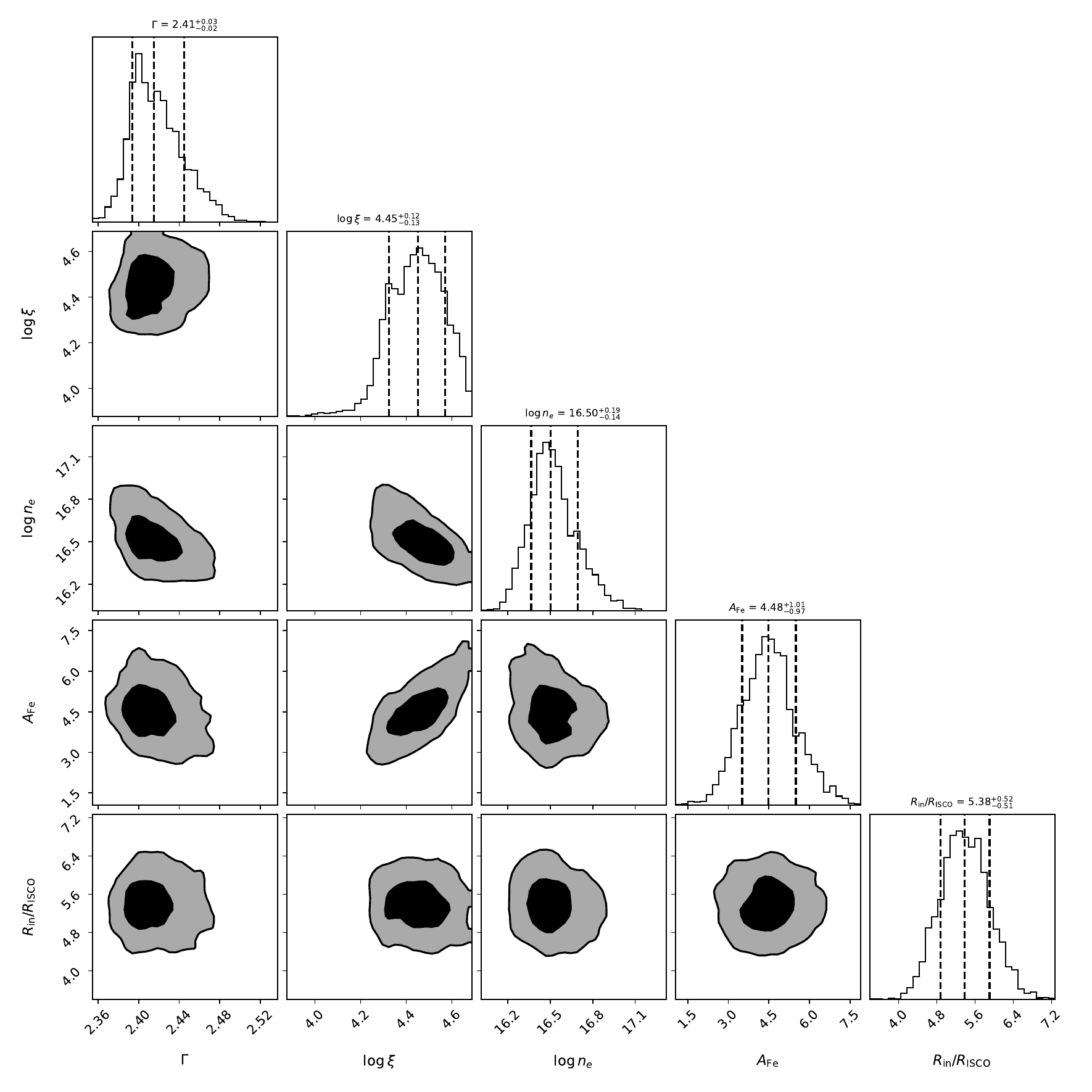}\\
{\footnotesize ObsID 10014001001 --- soft, $\Gamma=2.41$, $\chi^2_\nu=1.10$}
\end{minipage}\hfill\begin{minipage}[t]{0.48\textwidth}\centering
\includegraphics[width=\linewidth,height=0.27\textheight,keepaspectratio]{corners/30001011002}\\
{\footnotesize ObsID 30001011002 --- soft, $\Gamma=2.48$, $\chi^2_\nu=0.95$}
\end{minipage}
\caption{Posterior corner plots for the complete sample (7 of 7).}
\end{figure*}

\bibliographystyle{aasjournalv7}
\bibliography{references}

\end{document}